\documentclass[twocolumn,tighten]{aastex61}
\usepackage[]{txfonts}
\usepackage{natbib, twoopt}
\usepackage{float}
\usepackage{graphicx}
\usepackage{epstopdf}
\usepackage{enumitem}
\usepackage[caption=false]{subfig}
\usepackage{color}\definecolor{AliceBlue}{rgb}{0.94,0.97,1.00}
\definecolor{AntiqueWhite1}{rgb}{1.00,0.94,0.86}
\definecolor{AntiqueWhite2}{rgb}{0.93,0.87,0.80}
\definecolor{AntiqueWhite3}{rgb}{0.80,0.75,0.69}
\definecolor{AntiqueWhite4}{rgb}{0.55,0.51,0.47}
\definecolor{AntiqueWhite}{rgb}{0.98,0.92,0.84}
\definecolor{BlanchedAlmond}{rgb}{1.00,0.92,0.80}
\definecolor{BlueViolet}{rgb}{0.54,0.17,0.89}
\definecolor{CadetBlue1}{rgb}{0.60,0.96,1.00}
\definecolor{CadetBlue2}{rgb}{0.56,0.90,0.93}
\definecolor{CadetBlue3}{rgb}{0.48,0.77,0.80}
\definecolor{CadetBlue4}{rgb}{0.33,0.53,0.55}
\definecolor{CadetBlue}{rgb}{0.37,0.62,0.63}
\definecolor{CornflowerBlue}{rgb}{0.39,0.58,0.93}
\definecolor{DarkBlue}{rgb}{0.00,0.00,0.55}
\definecolor{DarkCyan}{rgb}{0.00,0.55,0.55}
\definecolor{DarkGoldenrod1}{rgb}{1.00,0.73,0.06}
\definecolor{DarkGoldenrod2}{rgb}{0.93,0.68,0.05}
\definecolor{DarkGoldenrod3}{rgb}{0.80,0.58,0.05}
\definecolor{DarkGoldenrod4}{rgb}{0.55,0.40,0.03}
\definecolor{DarkGoldenrod}{rgb}{0.72,0.53,0.04}
\definecolor{DarkGray}{rgb}{0.66,0.66,0.66}
\definecolor{DarkGreen}{rgb}{0.00,0.39,0.00}
\definecolor{DarkGrey}{rgb}{0.66,0.66,0.66}
\definecolor{DarkKhaki}{rgb}{0.74,0.72,0.42}
\definecolor{DarkMagenta}{rgb}{0.55,0.00,0.55}
\definecolor{DarkOliveGreen1}{rgb}{0.79,1.00,0.44}
\definecolor{DarkOliveGreen2}{rgb}{0.74,0.93,0.41}
\definecolor{DarkOliveGreen3}{rgb}{0.64,0.80,0.35}
\definecolor{DarkOliveGreen4}{rgb}{0.43,0.55,0.24}
\definecolor{DarkOliveGreen}{rgb}{0.33,0.42,0.18}
\definecolor{DarkOrange1}{rgb}{1.00,0.50,0.00}
\definecolor{DarkOrange2}{rgb}{0.93,0.46,0.00}
\definecolor{DarkOrange3}{rgb}{0.80,0.40,0.00}
\definecolor{DarkOrange4}{rgb}{0.55,0.27,0.00}
\definecolor{DarkOrange}{rgb}{1.00,0.55,0.00}
\definecolor{DarkOrchid1}{rgb}{0.75,0.24,1.00}
\definecolor{DarkOrchid2}{rgb}{0.70,0.23,0.93}
\definecolor{DarkOrchid3}{rgb}{0.60,0.20,0.80}
\definecolor{DarkOrchid4}{rgb}{0.41,0.13,0.55}
\definecolor{DarkOrchid}{rgb}{0.60,0.20,0.80}
\definecolor{DarkRed}{rgb}{0.55,0.00,0.00}
\definecolor{DarkSalmon}{rgb}{0.91,0.59,0.48}
\definecolor{DarkSeaGreen1}{rgb}{0.76,1.00,0.76}
\definecolor{DarkSeaGreen2}{rgb}{0.71,0.93,0.71}
\definecolor{DarkSeaGreen3}{rgb}{0.61,0.80,0.61}
\definecolor{DarkSeaGreen4}{rgb}{0.41,0.55,0.41}
\definecolor{DarkSeaGreen}{rgb}{0.56,0.74,0.56}
\definecolor{DarkSlateBlue}{rgb}{0.28,0.24,0.55}
\definecolor{DarkSlateGray1}{rgb}{0.59,1.00,1.00}
\definecolor{DarkSlateGray2}{rgb}{0.55,0.93,0.93}
\definecolor{DarkSlateGray3}{rgb}{0.47,0.80,0.80}
\definecolor{DarkSlateGray4}{rgb}{0.32,0.55,0.55}
\definecolor{DarkSlateGray}{rgb}{0.18,0.31,0.31}
\definecolor{DarkSlateGrey}{rgb}{0.18,0.31,0.31}
\definecolor{DarkTurquoise}{rgb}{0.00,0.81,0.82}
\definecolor{DarkViolet}{rgb}{0.58,0.00,0.83}
\definecolor{DeepPink1}{rgb}{1.00,0.08,0.58}
\definecolor{DeepPink2}{rgb}{0.93,0.07,0.54}
\definecolor{DeepPink3}{rgb}{0.80,0.06,0.46}
\definecolor{DeepPink4}{rgb}{0.55,0.04,0.31}
\definecolor{DeepPink}{rgb}{1.00,0.08,0.58}
\definecolor{DeepSkyBlue1}{rgb}{0.00,0.75,1.00}
\definecolor{DeepSkyBlue2}{rgb}{0.00,0.70,0.93}
\definecolor{DeepSkyBlue3}{rgb}{0.00,0.60,0.80}
\definecolor{DeepSkyBlue4}{rgb}{0.00,0.41,0.55}
\definecolor{DeepSkyBlue}{rgb}{0.00,0.75,1.00}
\definecolor{DimGray}{rgb}{0.41,0.41,0.41}
\definecolor{DimGrey}{rgb}{0.41,0.41,0.41}
\definecolor{DodgerBlue1}{rgb}{0.12,0.56,1.00}
\definecolor{DodgerBlue2}{rgb}{0.11,0.53,0.93}
\definecolor{DodgerBlue3}{rgb}{0.09,0.45,0.80}
\definecolor{DodgerBlue4}{rgb}{0.06,0.31,0.55}
\definecolor{DodgerBlue}{rgb}{0.12,0.56,1.00}
\definecolor{FloralWhite}{rgb}{1.00,0.98,0.94}
\definecolor{ForestGreen}{rgb}{0.13,0.55,0.13}
\definecolor{GhostWhite}{rgb}{0.97,0.97,1.00}
\definecolor{GreenYellow}{rgb}{0.68,1.00,0.18}
\definecolor{HotPink1}{rgb}{1.00,0.43,0.71}
\definecolor{HotPink2}{rgb}{0.93,0.42,0.65}
\definecolor{HotPink3}{rgb}{0.80,0.38,0.56}
\definecolor{HotPink4}{rgb}{0.55,0.23,0.38}
\definecolor{HotPink}{rgb}{1.00,0.41,0.71}
\definecolor{IndianRed1}{rgb}{1.00,0.42,0.42}
\definecolor{IndianRed2}{rgb}{0.93,0.39,0.39}
\definecolor{IndianRed3}{rgb}{0.80,0.33,0.33}
\definecolor{IndianRed4}{rgb}{0.55,0.23,0.23}
\definecolor{IndianRed}{rgb}{0.80,0.36,0.36}
\definecolor{LavenderBlush1}{rgb}{1.00,0.94,0.96}
\definecolor{LavenderBlush2}{rgb}{0.93,0.88,0.90}
\definecolor{LavenderBlush3}{rgb}{0.80,0.76,0.77}
\definecolor{LavenderBlush4}{rgb}{0.55,0.51,0.53}
\definecolor{LavenderBlush}{rgb}{1.00,0.94,0.96}
\definecolor{LawnGreen}{rgb}{0.49,0.99,0.00}
\definecolor{LemonChiffon1}{rgb}{1.00,0.98,0.80}
\definecolor{LemonChiffon2}{rgb}{0.93,0.91,0.75}
\definecolor{LemonChiffon3}{rgb}{0.80,0.79,0.65}
\definecolor{LemonChiffon4}{rgb}{0.55,0.54,0.44}
\definecolor{LemonChiffon}{rgb}{1.00,0.98,0.80}
\definecolor{LightBlue1}{rgb}{0.75,0.94,1.00}
\definecolor{LightBlue2}{rgb}{0.70,0.87,0.93}
\definecolor{LightBlue3}{rgb}{0.60,0.75,0.80}
\definecolor{LightBlue4}{rgb}{0.41,0.51,0.55}
\definecolor{LightBlue}{rgb}{0.68,0.85,0.90}
\definecolor{LightCoral}{rgb}{0.94,0.50,0.50}
\definecolor{LightCyan1}{rgb}{0.88,1.00,1.00}
\definecolor{LightCyan2}{rgb}{0.82,0.93,0.93}
\definecolor{LightCyan3}{rgb}{0.71,0.80,0.80}
\definecolor{LightCyan4}{rgb}{0.48,0.55,0.55}
\definecolor{LightCyan}{rgb}{0.88,1.00,1.00}
\definecolor{LightGoldenrod1}{rgb}{1.00,0.93,0.55}
\definecolor{LightGoldenrod2}{rgb}{0.93,0.86,0.51}
\definecolor{LightGoldenrod3}{rgb}{0.80,0.75,0.44}
\definecolor{LightGoldenrod4}{rgb}{0.55,0.51,0.30}
\definecolor{LightGoldenrodYellow}{rgb}{0.98,0.98,0.82}
\definecolor{LightGoldenrod}{rgb}{0.93,0.87,0.51}
\definecolor{LightGray}{rgb}{0.83,0.83,0.83}
\definecolor{LightGreen}{rgb}{0.56,0.93,0.56}
\definecolor{LightGrey}{rgb}{0.83,0.83,0.83}
\definecolor{LightPink1}{rgb}{1.00,0.68,0.73}
\definecolor{LightPink2}{rgb}{0.93,0.64,0.68}
\definecolor{LightPink3}{rgb}{0.80,0.55,0.58}
\definecolor{LightPink4}{rgb}{0.55,0.37,0.40}
\definecolor{LightPink}{rgb}{1.00,0.71,0.76}
\definecolor{LightSalmon1}{rgb}{1.00,0.63,0.48}
\definecolor{LightSalmon2}{rgb}{0.93,0.58,0.45}
\definecolor{LightSalmon3}{rgb}{0.80,0.51,0.38}
\definecolor{LightSalmon4}{rgb}{0.55,0.34,0.26}
\definecolor{LightSalmon}{rgb}{1.00,0.63,0.48}
\definecolor{LightSeaGreen}{rgb}{0.13,0.70,0.67}
\definecolor{LightSkyBlue1}{rgb}{0.69,0.89,1.00}
\definecolor{LightSkyBlue2}{rgb}{0.64,0.83,0.93}
\definecolor{LightSkyBlue3}{rgb}{0.55,0.71,0.80}
\definecolor{LightSkyBlue4}{rgb}{0.38,0.48,0.55}
\definecolor{LightSkyBlue}{rgb}{0.53,0.81,0.98}
\definecolor{LightSlateBlue}{rgb}{0.52,0.44,1.00}
\definecolor{LightSlateGray}{rgb}{0.47,0.53,0.60}
\definecolor{LightSlateGrey}{rgb}{0.47,0.53,0.60}
\definecolor{LightSteelBlue1}{rgb}{0.79,0.88,1.00}
\definecolor{LightSteelBlue2}{rgb}{0.74,0.82,0.93}
\definecolor{LightSteelBlue3}{rgb}{0.64,0.71,0.80}
\definecolor{LightSteelBlue4}{rgb}{0.43,0.48,0.55}
\definecolor{LightSteelBlue}{rgb}{0.69,0.77,0.87}
\definecolor{LightYellow1}{rgb}{1.00,1.00,0.88}
\definecolor{LightYellow2}{rgb}{0.93,0.93,0.82}
\definecolor{LightYellow3}{rgb}{0.80,0.80,0.71}
\definecolor{LightYellow4}{rgb}{0.55,0.55,0.48}
\definecolor{LightYellow}{rgb}{1.00,1.00,0.88}
\definecolor{LimeGreen}{rgb}{0.20,0.80,0.20}
\definecolor{MediumAquamarine}{rgb}{0.40,0.80,0.67}
\definecolor{MediumBlue}{rgb}{0.00,0.00,0.80}
\definecolor{MediumOrchid1}{rgb}{0.88,0.40,1.00}
\definecolor{MediumOrchid2}{rgb}{0.82,0.37,0.93}
\definecolor{MediumOrchid3}{rgb}{0.71,0.32,0.80}
\definecolor{MediumOrchid4}{rgb}{0.48,0.22,0.55}
\definecolor{MediumOrchid}{rgb}{0.73,0.33,0.83}
\definecolor{MediumPurple1}{rgb}{0.67,0.51,1.00}
\definecolor{MediumPurple2}{rgb}{0.62,0.47,0.93}
\definecolor{MediumPurple3}{rgb}{0.54,0.41,0.80}
\definecolor{MediumPurple4}{rgb}{0.36,0.28,0.55}
\definecolor{MediumPurple}{rgb}{0.58,0.44,0.86}
\definecolor{MediumSeaGreen}{rgb}{0.24,0.70,0.44}
\definecolor{MediumSlateBlue}{rgb}{0.48,0.41,0.93}
\definecolor{MediumSpringGreen}{rgb}{0.00,0.98,0.60}
\definecolor{MediumTurquoise}{rgb}{0.28,0.82,0.80}
\definecolor{MediumVioletRed}{rgb}{0.78,0.08,0.52}
\definecolor{MidnightBlue}{rgb}{0.10,0.10,0.44}
\definecolor{MintCream}{rgb}{0.96,1.00,0.98}
\definecolor{MistyRose1}{rgb}{1.00,0.89,0.88}
\definecolor{MistyRose2}{rgb}{0.93,0.84,0.82}
\definecolor{MistyRose3}{rgb}{0.80,0.72,0.71}
\definecolor{MistyRose4}{rgb}{0.55,0.49,0.48}
\definecolor{MistyRose}{rgb}{1.00,0.89,0.88}
\definecolor{NavajoWhite1}{rgb}{1.00,0.87,0.68}
\definecolor{NavajoWhite2}{rgb}{0.93,0.81,0.63}
\definecolor{NavajoWhite3}{rgb}{0.80,0.70,0.55}
\definecolor{NavajoWhite4}{rgb}{0.55,0.47,0.37}
\definecolor{NavajoWhite}{rgb}{1.00,0.87,0.68}
\definecolor{NavyBlue}{rgb}{0.00,0.00,0.50}
\definecolor{OldLace}{rgb}{0.99,0.96,0.90}
\definecolor{OliveDrab1}{rgb}{0.75,1.00,0.24}
\definecolor{OliveDrab2}{rgb}{0.70,0.93,0.23}
\definecolor{OliveDrab3}{rgb}{0.60,0.80,0.20}
\definecolor{OliveDrab4}{rgb}{0.41,0.55,0.13}
\definecolor{OliveDrab}{rgb}{0.42,0.56,0.14}
\definecolor{OrangeRed1}{rgb}{1.00,0.27,0.00}
\definecolor{OrangeRed2}{rgb}{0.93,0.25,0.00}
\definecolor{OrangeRed3}{rgb}{0.80,0.22,0.00}
\definecolor{OrangeRed4}{rgb}{0.55,0.15,0.00}
\definecolor{OrangeRed}{rgb}{1.00,0.27,0.00}
\definecolor{PaleGoldenrod}{rgb}{0.93,0.91,0.67}
\definecolor{PaleGreen1}{rgb}{0.60,1.00,0.60}
\definecolor{PaleGreen2}{rgb}{0.56,0.93,0.56}
\definecolor{PaleGreen3}{rgb}{0.49,0.80,0.49}
\definecolor{PaleGreen4}{rgb}{0.33,0.55,0.33}
\definecolor{PaleGreen}{rgb}{0.60,0.98,0.60}
\definecolor{PaleTurquoise1}{rgb}{0.73,1.00,1.00}
\definecolor{PaleTurquoise2}{rgb}{0.68,0.93,0.93}
\definecolor{PaleTurquoise3}{rgb}{0.59,0.80,0.80}
\definecolor{PaleTurquoise4}{rgb}{0.40,0.55,0.55}
\definecolor{PaleTurquoise}{rgb}{0.69,0.93,0.93}
\definecolor{PaleVioletRed1}{rgb}{1.00,0.51,0.67}
\definecolor{PaleVioletRed2}{rgb}{0.93,0.47,0.62}
\definecolor{PaleVioletRed3}{rgb}{0.80,0.41,0.54}
\definecolor{PaleVioletRed4}{rgb}{0.55,0.28,0.36}
\definecolor{PaleVioletRed}{rgb}{0.86,0.44,0.58}
\definecolor{PapayaWhip}{rgb}{1.00,0.94,0.84}
\definecolor{PeachPuff1}{rgb}{1.00,0.85,0.73}
\definecolor{PeachPuff2}{rgb}{0.93,0.80,0.68}
\definecolor{PeachPuff3}{rgb}{0.80,0.69,0.58}
\definecolor{PeachPuff4}{rgb}{0.55,0.47,0.40}
\definecolor{PeachPuff}{rgb}{1.00,0.85,0.73}
\definecolor{PowderBlue}{rgb}{0.69,0.88,0.90}
\definecolor{RosyBrown1}{rgb}{1.00,0.76,0.76}
\definecolor{RosyBrown2}{rgb}{0.93,0.71,0.71}
\definecolor{RosyBrown3}{rgb}{0.80,0.61,0.61}
\definecolor{RosyBrown4}{rgb}{0.55,0.41,0.41}
\definecolor{RosyBrown}{rgb}{0.74,0.56,0.56}
\definecolor{RoyalBlue1}{rgb}{0.28,0.46,1.00}
\definecolor{RoyalBlue2}{rgb}{0.26,0.43,0.93}
\definecolor{RoyalBlue3}{rgb}{0.23,0.37,0.80}
\definecolor{RoyalBlue4}{rgb}{0.15,0.25,0.55}
\definecolor{RoyalBlue}{rgb}{0.25,0.41,0.88}
\definecolor{SaddleBrown}{rgb}{0.55,0.27,0.07}
\definecolor{SandyBrown}{rgb}{0.96,0.64,0.38}
\definecolor{SeaGreen1}{rgb}{0.33,1.00,0.62}
\definecolor{SeaGreen2}{rgb}{0.31,0.93,0.58}
\definecolor{SeaGreen3}{rgb}{0.26,0.80,0.50}
\definecolor{SeaGreen4}{rgb}{0.18,0.55,0.34}
\definecolor{SeaGreen}{rgb}{0.18,0.55,0.34}
\definecolor{SkyBlue1}{rgb}{0.53,0.81,1.00}
\definecolor{SkyBlue2}{rgb}{0.49,0.75,0.93}
\definecolor{SkyBlue3}{rgb}{0.42,0.65,0.80}
\definecolor{SkyBlue4}{rgb}{0.29,0.44,0.55}
\definecolor{SkyBlue}{rgb}{0.53,0.81,0.92}
\definecolor{SlateBlue1}{rgb}{0.51,0.44,1.00}
\definecolor{SlateBlue2}{rgb}{0.48,0.40,0.93}
\definecolor{SlateBlue3}{rgb}{0.41,0.35,0.80}
\definecolor{SlateBlue4}{rgb}{0.28,0.24,0.55}
\definecolor{SlateBlue}{rgb}{0.42,0.35,0.80}
\definecolor{SlateGray1}{rgb}{0.78,0.89,1.00}
\definecolor{SlateGray2}{rgb}{0.73,0.83,0.93}
\definecolor{SlateGray3}{rgb}{0.62,0.71,0.80}
\definecolor{SlateGray4}{rgb}{0.42,0.48,0.55}
\definecolor{SlateGray}{rgb}{0.44,0.50,0.56}
\definecolor{SlateGrey}{rgb}{0.44,0.50,0.56}
\definecolor{SpringGreen1}{rgb}{0.00,1.00,0.50}
\definecolor{SpringGreen2}{rgb}{0.00,0.93,0.46}
\definecolor{SpringGreen3}{rgb}{0.00,0.80,0.40}
\definecolor{SpringGreen4}{rgb}{0.00,0.55,0.27}
\definecolor{SpringGreen}{rgb}{0.00,1.00,0.50}
\definecolor{SteelBlue1}{rgb}{0.39,0.72,1.00}
\definecolor{SteelBlue2}{rgb}{0.36,0.67,0.93}
\definecolor{SteelBlue3}{rgb}{0.31,0.58,0.80}
\definecolor{SteelBlue4}{rgb}{0.21,0.39,0.55}
\definecolor{SteelBlue}{rgb}{0.27,0.51,0.71}
\definecolor{VioletRed1}{rgb}{1.00,0.24,0.59}
\definecolor{VioletRed2}{rgb}{0.93,0.23,0.55}
\definecolor{VioletRed3}{rgb}{0.80,0.20,0.47}
\definecolor{VioletRed4}{rgb}{0.55,0.13,0.32}
\definecolor{VioletRed}{rgb}{0.82,0.13,0.56}
\definecolor{WhiteSmoke}{rgb}{0.96,0.96,0.96}
\definecolor{YellowGreen}{rgb}{0.60,0.80,0.20}
\definecolor{aliceblue}{rgb}{0.94,0.97,1.00}
\definecolor{antiquewhite}{rgb}{0.98,0.92,0.84}
\definecolor{aquamarine1}{rgb}{0.50,1.00,0.83}
\definecolor{aquamarine2}{rgb}{0.46,0.93,0.78}
\definecolor{aquamarine3}{rgb}{0.40,0.80,0.67}
\definecolor{aquamarine4}{rgb}{0.27,0.55,0.45}
\definecolor{aquamarine}{rgb}{0.50,1.00,0.83}
\definecolor{azure1}{rgb}{0.94,1.00,1.00}
\definecolor{azure2}{rgb}{0.88,0.93,0.93}
\definecolor{azure3}{rgb}{0.76,0.80,0.80}
\definecolor{azure4}{rgb}{0.51,0.55,0.55}
\definecolor{azure}{rgb}{0.94,1.00,1.00}
\definecolor{beige}{rgb}{0.96,0.96,0.86}
\definecolor{bisque1}{rgb}{1.00,0.89,0.77}
\definecolor{bisque2}{rgb}{0.93,0.84,0.72}
\definecolor{bisque3}{rgb}{0.80,0.72,0.62}
\definecolor{bisque4}{rgb}{0.55,0.49,0.42}
\definecolor{bisque}{rgb}{1.00,0.89,0.77}
\definecolor{black}{rgb}{0.00,0.00,0.00}
\definecolor{blanchedalmond}{rgb}{1.00,0.92,0.80}
\definecolor{blue1}{rgb}{0.00,0.00,1.00}
\definecolor{blue2}{rgb}{0.00,0.00,0.93}
\definecolor{blue3}{rgb}{0.00,0.00,0.80}
\definecolor{blue4}{rgb}{0.00,0.00,0.55}
\definecolor{blueviolet}{rgb}{0.54,0.17,0.89}
\definecolor{blue}{rgb}{0.00,0.00,1.00}
\definecolor{brown1}{rgb}{1.00,0.25,0.25}
\definecolor{brown2}{rgb}{0.93,0.23,0.23}
\definecolor{brown3}{rgb}{0.80,0.20,0.20}
\definecolor{brown4}{rgb}{0.55,0.14,0.14}
\definecolor{brown}{rgb}{0.65,0.16,0.16}
\definecolor{burlywood1}{rgb}{1.00,0.83,0.61}
\definecolor{burlywood2}{rgb}{0.93,0.77,0.57}
\definecolor{burlywood3}{rgb}{0.80,0.67,0.49}
\definecolor{burlywood4}{rgb}{0.55,0.45,0.33}
\definecolor{burlywood}{rgb}{0.87,0.72,0.53}
\definecolor{cadetblue}{rgb}{0.37,0.62,0.63}
\definecolor{chartreuse1}{rgb}{0.50,1.00,0.00}
\definecolor{chartreuse2}{rgb}{0.46,0.93,0.00}
\definecolor{chartreuse3}{rgb}{0.40,0.80,0.00}
\definecolor{chartreuse4}{rgb}{0.27,0.55,0.00}
\definecolor{chartreuse}{rgb}{0.50,1.00,0.00}
\definecolor{chocolate1}{rgb}{1.00,0.50,0.14}
\definecolor{chocolate2}{rgb}{0.93,0.46,0.13}
\definecolor{chocolate3}{rgb}{0.80,0.40,0.11}
\definecolor{chocolate4}{rgb}{0.55,0.27,0.07}
\definecolor{chocolate}{rgb}{0.82,0.41,0.12}
\definecolor{coral1}{rgb}{1.00,0.45,0.34}
\definecolor{coral2}{rgb}{0.93,0.42,0.31}
\definecolor{coral3}{rgb}{0.80,0.36,0.27}
\definecolor{coral4}{rgb}{0.55,0.24,0.18}
\definecolor{coral}{rgb}{1.00,0.50,0.31}
\definecolor{cornflowerblue}{rgb}{0.39,0.58,0.93}
\definecolor{cornsilk1}{rgb}{1.00,0.97,0.86}
\definecolor{cornsilk2}{rgb}{0.93,0.91,0.80}
\definecolor{cornsilk3}{rgb}{0.80,0.78,0.69}
\definecolor{cornsilk4}{rgb}{0.55,0.53,0.47}
\definecolor{cornsilk}{rgb}{1.00,0.97,0.86}
\definecolor{cyan1}{rgb}{0.00,1.00,1.00}
\definecolor{cyan2}{rgb}{0.00,0.93,0.93}
\definecolor{cyan3}{rgb}{0.00,0.80,0.80}
\definecolor{cyan4}{rgb}{0.00,0.55,0.55}
\definecolor{cyan}{rgb}{0.00,1.00,1.00}
\definecolor{darkblue}{rgb}{0.00,0.00,0.55}
\definecolor{darkcyan}{rgb}{0.00,0.55,0.55}
\definecolor{darkgoldenrod}{rgb}{0.72,0.53,0.04}
\definecolor{darkgray}{rgb}{0.66,0.66,0.66}
\definecolor{darkgreen}{rgb}{0.00,0.39,0.00}
\definecolor{darkgrey}{rgb}{0.66,0.66,0.66}
\definecolor{darkkhaki}{rgb}{0.74,0.72,0.42}
\definecolor{darkmagenta}{rgb}{0.55,0.00,0.55}
\definecolor{darkolive}{rgb}{0.33,0.42,0.18}
\definecolor{darkorange}{rgb}{1.00,0.55,0.00}
\definecolor{darkorchid}{rgb}{0.60,0.20,0.80}
\definecolor{darkred}{rgb}{0.55,0.00,0.00}
\definecolor{darksalmon}{rgb}{0.91,0.59,0.48}
\definecolor{darksea}{rgb}{0.56,0.74,0.56}
\definecolor{darkslate}{rgb}{0.18,0.31,0.31}
\definecolor{darkslate}{rgb}{0.18,0.31,0.31}
\definecolor{darkslate}{rgb}{0.28,0.24,0.55}
\definecolor{darkturquoise}{rgb}{0.00,0.81,0.82}
\definecolor{darkviolet}{rgb}{0.58,0.00,0.83}
\definecolor{deeppink}{rgb}{1.00,0.08,0.58}
\definecolor{deepsky}{rgb}{0.00,0.75,1.00}
\definecolor{dimgray}{rgb}{0.41,0.41,0.41}
\definecolor{dimgrey}{rgb}{0.41,0.41,0.41}
\definecolor{dodgerblue}{rgb}{0.12,0.56,1.00}
\definecolor{firebrick1}{rgb}{1.00,0.19,0.19}
\definecolor{firebrick2}{rgb}{0.93,0.17,0.17}
\definecolor{firebrick3}{rgb}{0.80,0.15,0.15}
\definecolor{firebrick4}{rgb}{0.55,0.10,0.10}
\definecolor{firebrick}{rgb}{0.70,0.13,0.13}
\definecolor{floralwhite}{rgb}{1.00,0.98,0.94}
\definecolor{forestgreen}{rgb}{0.13,0.55,0.13}
\definecolor{gainsboro}{rgb}{0.86,0.86,0.86}
\definecolor{ghostwhite}{rgb}{0.97,0.97,1.00}
\definecolor{gold1}{rgb}{1.00,0.84,0.00}
\definecolor{gold2}{rgb}{0.93,0.79,0.00}
\definecolor{gold3}{rgb}{0.80,0.68,0.00}
\definecolor{gold4}{rgb}{0.55,0.46,0.00}
\definecolor{goldenrod1}{rgb}{1.00,0.76,0.15}
\definecolor{goldenrod2}{rgb}{0.93,0.71,0.13}
\definecolor{goldenrod3}{rgb}{0.80,0.61,0.11}
\definecolor{goldenrod4}{rgb}{0.55,0.41,0.08}
\definecolor{goldenrod}{rgb}{0.85,0.65,0.13}
\definecolor{gold}{rgb}{1.00,0.84,0.00}
\definecolor{gray0}{rgb}{0.00,0.00,0.00}
\definecolor{gray100}{rgb}{1.00,1.00,1.00}
\definecolor{gray10}{rgb}{0.10,0.10,0.10}
\definecolor{gray11}{rgb}{0.11,0.11,0.11}
\definecolor{gray12}{rgb}{0.12,0.12,0.12}
\definecolor{gray13}{rgb}{0.13,0.13,0.13}
\definecolor{gray14}{rgb}{0.14,0.14,0.14}
\definecolor{gray15}{rgb}{0.15,0.15,0.15}
\definecolor{gray16}{rgb}{0.16,0.16,0.16}
\definecolor{gray17}{rgb}{0.17,0.17,0.17}
\definecolor{gray18}{rgb}{0.18,0.18,0.18}
\definecolor{gray19}{rgb}{0.19,0.19,0.19}
\definecolor{gray1}{rgb}{0.01,0.01,0.01}
\definecolor{gray20}{rgb}{0.20,0.20,0.20}
\definecolor{gray21}{rgb}{0.21,0.21,0.21}
\definecolor{gray22}{rgb}{0.22,0.22,0.22}
\definecolor{gray23}{rgb}{0.23,0.23,0.23}
\definecolor{gray24}{rgb}{0.24,0.24,0.24}
\definecolor{gray25}{rgb}{0.25,0.25,0.25}
\definecolor{gray26}{rgb}{0.26,0.26,0.26}
\definecolor{gray27}{rgb}{0.27,0.27,0.27}
\definecolor{gray28}{rgb}{0.28,0.28,0.28}
\definecolor{gray29}{rgb}{0.29,0.29,0.29}
\definecolor{gray2}{rgb}{0.02,0.02,0.02}
\definecolor{gray30}{rgb}{0.30,0.30,0.30}
\definecolor{gray31}{rgb}{0.31,0.31,0.31}
\definecolor{gray32}{rgb}{0.32,0.32,0.32}
\definecolor{gray33}{rgb}{0.33,0.33,0.33}
\definecolor{gray34}{rgb}{0.34,0.34,0.34}
\definecolor{gray35}{rgb}{0.35,0.35,0.35}
\definecolor{gray36}{rgb}{0.36,0.36,0.36}
\definecolor{gray37}{rgb}{0.37,0.37,0.37}
\definecolor{gray38}{rgb}{0.38,0.38,0.38}
\definecolor{gray39}{rgb}{0.39,0.39,0.39}
\definecolor{gray3}{rgb}{0.03,0.03,0.03}
\definecolor{gray40}{rgb}{0.40,0.40,0.40}
\definecolor{gray41}{rgb}{0.41,0.41,0.41}
\definecolor{gray42}{rgb}{0.42,0.42,0.42}
\definecolor{gray43}{rgb}{0.43,0.43,0.43}
\definecolor{gray44}{rgb}{0.44,0.44,0.44}
\definecolor{gray45}{rgb}{0.45,0.45,0.45}
\definecolor{gray46}{rgb}{0.46,0.46,0.46}
\definecolor{gray47}{rgb}{0.47,0.47,0.47}
\definecolor{gray48}{rgb}{0.48,0.48,0.48}
\definecolor{gray49}{rgb}{0.49,0.49,0.49}
\definecolor{gray4}{rgb}{0.04,0.04,0.04}
\definecolor{gray50}{rgb}{0.50,0.50,0.50}
\definecolor{gray51}{rgb}{0.51,0.51,0.51}
\definecolor{gray52}{rgb}{0.52,0.52,0.52}
\definecolor{gray53}{rgb}{0.53,0.53,0.53}
\definecolor{gray54}{rgb}{0.54,0.54,0.54}
\definecolor{gray55}{rgb}{0.55,0.55,0.55}
\definecolor{gray56}{rgb}{0.56,0.56,0.56}
\definecolor{gray57}{rgb}{0.57,0.57,0.57}
\definecolor{gray58}{rgb}{0.58,0.58,0.58}
\definecolor{gray59}{rgb}{0.59,0.59,0.59}
\definecolor{gray5}{rgb}{0.05,0.05,0.05}
\definecolor{gray60}{rgb}{0.60,0.60,0.60}
\definecolor{gray61}{rgb}{0.61,0.61,0.61}
\definecolor{gray62}{rgb}{0.62,0.62,0.62}
\definecolor{gray63}{rgb}{0.63,0.63,0.63}
\definecolor{gray64}{rgb}{0.64,0.64,0.64}
\definecolor{gray65}{rgb}{0.65,0.65,0.65}
\definecolor{gray66}{rgb}{0.66,0.66,0.66}
\definecolor{gray67}{rgb}{0.67,0.67,0.67}
\definecolor{gray68}{rgb}{0.68,0.68,0.68}
\definecolor{gray69}{rgb}{0.69,0.69,0.69}
\definecolor{gray6}{rgb}{0.06,0.06,0.06}
\definecolor{gray70}{rgb}{0.70,0.70,0.70}
\definecolor{gray71}{rgb}{0.71,0.71,0.71}
\definecolor{gray72}{rgb}{0.72,0.72,0.72}
\definecolor{gray73}{rgb}{0.73,0.73,0.73}
\definecolor{gray74}{rgb}{0.74,0.74,0.74}
\definecolor{gray75}{rgb}{0.75,0.75,0.75}
\definecolor{gray76}{rgb}{0.76,0.76,0.76}
\definecolor{gray77}{rgb}{0.77,0.77,0.77}
\definecolor{gray78}{rgb}{0.78,0.78,0.78}
\definecolor{gray79}{rgb}{0.79,0.79,0.79}
\definecolor{gray7}{rgb}{0.07,0.07,0.07}
\definecolor{gray80}{rgb}{0.80,0.80,0.80}
\definecolor{gray81}{rgb}{0.81,0.81,0.81}
\definecolor{gray82}{rgb}{0.82,0.82,0.82}
\definecolor{gray83}{rgb}{0.83,0.83,0.83}
\definecolor{gray84}{rgb}{0.84,0.84,0.84}
\definecolor{gray85}{rgb}{0.85,0.85,0.85}
\definecolor{gray86}{rgb}{0.86,0.86,0.86}
\definecolor{gray87}{rgb}{0.87,0.87,0.87}
\definecolor{gray88}{rgb}{0.88,0.88,0.88}
\definecolor{gray89}{rgb}{0.89,0.89,0.89}
\definecolor{gray8}{rgb}{0.08,0.08,0.08}
\definecolor{gray90}{rgb}{0.90,0.90,0.90}
\definecolor{gray91}{rgb}{0.91,0.91,0.91}
\definecolor{gray92}{rgb}{0.92,0.92,0.92}
\definecolor{gray93}{rgb}{0.93,0.93,0.93}
\definecolor{gray94}{rgb}{0.94,0.94,0.94}
\definecolor{gray95}{rgb}{0.95,0.95,0.95}
\definecolor{gray96}{rgb}{0.96,0.96,0.96}
\definecolor{gray97}{rgb}{0.97,0.97,0.97}
\definecolor{gray98}{rgb}{0.98,0.98,0.98}
\definecolor{gray99}{rgb}{0.99,0.99,0.99}
\definecolor{gray9}{rgb}{0.09,0.09,0.09}
\definecolor{gray}{rgb}{0.75,0.75,0.75}
\definecolor{green1}{rgb}{0.00,1.00,0.00}
\definecolor{green2}{rgb}{0.00,0.93,0.00}
\definecolor{green3}{rgb}{0.00,0.80,0.00}
\definecolor{green4}{rgb}{0.00,0.55,0.00}
\definecolor{greenyellow}{rgb}{0.68,1.00,0.18}
\definecolor{green}{rgb}{0.00,1.00,0.00}
\definecolor{grey0}{rgb}{0.00,0.00,0.00}
\definecolor{grey100}{rgb}{1.00,1.00,1.00}
\definecolor{grey10}{rgb}{0.10,0.10,0.10}
\definecolor{grey11}{rgb}{0.11,0.11,0.11}
\definecolor{grey12}{rgb}{0.12,0.12,0.12}
\definecolor{grey13}{rgb}{0.13,0.13,0.13}
\definecolor{grey14}{rgb}{0.14,0.14,0.14}
\definecolor{grey15}{rgb}{0.15,0.15,0.15}
\definecolor{grey16}{rgb}{0.16,0.16,0.16}
\definecolor{grey17}{rgb}{0.17,0.17,0.17}
\definecolor{grey18}{rgb}{0.18,0.18,0.18}
\definecolor{grey19}{rgb}{0.19,0.19,0.19}
\definecolor{grey1}{rgb}{0.01,0.01,0.01}
\definecolor{grey20}{rgb}{0.20,0.20,0.20}
\definecolor{grey21}{rgb}{0.21,0.21,0.21}
\definecolor{grey22}{rgb}{0.22,0.22,0.22}
\definecolor{grey23}{rgb}{0.23,0.23,0.23}
\definecolor{grey24}{rgb}{0.24,0.24,0.24}
\definecolor{grey25}{rgb}{0.25,0.25,0.25}
\definecolor{grey26}{rgb}{0.26,0.26,0.26}
\definecolor{grey27}{rgb}{0.27,0.27,0.27}
\definecolor{grey28}{rgb}{0.28,0.28,0.28}
\definecolor{grey29}{rgb}{0.29,0.29,0.29}
\definecolor{grey2}{rgb}{0.02,0.02,0.02}
\definecolor{grey30}{rgb}{0.30,0.30,0.30}
\definecolor{grey31}{rgb}{0.31,0.31,0.31}
\definecolor{grey32}{rgb}{0.32,0.32,0.32}
\definecolor{grey33}{rgb}{0.33,0.33,0.33}
\definecolor{grey34}{rgb}{0.34,0.34,0.34}
\definecolor{grey35}{rgb}{0.35,0.35,0.35}
\definecolor{grey36}{rgb}{0.36,0.36,0.36}
\definecolor{grey37}{rgb}{0.37,0.37,0.37}
\definecolor{grey38}{rgb}{0.38,0.38,0.38}
\definecolor{grey39}{rgb}{0.39,0.39,0.39}
\definecolor{grey3}{rgb}{0.03,0.03,0.03}
\definecolor{grey40}{rgb}{0.40,0.40,0.40}
\definecolor{grey41}{rgb}{0.41,0.41,0.41}
\definecolor{grey42}{rgb}{0.42,0.42,0.42}
\definecolor{grey43}{rgb}{0.43,0.43,0.43}
\definecolor{grey44}{rgb}{0.44,0.44,0.44}
\definecolor{grey45}{rgb}{0.45,0.45,0.45}
\definecolor{grey46}{rgb}{0.46,0.46,0.46}
\definecolor{grey47}{rgb}{0.47,0.47,0.47}
\definecolor{grey48}{rgb}{0.48,0.48,0.48}
\definecolor{grey49}{rgb}{0.49,0.49,0.49}
\definecolor{grey4}{rgb}{0.04,0.04,0.04}
\definecolor{grey50}{rgb}{0.50,0.50,0.50}
\definecolor{grey51}{rgb}{0.51,0.51,0.51}
\definecolor{grey52}{rgb}{0.52,0.52,0.52}
\definecolor{grey53}{rgb}{0.53,0.53,0.53}
\definecolor{grey54}{rgb}{0.54,0.54,0.54}
\definecolor{grey55}{rgb}{0.55,0.55,0.55}
\definecolor{grey56}{rgb}{0.56,0.56,0.56}
\definecolor{grey57}{rgb}{0.57,0.57,0.57}
\definecolor{grey58}{rgb}{0.58,0.58,0.58}
\definecolor{grey59}{rgb}{0.59,0.59,0.59}
\definecolor{grey5}{rgb}{0.05,0.05,0.05}
\definecolor{grey60}{rgb}{0.60,0.60,0.60}
\definecolor{grey61}{rgb}{0.61,0.61,0.61}
\definecolor{grey62}{rgb}{0.62,0.62,0.62}
\definecolor{grey63}{rgb}{0.63,0.63,0.63}
\definecolor{grey64}{rgb}{0.64,0.64,0.64}
\definecolor{grey65}{rgb}{0.65,0.65,0.65}
\definecolor{grey66}{rgb}{0.66,0.66,0.66}
\definecolor{grey67}{rgb}{0.67,0.67,0.67}
\definecolor{grey68}{rgb}{0.68,0.68,0.68}
\definecolor{grey69}{rgb}{0.69,0.69,0.69}
\definecolor{grey6}{rgb}{0.06,0.06,0.06}
\definecolor{grey70}{rgb}{0.70,0.70,0.70}
\definecolor{grey71}{rgb}{0.71,0.71,0.71}
\definecolor{grey72}{rgb}{0.72,0.72,0.72}
\definecolor{grey73}{rgb}{0.73,0.73,0.73}
\definecolor{grey74}{rgb}{0.74,0.74,0.74}
\definecolor{grey75}{rgb}{0.75,0.75,0.75}
\definecolor{grey76}{rgb}{0.76,0.76,0.76}
\definecolor{grey77}{rgb}{0.77,0.77,0.77}
\definecolor{grey78}{rgb}{0.78,0.78,0.78}
\definecolor{grey79}{rgb}{0.79,0.79,0.79}
\definecolor{grey7}{rgb}{0.07,0.07,0.07}
\definecolor{grey80}{rgb}{0.80,0.80,0.80}
\definecolor{grey81}{rgb}{0.81,0.81,0.81}
\definecolor{grey82}{rgb}{0.82,0.82,0.82}
\definecolor{grey83}{rgb}{0.83,0.83,0.83}
\definecolor{grey84}{rgb}{0.84,0.84,0.84}
\definecolor{grey85}{rgb}{0.85,0.85,0.85}
\definecolor{grey86}{rgb}{0.86,0.86,0.86}
\definecolor{grey87}{rgb}{0.87,0.87,0.87}
\definecolor{grey88}{rgb}{0.88,0.88,0.88}
\definecolor{grey89}{rgb}{0.89,0.89,0.89}
\definecolor{grey8}{rgb}{0.08,0.08,0.08}
\definecolor{grey90}{rgb}{0.90,0.90,0.90}
\definecolor{grey91}{rgb}{0.91,0.91,0.91}
\definecolor{grey92}{rgb}{0.92,0.92,0.92}
\definecolor{grey93}{rgb}{0.93,0.93,0.93}
\definecolor{grey94}{rgb}{0.94,0.94,0.94}
\definecolor{grey95}{rgb}{0.95,0.95,0.95}
\definecolor{grey96}{rgb}{0.96,0.96,0.96}
\definecolor{grey97}{rgb}{0.97,0.97,0.97}
\definecolor{grey98}{rgb}{0.98,0.98,0.98}
\definecolor{grey99}{rgb}{0.99,0.99,0.99}
\definecolor{grey9}{rgb}{0.09,0.09,0.09}
\definecolor{grey}{rgb}{0.75,0.75,0.75}
\definecolor{honeydew1}{rgb}{0.94,1.00,0.94}
\definecolor{honeydew2}{rgb}{0.88,0.93,0.88}
\definecolor{honeydew3}{rgb}{0.76,0.80,0.76}
\definecolor{honeydew4}{rgb}{0.51,0.55,0.51}
\definecolor{honeydew}{rgb}{0.94,1.00,0.94}
\definecolor{hotpink}{rgb}{1.00,0.41,0.71}
\definecolor{indianred}{rgb}{0.80,0.36,0.36}
\definecolor{ivory1}{rgb}{1.00,1.00,0.94}
\definecolor{ivory2}{rgb}{0.93,0.93,0.88}
\definecolor{ivory3}{rgb}{0.80,0.80,0.76}
\definecolor{ivory4}{rgb}{0.55,0.55,0.51}
\definecolor{ivory}{rgb}{1.00,1.00,0.94}
\definecolor{khaki1}{rgb}{1.00,0.96,0.56}
\definecolor{khaki2}{rgb}{0.93,0.90,0.52}
\definecolor{khaki3}{rgb}{0.80,0.78,0.45}
\definecolor{khaki4}{rgb}{0.55,0.53,0.31}
\definecolor{khaki}{rgb}{0.94,0.90,0.55}
\definecolor{lavenderblush}{rgb}{1.00,0.94,0.96}
\definecolor{lavender}{rgb}{0.90,0.90,0.98}
\definecolor{lawngreen}{rgb}{0.49,0.99,0.00}
\definecolor{lemonchiffon}{rgb}{1.00,0.98,0.80}
\definecolor{lightblue}{rgb}{0.68,0.85,0.90}
\definecolor{lightcoral}{rgb}{0.94,0.50,0.50}
\definecolor{lightcyan}{rgb}{0.88,1.00,1.00}
\definecolor{lightgoldenrod}{rgb}{0.93,0.87,0.51}
\definecolor{lightgoldenrod}{rgb}{0.98,0.98,0.82}
\definecolor{lightgray}{rgb}{0.83,0.83,0.83}
\definecolor{lightgreen}{rgb}{0.56,0.93,0.56}
\definecolor{lightgrey}{rgb}{0.83,0.83,0.83}
\definecolor{lightpink}{rgb}{1.00,0.71,0.76}
\definecolor{lightsalmon}{rgb}{1.00,0.63,0.48}
\definecolor{lightsea}{rgb}{0.13,0.70,0.67}
\definecolor{lightsky}{rgb}{0.53,0.81,0.98}
\definecolor{lightslate}{rgb}{0.47,0.53,0.60}
\definecolor{lightslate}{rgb}{0.47,0.53,0.60}
\definecolor{lightslate}{rgb}{0.52,0.44,1.00}
\definecolor{lightsteel}{rgb}{0.69,0.77,0.87}
\definecolor{lightyellow}{rgb}{1.00,1.00,0.88}
\definecolor{limegreen}{rgb}{0.20,0.80,0.20}
\definecolor{linen}{rgb}{0.98,0.94,0.90}
\definecolor{magenta1}{rgb}{1.00,0.00,1.00}
\definecolor{magenta2}{rgb}{0.93,0.00,0.93}
\definecolor{magenta3}{rgb}{0.80,0.00,0.80}
\definecolor{magenta4}{rgb}{0.55,0.00,0.55}
\definecolor{magenta}{rgb}{1.00,0.00,1.00}
\definecolor{maroon1}{rgb}{1.00,0.20,0.70}
\definecolor{maroon2}{rgb}{0.93,0.19,0.65}
\definecolor{maroon3}{rgb}{0.80,0.16,0.56}
\definecolor{maroon4}{rgb}{0.55,0.11,0.38}
\definecolor{maroon}{rgb}{0.69,0.19,0.38}
\definecolor{mediumaquamarine}{rgb}{0.40,0.80,0.67}
\definecolor{mediumblue}{rgb}{0.00,0.00,0.80}
\definecolor{mediumorchid}{rgb}{0.73,0.33,0.83}
\definecolor{mediumpurple}{rgb}{0.58,0.44,0.86}
\definecolor{mediumsea}{rgb}{0.24,0.70,0.44}
\definecolor{mediumslate}{rgb}{0.48,0.41,0.93}
\definecolor{mediumspring}{rgb}{0.00,0.98,0.60}
\definecolor{mediumturquoise}{rgb}{0.28,0.82,0.80}
\definecolor{mediumviolet}{rgb}{0.78,0.08,0.52}
\definecolor{midnightblue}{rgb}{0.10,0.10,0.44}
\definecolor{mintcream}{rgb}{0.96,1.00,0.98}
\definecolor{mistyrose}{rgb}{1.00,0.89,0.88}
\definecolor{moccasin}{rgb}{1.00,0.89,0.71}
\definecolor{navajowhite}{rgb}{1.00,0.87,0.68}
\definecolor{navyblue}{rgb}{0.00,0.00,0.50}
\definecolor{navy}{rgb}{0.00,0.00,0.50}
\definecolor{oldlace}{rgb}{0.99,0.96,0.90}
\definecolor{olivedrab}{rgb}{0.42,0.56,0.14}
\definecolor{orange1}{rgb}{1.00,0.65,0.00}
\definecolor{orange2}{rgb}{0.93,0.60,0.00}
\definecolor{orange3}{rgb}{0.80,0.52,0.00}
\definecolor{orange4}{rgb}{0.55,0.35,0.00}
\definecolor{orangered}{rgb}{1.00,0.27,0.00}
\definecolor{orange}{rgb}{1.00,0.65,0.00}
\definecolor{orchid1}{rgb}{1.00,0.51,0.98}
\definecolor{orchid2}{rgb}{0.93,0.48,0.91}
\definecolor{orchid3}{rgb}{0.80,0.41,0.79}
\definecolor{orchid4}{rgb}{0.55,0.28,0.54}
\definecolor{orchid}{rgb}{0.85,0.44,0.84}
\definecolor{palegoldenrod}{rgb}{0.93,0.91,0.67}
\definecolor{palegreen}{rgb}{0.60,0.98,0.60}
\definecolor{paleturquoise}{rgb}{0.69,0.93,0.93}
\definecolor{paleviolet}{rgb}{0.86,0.44,0.58}
\definecolor{papayawhip}{rgb}{1.00,0.94,0.84}
\definecolor{peachpuff}{rgb}{1.00,0.85,0.73}
\definecolor{peru}{rgb}{0.80,0.52,0.25}
\definecolor{pink1}{rgb}{1.00,0.71,0.77}
\definecolor{pink2}{rgb}{0.93,0.66,0.72}
\definecolor{pink3}{rgb}{0.80,0.57,0.62}
\definecolor{pink4}{rgb}{0.55,0.39,0.42}
\definecolor{pink}{rgb}{1.00,0.75,0.80}
\definecolor{plum1}{rgb}{1.00,0.73,1.00}
\definecolor{plum2}{rgb}{0.93,0.68,0.93}
\definecolor{plum3}{rgb}{0.80,0.59,0.80}
\definecolor{plum4}{rgb}{0.55,0.40,0.55}
\definecolor{plum}{rgb}{0.87,0.63,0.87}
\definecolor{powderblue}{rgb}{0.69,0.88,0.90}
\definecolor{purple1}{rgb}{0.61,0.19,1.00}
\definecolor{purple2}{rgb}{0.57,0.17,0.93}
\definecolor{purple3}{rgb}{0.49,0.15,0.80}
\definecolor{purple4}{rgb}{0.33,0.10,0.55}
\definecolor{purple}{rgb}{0.63,0.13,0.94}
\definecolor{red1}{rgb}{1.00,0.00,0.00}
\definecolor{red2}{rgb}{0.93,0.00,0.00}
\definecolor{red3}{rgb}{0.80,0.00,0.00}
\definecolor{red4}{rgb}{0.55,0.00,0.00}
\definecolor{red}{rgb}{1.00,0.00,0.00}
\definecolor{rosybrown}{rgb}{0.74,0.56,0.56}
\definecolor{royalblue}{rgb}{0.25,0.41,0.88}
\definecolor{saddlebrown}{rgb}{0.55,0.27,0.07}
\definecolor{salmon1}{rgb}{1.00,0.55,0.41}
\definecolor{salmon2}{rgb}{0.93,0.51,0.38}
\definecolor{salmon3}{rgb}{0.80,0.44,0.33}
\definecolor{salmon4}{rgb}{0.55,0.30,0.22}
\definecolor{salmon}{rgb}{0.98,0.50,0.45}
\definecolor{sandybrown}{rgb}{0.96,0.64,0.38}
\definecolor{seagreen}{rgb}{0.18,0.55,0.34}
\definecolor{seashell1}{rgb}{1.00,0.96,0.93}
\definecolor{seashell2}{rgb}{0.93,0.90,0.87}
\definecolor{seashell3}{rgb}{0.80,0.77,0.75}
\definecolor{seashell4}{rgb}{0.55,0.53,0.51}
\definecolor{seashell}{rgb}{1.00,0.96,0.93}
\definecolor{sienna1}{rgb}{1.00,0.51,0.28}
\definecolor{sienna2}{rgb}{0.93,0.47,0.26}
\definecolor{sienna3}{rgb}{0.80,0.41,0.22}
\definecolor{sienna4}{rgb}{0.55,0.28,0.15}
\definecolor{sienna}{rgb}{0.63,0.32,0.18}
\definecolor{skyblue}{rgb}{0.53,0.81,0.92}
\definecolor{slateblue}{rgb}{0.42,0.35,0.80}
\definecolor{slategray}{rgb}{0.44,0.50,0.56}
\definecolor{slategrey}{rgb}{0.44,0.50,0.56}
\definecolor{snow1}{rgb}{1.00,0.98,0.98}
\definecolor{snow2}{rgb}{0.93,0.91,0.91}
\definecolor{snow3}{rgb}{0.80,0.79,0.79}
\definecolor{snow4}{rgb}{0.55,0.54,0.54}
\definecolor{snow}{rgb}{1.00,0.98,0.98}
\definecolor{springgreen}{rgb}{0.00,1.00,0.50}
\definecolor{steelblue}{rgb}{0.27,0.51,0.71}
\definecolor{tan1}{rgb}{1.00,0.65,0.31}
\definecolor{tan2}{rgb}{0.93,0.60,0.29}
\definecolor{tan3}{rgb}{0.80,0.52,0.25}
\definecolor{tan4}{rgb}{0.55,0.35,0.17}
\definecolor{tan}{rgb}{0.82,0.71,0.55}
\definecolor{thistle1}{rgb}{1.00,0.88,1.00}
\definecolor{thistle2}{rgb}{0.93,0.82,0.93}
\definecolor{thistle3}{rgb}{0.80,0.71,0.80}
\definecolor{thistle4}{rgb}{0.55,0.48,0.55}
\definecolor{thistle}{rgb}{0.85,0.75,0.85}
\definecolor{tomato1}{rgb}{1.00,0.39,0.28}
\definecolor{tomato2}{rgb}{0.93,0.36,0.26}
\definecolor{tomato3}{rgb}{0.80,0.31,0.22}
\definecolor{tomato4}{rgb}{0.55,0.21,0.15}
\definecolor{tomato}{rgb}{1.00,0.39,0.28}
\definecolor{turquoise1}{rgb}{0.00,0.96,1.00}
\definecolor{turquoise2}{rgb}{0.00,0.90,0.93}
\definecolor{turquoise3}{rgb}{0.00,0.77,0.80}
\definecolor{turquoise4}{rgb}{0.00,0.53,0.55}
\definecolor{turquoise}{rgb}{0.25,0.88,0.82}
\definecolor{violetred}{rgb}{0.82,0.13,0.56}
\definecolor{violet}{rgb}{0.93,0.51,0.93}
\definecolor{wheat1}{rgb}{1.00,0.91,0.73}
\definecolor{wheat2}{rgb}{0.93,0.85,0.68}
\definecolor{wheat3}{rgb}{0.80,0.73,0.59}
\definecolor{wheat4}{rgb}{0.55,0.49,0.40}
\definecolor{wheat}{rgb}{0.96,0.87,0.70}
\definecolor{whitesmoke}{rgb}{0.96,0.96,0.96}
\definecolor{white}{rgb}{1.00,1.00,1.00}
\definecolor{yellow1}{rgb}{1.00,1.00,0.00}
\definecolor{yellow2}{rgb}{0.93,0.93,0.00}
\definecolor{yellow3}{rgb}{0.80,0.80,0.00}
\definecolor{yellow4}{rgb}{0.55,0.55,0.00}
\definecolor{yellowgreen}{rgb}{0.60,0.80,0.20}
\definecolor{yellow}{rgb}{1.00,1.00,0.00}
\usepackage{amsfonts}
\usepackage{bm}
\usepackage{longtable}
\usepackage{multirow}
\usepackage{pifont}
\usepackage{gensymb}

\usepackage{booktabs} 
 
\newcommand{\gskfont}{
  \bfseries 
  \color{red}
}

\newcommand{\yxfont}{
  \bfseries 
  \color{blue}
}

\newcommand{\jcafont}{
  \bfseries 
  \color{green}
}

\newcommand{\vsfont}{\bfseries{}\color{orange}}

\newcommand{\nhfont}{
  \bfseries 
  \color{grey}
}

\newcommand{\vpfont}{
  \bfseries 
  \color{magenta}
}

\DeclareTextFontCommand{\gsk}{\gskfont}
\DeclareTextFontCommand{\yx}{\yxfont}
\DeclareTextFontCommand{\jca}{\jcafont}
\DeclareTextFontCommand{\vs}{\vsfont}
\DeclareTextFontCommand{\nh}{\nhfont}
\DeclareTextFontCommand{\vp}{\vpfont}
\DeclareTextFontCommand{\hw}{\vpfont}

\shorttitle{{Non-Thermal Collisional Ionisation of Helium}}
\shortauthors{Kerr, Xu, Allred, Polito, Sadykov, Huang \& Wang}

\begin{document}

%%%%%%%%%%%%%%%%%%%%%%%%%%%%%%%%%%%%%%%%%%%%%%%%%%%%%
%%%%%%%%%%%%%%%%%%%%     TITLE & ABSTRACT     %%%%%%%%%%%%%%%%%%%%
%%%%%%%%%%%%%%%%%%%%%%%%%%%%%%%%%%%%%%%%%%%%%%%%%%%%%

	\title{He~\textsc{i} 10830~\AA\ Dimming During Solar Flares, I: The Crucial Role of Non-Thermal Collisional Ionisations}
	\author{Graham~S. Kerr}
	\email{graham.s.kerr@nasa.gov}
	\email{kerrg@cua.edu}
	\affil{NASA Goddard Space Flight Center, Heliophysics Sciences Division, Code 671, 8800 Greenbelt Rd., Greenbelt, MD 20771, USA}
 	\affil{Department of Physics, Catholic University of America, 620 Michigan Avenue, Northeast, Washington, DC 20064, USA}
 
	 \author{Yan Xu}
	 \affil{Institute for Space Weather Sciences, New Jersey Institute of Technology, 323 Martin Luther King Boulevard, Newark, NJ 07102-1982}
	 \affil{Big Bear Solar Observatory, New Jersey Institute of Technology, 40386 North Shore Lane, Big Bear City, CA 92314-9672, USA}
	
	\author{Joel~C. Allred}
	\affil{NASA Goddard Space Flight Center, Heliophysics Sciences Division, Code 671, 8800 Greenbelt Rd., Greenbelt, MD 20771, USA}

        \author{Vanessa Polito}
	\affiliation{Bay Area Environmental Research Institute, NASA Research Park,  Moffett Field, CA 94035-0001, USA}
        \affiliation{Lockheed Martin Solar and Astrophysics Laboratory, Building 252, 3251 Hanover Street, Palo Alto, CA 94304, USA}

	\author{Viacheslav~M. Sadykov}
	\affil{Physics \& Astronomy Department, Georgia State University, 25 Park Place NE, Atlanta, GA 30303, USA}
	
	\author{Nengyi Huang}
	\affil{Institute for Space Weather Sciences, New Jersey Institute of Technology, 323 Martin Luther King Boulevard, Newark, NJ 07102-1982}
  
	\author{Haimin Wang}
	 \affil{Institute for Space Weather Sciences, New Jersey Institute of Technology, 323 Martin Luther King Boulevard, Newark, NJ 07102-1982}
	 \affil{Big Bear Solar Observatory, New Jersey Institute of Technology, 40386 North Shore Lane, Big Bear City, CA 92314-9672, USA}
	
	\date{Received / Accepted}
	
	
	\begin{abstract}	
	While solar flares are predominantly characterised by an intense broadband enhancement to the solar radiative output, certain spectral lines and continua will, in theory, exhibit flare-induced \textsl{dimmings}. Observations of transitions of orthohelium He~\textsc{i} $\lambda\lambda 10830$~\AA\ and the He~\textsc{i} D3 lines have shown evidence of such dimming, usually followed by enhanced emission. It has been suggested that non-thermal collisional ionisation of helium by an electron beam, followed by recombinations to orthohelium, is responsible for overpopulating the those levels, leading to stronger absorption. However it has not been possible observationally to preclude the possibility of overpopulating orthohelium via enhanced photoionisation of He~\textsc{i} by EUV irradiance from the flaring corona followed by recombinations. Here we present radiation hydrodynamics simulations of non-thermal electron beam-driven flares where (1) both non-thermal collisional ionisation of Helium and coronal irradiance are included, and (2) only coronal irradiance is included. A grid of simulations covering a range of total energies deposited by the electron beam, and a range of non-thermal electron beam low-energy cutoff values, were simulated. In order to obtain flare-induced dimming of the He~\textsc{i} 10830~\AA\ line it was necessary for non-thermal collisional ionisations to be present. The effect was more prominent in flares with larger low-energy cutoff values and longer lived in weaker flares and flares with a more gradual energy deposition timescale. These results demonstrate the usefulness of orthohelium line emission as a diagnostic of flare energy transport.
 	\end{abstract}

%%%%%%%%%%%%%%%%%%%%%%%%%%%%%%%%%%%%%%%%%%%%%%%%%%%%%
%%%%%%%%%%%%%%%%%%%%     INTRODUCTION      %%%%%%%%%%%%%%%%%%%%%
%%%%%%%%%%%%%%%%%%%%%%%%%%%%%%%%%%%%%%%%%%%%%%%%%%%%%

\section{Introduction}\label{sec:intro}

The intense broadband enhancement to the solar radiative output, following deposition of tremendous amounts of energy released during magnetic reconnection, is what characterises solar flares. As well as these brightenings, models of flare radiation have suggested that there may be flare-induced dimmings at certain wavelengths. Typically these concern the Balmer and Paschen continua and are known as `black-light flares' (BLF), or `negative flares'. Despite their modest magnitude and transient nature, BLFs may be a powerful means to constrain the magnitude, temporal profile, spectral properties, and mechanism of energy transported from the coronal release site through the transition region (TR) and chromosphere. The strength and duration of such dimmings depend on the magnitude of energy deposited, but they are typically thought to be transient and difficult to observe due to their small contrast. Indeed no unambiguous continuum BLF has been observed \citep[e.g.][]{1990A&A...233..577H,1994SoPh..152..145V,2003A&A...403.1151D}. On the other hand, a similar phenomenon \textsl{has} been convincingly observed in solar and stellar flare spectra of orthohelium (see discussion below). That is, flare-induced dimmings of the He~\textsc{i} 10830~\AA\ and He~\textsc{i} D3 lines. 

In the standard model of solar flares the bulk of the flare energy is transported by a directed beam of non-thermal electrons, accelerated out of the ambient thermal background during magnetic reconnection. As these electrons propagate at relativistic speeds through the solar atmosphere they lose energy primarily via Coulomb collisions, ultimately thermalising in the chromosphere or TR. Hard X-ray \textsl{bremsstrahlung} is observed at the site of thermalisation, in compact footpoint sources, from which the properties of the non-thermal electron distribution can be inferred \citep[e.g.][]{2011SSRv..159..107H}. The chromospheric and TR plasma rapidly heats and ionises, and strong mass flows develop: chromospheric ablations (upward flows, also referred to as evaporation) and condensations (dense, downward directed flows). 

In addition to heating the plasma, the precipitating non-thermal electrons (or protons, or other ions) can collisionally excite or ionise the target atoms or ions, referred to as non-thermal collisional ionisation/excitation, $C_{nt}$. It has been known for some time that non-thermal effects are important for the response of hydrogen continua and spectral line transitions \citep[e.g.][and references therein]{1986A&A...168..301A,1993A&A...274..917F,2009A&A...499..923K}. Forward models of hydrogen continua in flares have shown that brief dimmings may result due to the increased population of excited states following non-thermal collisional excitation of hydrogen, increasing absorption of photospheric photons \citep[e.g.][]{1990A&A...233..577H,1999ApJ...521..906A}. 

While there is a lack of convincing observations of continuum BLFs, spectral transitions of orthohelium have shown convincing evidence of dimming, with the appearance of `negative' flare ribbons. That is, the background subtracted emission in flares is weaker than the pre-flare emission, with an enhanced absorption line profile. A remarkable flare observation in the He D3 line at 5876~\AA\ was carried out by Dr. Zirin \citep{1980ApJ...235..618Z} on 1978-July-10 at Big Bear Solar Observatory (BBSO).  A dark shell in the He~\textsc{i}~D3 line was observed at the very beginning of the flare and eventually turned to emission. Later on, using the digitized film data, \cite{2013ApJ...774...60L} reported an M6.3 flare on 1984-May-22 with an initial dimming about $5$\% in He~\textsc{i}~D3 that persisted for four minutes. 

In the He~\textsc{i}~10830~\AA\ line, emission has usually been observed during flares \citep[e.g.][]{1992ApJ...389L..33Y, 1995ApJ...441L..51P, 2000SoPh..197..313P,2006SoPh..235..107L,2013ApJ...774...60L,2014ApJ...793...87Z}. There is a correlation between the intensity of the line and the X-ray flux, and emission only appears in moderate-to-strong flares. This threshold is not definitively known, but seems to be around mid-C class  \citep{2007SoPh..241..301L,Du_2008}. 

Though rare, He~\textsc{i} 10830~\AA\ dimmings have now been observed in several flares. For example, \cite{1984SoPh...91..127H} presented observations of persistent dark flare ribbons during a long-duration event. More recently, \citet{2016ApJ...819...89X} found dimmings on the leading edge of the propagating ribbons in two M-class flares, using the 1.6~m GST/BBSO \citep{2012ASPC..463..357G} observations that have provided the highest spatial resolutions so far. Their observations were carried out using a Lyot filter \citep{2010AN....331..636C} tuned to the blue wing (10830.05~\AA~$\pm$~0.25~\AA), and showed dimming of the leading edge of the flare ribbon. The dark ribbon front observed in He~\textsc{i}~10830~\AA\ had a characteristic width $\approx350-500$~km, with a dimming up to $13.7$\%. Simultaneously, blue-wing enhanced H$\alpha$ emission and significantly broadened Mg~\textsc{ii} h \& k lines, which retained their central reversal, were observed at the same location where He~\textsc{i}~10830~\AA\ darkening is seen. Enhanced absorption was confined to the propagating ribbon front, the presumed site of energy injection at the base of newly reconnected magnetic flux loops, and persisted for several tens to over a hundred seconds before turning into a positive contrast. This positive contrast was associated with the trailing region of the flare ribbons. Spectroscopic observations of a small flare (of B2 class) were analyzed by \cite{2018JASTP.173...50K}, who saw only enhanced absorption of up to $25$\%, appearing in two minima but with no subsequent increase in emission. Asymmetries and Doppler shifts were detected along with increases in line width.

Stellar flare observations of 10830~\AA\  are even rarer than the solar case, but demonstrate similar patterns. On M Dwarf flare stars He~\textsc{i} 10830~\AA\ is generally in emission \citep{2012ApJ...745...14S}, strongly broadened and asymmetric \citep{2020A&A...640A..52F}. A strong flare response in He~\textsc{i} 10830~\AA\ was typically accompanied by broad, asymmetric H$\alpha$ profiles \citep{2020A&A...640A..52F}. There has now been a detection of He~\textsc{i} 10830~\AA\ enhanced absorption observed at the onset of M Dwarf flares using data from the CARMENES study \citep{2020A&A...640A..52F}. When He~\textsc{i} 10830~\AA\ was undergoing flare-induced dimming, the H$\alpha$ showed a modest response coupled with a blue wing asymmetry. 

Transitions in the triplet state of He~\textsc{i} (orthohelium) form the 10830 and D3 lines. He~\textsc{i} 10830~\AA\ is a self-blend of three transitions closely spaced in wavelength, from He~\textsc{i} 2s $^{3}$S$_{1}$$\rightarrow$2p$^{3}$P$_{2,1,0}$ transitions. The He~\textsc{i} D3 lines are a multiplet of lines near 5875~\AA, from the He~\textsc{i} 2p$^{3}$P$_{2,1,0}$$\rightarrow$3d$^{3}$D$_{3,2,1}$ transitions. They generally appear in absorption in the quiet Sun \citep{1997ApJ...489..375A}, but have been observed to go into emission during flares as described above, or in more localised heating events \citep[see][and references therein]{2019A&A...621A..35L,2020arXiv201015946L}. The excitation energy between the ground and lower level of the triplet is $\sim20$~eV, meaning that thermal electron collisions in the chromosphere are unable to directly populate orthohelium (we refer to this as the thermal collisional mechanism, CM). Without a sufficiently large population of orthohelium to absorb photospheric 10830~\AA\ photons, the strong absorption line that is observed cannot form. This led \cite{1939ApJ....89..673G} to propose  photoionisation of He~\textsc{i} followed by recombinations to orthohelium as a population mechanism. This is known as the photoionisation-recombination mechanism (PRM). The source of $\lambda < 504$~\AA\ photons is the extreme ultraviolet (EUV) and soft X-ray emission from the hot overlying corona, which irradiates the chromosphere. He~\textsc{i} 10830~\AA\ is therefore a good tracer of coronal activity. Extensive modelling and observations have confirmed the PRM as the dominant pathway to populating orthohelium in the quiet Sun \citep[e.g.][]{1975ApJ...199L..63Z,1994IAUS..154...35A,1997ApJ...489..375A,2008ApJ...677..742C,2016A&A...594A.104L}.

To explain flare observations of He~\textsc{i} 10830~\AA\ and He~\textsc{i} D3 dimming, it has been speculated that the PRM becomes enhanced during the flare. Coronal EUV irradiance would increase due to heating and chromospheric ablation which substantially raises the emission measure, leading to excess helium ionisation over the pre-flare state. Overpopulated orthohelium would proceed to absorb more 10830~\AA\ photons, strengthening the absorption line. Once the plasma becomes sufficiently hot and dense, thermal collisions drive the line into emission. This mechanism has been used to interpret solar and stellar flare observations of orthohelium \citep[e.g.][]{2014ApJ...793...87Z,2018JASTP.173...50K,2020A&A...640A..52F}. 

An alternative idea to the PRM is that non-thermal collisional ionisation of Helium is responsible for the negative flares \citep{2005A&A...432..699D}. Non-thermal ionisation of He~\textsc{i} $\rightarrow$ He~\textsc{ii} followed by recombinations to orthohelium would overpopulate those levels relative to the pre-flare, with absorption and emission of orthohelium lines occurring as described before. We refer to this mechanism as the non-thermal collisional-recombination mechanism (CRM). CRM would demand the presence of a non-thermal electron distribution in the chromosphere during the flare (e.g. from bombardment by a beam of accelerated electrons), the properties of which would affect the magnitude and duration of dimming and enhancement. PRM would not require accelerated electrons but simply a flare heated, mass loaded corona. It is important to note that coronal irradiance would occur in both models, as regardless of the flare energy transport mechanism the corona is heated.

There has been a relatively little number of forward modelling attempts of orthohelium in solar or stellar flares, but past modelling has yielded interesting results. Using semi-empirical flare atmospheres \cite{2005A&A...432..699D} simulated the effect of inclusion of non-thermal collisional ionisation on He~\textsc{i} 10830~\AA. Applied to the the VAL-C atmosphere \citep{1981ApJS...45..635V} these non-thermal collisions led to enhanced absorption, and in the semi-empircal flare atmospheres F1 \& F2 \citep{1980ApJ...242..336M} the simulations with non-thermal collisions resulted in increased emission over the simulations without $C_{nt}$. However, while they included EUV radiation from the corona, this was not time-dependent, and those authors could not fully model both CRM and PRM in a realistic manner simultaneously. \cite{2020ApJ...897L...6H} considered both CRM and PRM in realistic time-dependent \texttt{RADYN} simulations of electron beam-driven flares produced by the F-CHROMA consortium\footnote{\url{https://star.pst.qub.ac.uk/wiki/doku.php/public/solarmodels/start}}. The implementation of these methods are described below.  They did find enhanced absorption in He~\textsc{i} 10830~\AA\ emission, and found some simulations with contrasts similar to observed albeit much shorter lived. However, \cite{2020ApJ...897L...6H} did not separate the effects of each mechanism, meaning it is not clear if the dominant role is CRM or PRM. 

Having previously confirmed that our state-of-the-art electron-beam driven flare radiation hydrodynamic models can exhibit He~\textsc{i} 10830~\AA\ dimming \citep{2020ApJ...897L...6H}, we can now determine if inclusion of CRM is strictly necessary, or if PRM alone can produce enhanced absorption. Determining if CRM is indeed the dominant mechanism is the first step in utilising Helium negative flares to diagnose energy transport in both solar and stellar flares. One set of experiments presented here includes both the CRM and PRM methods, the other only includes PRM. The impact on He~\textsc{i} 10380~\AA\ is determined in each case. 

%%%%%%%%%%%%%%%%%%%%%%%%%%%%%%%%%%%%%%%%%%%%%%%%%%%%%
%%%%%%%%%%%%%%     NUMERICAL EXPERIMENTS      %%%%%%%%%%%%%%%%%%%%%
%%%%%%%%%%%%%%%%%%%%%%%%%%%%%%%%%%%%%%%%%%%%%%%%%%%%%

%%%%%
\section{Numerical Experiments}\label{sec:numex}
%%%%%

%%%%%
\subsection{The \texttt{RADYN} code \& Modelling Helium in Flares}\label{sec:radyndescript}
%%%%%
To explore the response (this work) and formation properties (to appear in a forthcoming work) of the He~\textsc{i} 10830~\AA\ line during solar flares, we employ the radiation hydrodynamics (RHD) code \texttt{RADYN} \citep{1992ApJ...397L..59C,1995ApJ...440L..29C,1997ApJ...481..500C,2005ApJ...630..573A,2015ApJ...809..104A}, which is ideally suited for such a study. Non-equilibrium atomic level populations and ionisation of hydrogen, helium, and calcium are included, as are photoionisations from coronal irradiation of the chromosphere. Both of these processes are known to be important to accurately consider helium ionisation stratification and in populating orthohelium \citep[e.g.][]{1939ApJ....89..673G,1994IAUS..154...35A,1997ApJ...489..375A,2014ApJ...784...30G,2016ApJ...817..125G,2016A&A...594A.104L}. Non-thermal collisional ionisation of helium is also included in our simulations, so that both the PRM and CRM methods of populating orthohelium are considered. \texttt{RADYN} is a well-established resource that has been used to investigate flare atmospheric dynamics, the radiative response to flares, and to understand energy transport processes in both solar and stellar flares \citep[recent examples include][]{2015SoPh..290.3487K, 2017ApJ...836...12K, 2016ApJ...827...38R, 2016ApJ...827..101K, 2019ApJ...871...23K, 2020ApJ...900...18K, 2017A&A...605A.125S,2018ApJ...862...59B,2018ApJ...856..178P, 2019ApJ...879L..17P,2020ApJ...897L...6H}
\begin{figure}[ht]
	\centering 
	{\includegraphics[width = 0.46\textwidth, clip = true, trim = 0.cm 0.cm 0.cm 0.cm]{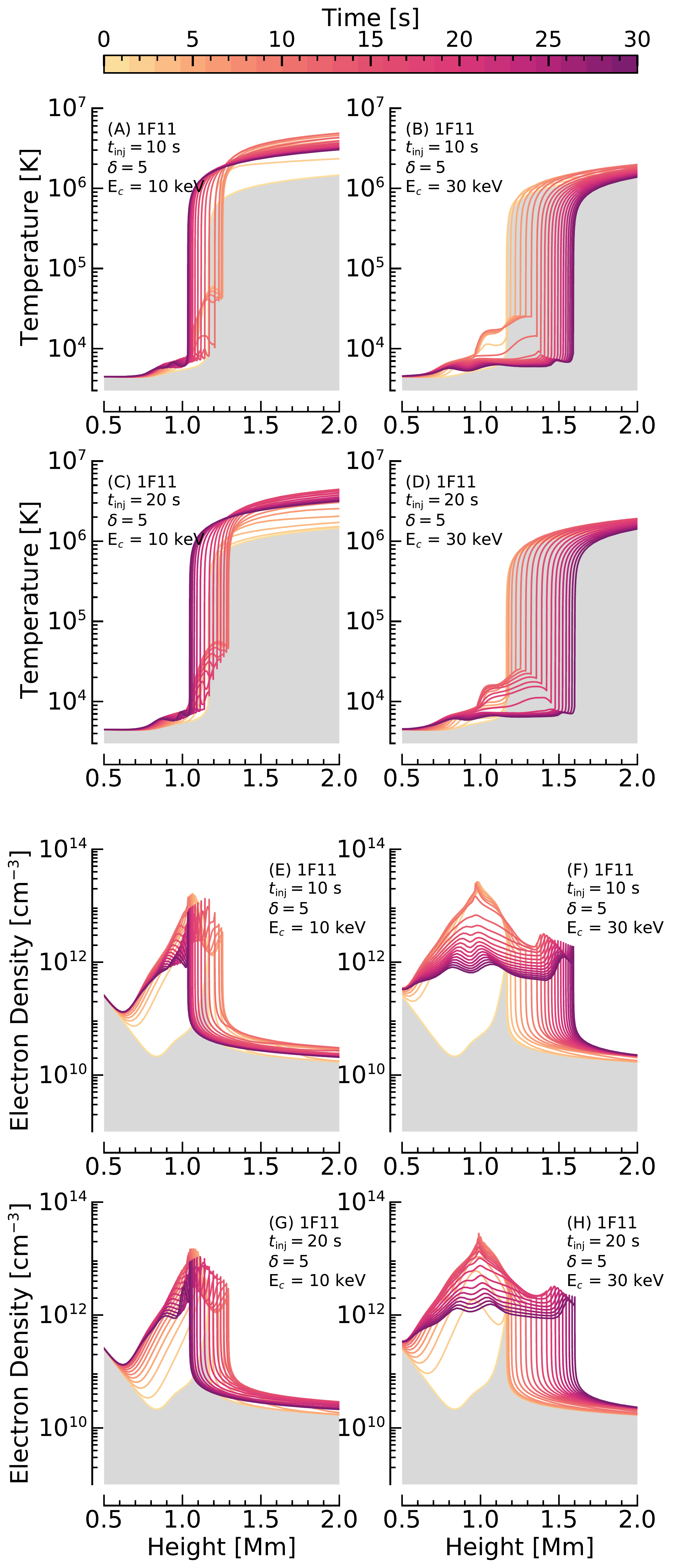}}
	\caption{\textsl{Temperature (top 4 panels) and electron density (bottom 4 panels) stratification as a function of time in the electron beam driven flare simulations. Total energy flux is $1\times10^{11}$~erg~cm$^{-2}$ with $\delta = 5$. Panels (A, B, E, F) show the $\tau_{\mathrm{inj}} = 10$~s scenario and panels (C, D, G, H) show $\tau_{\mathrm{inj}} = 20$~s. The simulations in the lefthand column have $E_{\mathrm{c}} = 10$~keV, and in the righthand column have $E_{\mathrm{c}} = 30$~keV. The greyscale shows the pre-flare stratification.}}
	\label{fig:atmos_beam}
\end{figure}
Some pertinent points are listed here, but for full details of \texttt{RADYN}, in particular the flare versions, consult \cite{1999ApJ...521..906A,2005ApJ...630..573A,2015ApJ...809..104A}. \texttt{RADYN} is a one-dimensional magnetic field field-aligned RHD numerical model, with an adaptive grid \citep{1987JCoPh..69..175D}, enabling resolution of shocks and strong gradients that typically form in flare simulations. 

Helium is included as a 9-level-with-continuum atom, with the ground states of He~\textsc{i}, He~\textsc{ii}, and He~\textsc{iii}, and  various excited states. The upper and lower levels of the He~\textsc{i} 10830~\AA\ line are included, but the sublevels of the $^{3}P$ state are modelled as a single level, with the statistical weight equal to the sum of the weights of the sublevels. This means only one component of the triplet is forward modelled, which is sufficient for this study. A future study will investigate the full triplet as well as the He~\textsc{i} D3 lines. Photoionisations-recombinations (bound-free transitions) from both the ground and excited states are included. \texttt{RADYN} does not include overlapping bound-bound transitions.% so that He~\textsc{ii} 304~\AA\ photons do not photoionise He~\textsc{i}.

Optically thin losses are included by summing the emissivity of all transitions from the CHIANTI \citep[V7.1.3;][]{1997A&AS..125..149D,2013ApJ...763...86L} atomic database (apart from those transitions included in the detailed radiation transfer). Half of the optically thin losses are directed downwards, and act to heat or ionise the denser layers of the atmosphere. Emissivities from CHIANTI were tabulated on a wavelength and temperature grid, and are the product of the emissivities and emission measures integrated through the TR and coronal portion of the loop. This is then included as a downward directed source of irradiation when solving the radiation transfer problem, in the manner described by \cite{1994ApJ...433..417W} and in \cite{2015ApJ...809..104A}. We can therefore account for the time-varying EUV irradiation of the chromosphere during the flare, which affects the formation of He~\textsc{i} 10830~\AA.

To model solar flares, a non-thermal electron distribution (described by a power-law energy spectrum with total instantaneous flux $F_{\mathrm{ins}}$ above a low-energy cutoff $E_{c}$, with spectral slope $\delta$) is injected at the apex of the loop, which propagates at relativistic speeds through the atmosphere. The atmosphere consists of sub-photosphere, photosphere, chromosphere, TR, and corona, with a total height (a loop half-length) of $z\approx11$~Mm.  The pre-flare atmosphere was in radiative equilibrium, with a coronal temperature and electron density of $T\approx3.15$~MK and $n_{e}\approx7.6\times10^{9}$~cm$^{-3}$. This version of \texttt{RADYN} uses a new code, \texttt{FP} \citep{2020ApJ...902...16A}, to model the evolution of this distribution as it is transported through the plasma, by solving the Fokker-Planck equations, an update from \cite{2015ApJ...809..104A}.  
Importantly, we make no assumption to the temperature of the target onto which the electrons precipitate. That is, we do not need to assume either a warm or a cold target, the actual temperature of the plasma is used in our solution. \texttt{FP} can act either as a standalone code or can be merged with RHD codes as we have done here with \texttt{RADYN}. It is a marked improvement over prior efforts and we encourage the reader to consult \cite{2020ApJ...902...16A} to learn more about this open-source code.
\begin{figure*}[ht]
	\centering 
	{\includegraphics[width = 0.85\textwidth, clip = true, trim = 0.cm 0.cm 0.cm 0.cm]{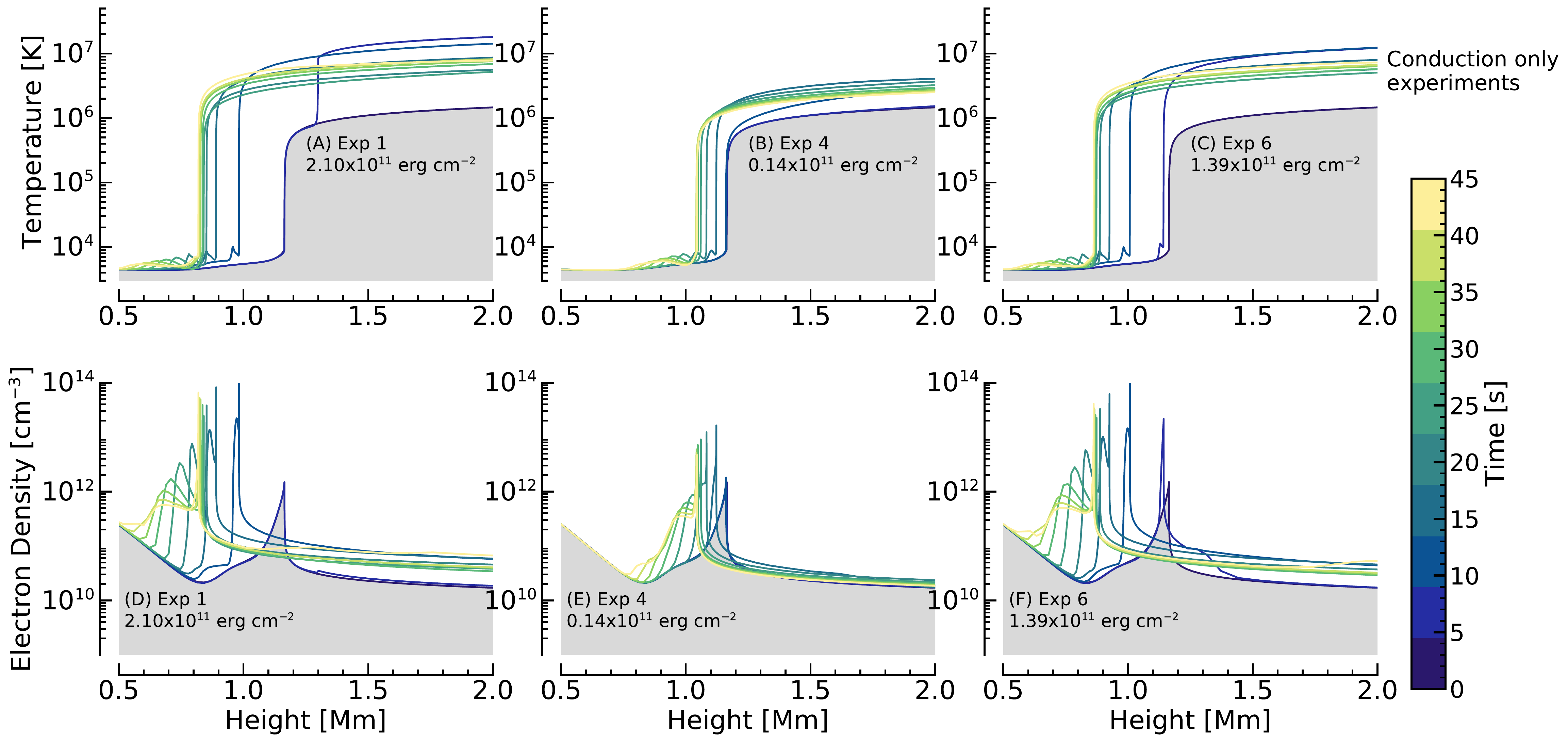}}
	\caption{\textsl{Temperature (top row) and electron density (bottom row) stratification as a function of time in the conduction-only flare simulations. Experiments 1, 4 \& 6 are shown (total energy flux indicated). The greyscale shows the pre-flare stratification.}}
	\label{fig:atmos_cond}
\end{figure*}
Non-thermal collisional ionisations and excitations from the ground state of hydrogen are included using the formalism of \cite{1993A&A...274..917F}. Non-thermal collisional ionisation from the ground state of He~\textsc{i}~$\rightarrow$~He~\textsc{ii}, and from ground state of He~\textsc{ii}~$\rightarrow$~He~\textsc{iii} are included by integrating the product of the collisional cross-sections $\sigma_{ij}$ and the non-thermal electron energy distribution function $v_{b}f$, over energy, $E$, and pitch angle $\mu$:

\begin{equation}\label{eq:cntrates}
C_{nt} = 2\pi \int\int \mu v_{b}f\sigma_{ij}\mathrm{d}E\mathrm{d}\mu.
\end{equation}

Parameterised values of $\sigma_{ij}$ as functions of electron energy were obtained from \cite{1985A&AS...60..425A}. This differs from the approach of \cite{2005A&A...432..699D} who instead assumed that the He~$C_{nt}$ could be obtained from the hydrogen rates $C_{nt,~hyd}$ multiplied by the ratio of helium to hydrogen cross-sections at $30$~keV: $C_{nt}$ = $\sigma_{He}/\sigma_{H}~C_{nt,~hyd} = 1.2C_{nt,~hyd}$. It is unclear to us which method is most appropriate, and so we elected to keep using the \cite{1985A&AS...60..425A} rates that have been employed by \texttt{RADYN} since \cite{2015ApJ...809..104A}. This also lets us model $C_{nt}$ of He~\textsc{i}~$\rightarrow$~He~\textsc{ii}~$\rightarrow$~He~\textsc{iii}, whereas  \cite{2005A&A...432..699D} only considered $C_{nt}$ of He~\textsc{i}~$\rightarrow$~He~\textsc{ii}.

%%%%%
\subsection{Flare Simulations}\label{sec:flaresims}
%%%%%

A grid of fifty non-thermal electron beam driven flare simulations were produced using \texttt{RADYN}, where we varied the magnitude of total energy flux ($F_{\mathrm{tot}}$ erg~cm$^{-2}$) injected, the timescale of energy injection ($\tau_{\mathrm{inj} s}$), so that the instantaneous energy flux ($F_{\mathrm{ins}}$ erg~cm$^{-2}$~s$^{-1}$) varied even for simulations with the same $F_{\mathrm{tot}}$, and the low-energy cutoff of the distribution ($E_{\mathrm{c}}$~keV). The spectral index $\delta = 5$ was fixed. This allowed a range of flare strengths, and spectral energy distributions to be studied. We vary the energy distribution by varying the low-energy cutoff of the distribution. A larger low-energy cutoff means that there are more high-energy electrons that can penetrate deeper into the chromosphere (since low-energy electrons thermalise at smaller column depths). We note that the descriptors of `hard' or `soft' non-thermal electron distributions generally refer to variations of $\delta$ since energy is distributed toward higher or lower energy electrons, respectively. However, in this work we refer to a `harder' spectrum as simulations with larger values of $E_{c}$ since there are more high-energy electrons compared to `softer' spectra with more low-energy electrons. We used five values of $E_{c}$ in our experiments: $E{c} = [10,15,20,25,30]$~keV. These experiments were repeated with non-thermal collisional ionisations of helium omitted, giving 100 simulations in total. 

The total injected energy fluxes were $F_{\mathrm{total}} = [1\times10^{10},~5\times10^{10},~1\times10^{11},~5\times10^{11},~1\times10^{12}]$~erg cm$^{-2}$. Hereafter the notation $X\mathrm{F}Y$ means $F_{\mathrm{total}} = X\times10^{Y}$~erg cm$^{-2}$ (this differs somewhat from other uses of $X\mathrm{F}Y$ in the flare modelling literature where this typically refers to the instantaneous energy flux, but since we are comparing injection timescales we focus on the total energy flux). Those fluxes were injected either over $\tau_\mathrm{inj} = [10,~20]$~s. The former means that the instantaneous energy flux $F_{\mathrm{ins}}$ was constant, the latter means that $F_{\mathrm{ins}}$ ramped up to a peak at $t = 10$~s then decreased in a triangular profile.  Once heating ceased the experiments ran to $t=50$~s. 

Examples illustrating the differences in the atmospheric response that arise from varying the injection timescale or spectral hardness of the electron distribution are shown in Figure~\ref{fig:atmos_beam}. The temperature and electron density stratification at 1.5~s cadence for the first 30~s of the 1F11 simulations are shown for $\tau_{\mathrm{inj}} = 10$~s (panels A,B,E,F) and $\tau_{\mathrm{inj}} = 20$~s (panels C,D,G,H). The lefthand column have $E_{\mathrm{c}} = 10$~keV (softest distribution) and the righthand are $E_{\mathrm{c}} = 30$~keV (hardest distribution).  While both heating durations (10~s or 20~s) ultimately produce fairly similar atmospheres by the end of their respective heating phases, the differences in the time taken to reach certain temperatures and electron densities will have a consequential impact on the radiative response. It is clear that the softer distributions strongly heat and ionise the upper chromosphere/lower TR due to the larger proportion of low energy electrons that more easily thermalise at the low column depth. The harder distribution can penetrate deeper, resulting in a lower magnitude of heating but much wider swathe of the chromosphere significantly involved in the flare. This shifts the peak electron density deeper. The electron density enhancement is governed not only by the temperature response but, particularly at greater depth, by hydrogen $C_{nt,hyd}$. Not shown are the velocities driven by energy injection, but in all experiments there is an upflow, that increases in magnitude with increasing flare energy, from a few~$\times10$~km~s$^{-1}$ to several~$\times100$~km~s$^{-1}$. The higher energy simulations also drive downflows of a few~$\times10$~km~s$^{-1}$, that occur more commonly in flares with softer non-thermal electron distributions. The experiments shown here all include He~$C_{nt}$. Those that omit He~$C_{nt}$ are very similar with only subtle differences, likely due to the importance of He~\textsc{ii} 304~\AA\ to radiative losses. 

\begin{table}[htb]
  \begin{center}
    \caption{Conduction-only Flare Simulations}
    \label{tab:conductexp}
    \begin{tabular}{{ c | c  | l |}}
   \toprule
& {\multirow{2}{*}{\textbf{Total Energy Flux}}} & {\multirow{2}{*}{\textbf{Energy Deposition Location}}}\\
 &[erg~s$^{-1}$~cm$^{-2}$] & \\
     \toprule
     \toprule
\textbf{Exp. 1} &  $2\times10^{11}$ & Equally over $11-9$~Mm  \\
\midrule
\textbf{Exp. 2} &  $2\times10^{10}$ & Equally over $11-9$~Mm  \\
\midrule
\textbf{Exp. 3} &  $2\times10^{9}$ & Equally over $11-9$~Mm  \\
\midrule
\textbf{Exp. 4} &  $1.4\times10^{10}$ & Constant over $11-5$~Mm then \\
& & decreasing to zero between $5-3$~Mm \\
\midrule
\textbf{Exp. 5} &  $7\times10^{9}$ & Constant over $11-5$~Mm then \\
& & decreasing to zero between $5-3$~Mm \\
\midrule
\textbf{Exp. 6} &  $1.4\times10^{11}$ & Constant over $11-5$~Mm then \\
& & decreasing to zero between $5-3$~Mm \\
\midrule
\textbf{Exp. 7} &  $1.6\times10^{11}$ & Equally over $11-3$~Mm \\
\bottomrule
    \end{tabular}
  \end{center}
\end{table}

While the majority of this work discusses electron beam driven flares we additionally performed seven experiments where energy was deposited directly in the corona and transported solely by the thermal conductive flux. Those experiments, listed in Table~\ref{tab:conductexp}, were not designed to be comprehensive. Rather they serve to be illustrative of how direct \textsl{in situ} heating at the loop apex resulting from reconnection might result in coronal heating, thereby increasing the irradiation of the chromosphere, before the chromosphere itself is directly heated. The atmospheric stratification for experiments 1, 4 \& 6 are shown in Figure~\ref{fig:atmos_cond}.  Gradients in the chromosphere are sharper than in the electron beam simulations, with the TR being pushed to greater depth in each case. The flaring corona precedes the flaring chromosphere by several seconds. While the coronal temperature near the top of the loop increases rapidly, the density (therefore emission measure) takes longer as the corona is mass loaded by evaporation of chromospheric material so it is only when the heat flux reaches the chromosphere that this commences.
 
\subsection{Modelling He~\textsc{i} 10830~\AA\ with \texttt{RADYN+RH}}\label{sec:rhdescript}
\texttt{RADYN} does not consider overlapping transitions, meaning that photoionisations of He~\textsc{i} by the He~\textsc{ii} 304~\AA\ line are not present. We can explore the importance of this effect using the \texttt{RH} radiation transfer code \citep{2001ApJ...557..389U}. \texttt{RH} is a stationary code, that solves the NLTE radiation transfer assuming statistical equilibrium for any species of interest. Overlapping transitions are included for those atomic species. Snapshots of \texttt{RADYN} flare atmospheres can be input to \texttt{RH} (using the NLTE non-equilibrium electron density from \texttt{RADYN} somewhat mitigates the use of statistical equilibrium) to forward model radiation either not modelled by \texttt{RADYN} or to explore additional physical processes. Similar experiments have been performed extensively \citep[e.g.][]{2016ApJ...827..101K,2016ApJ...827...38R,2017ApJ...842...82R,2019ApJ...879...19Z,2020ApJ...895....6G}.

As standard, \texttt{RH} does not include coronal irradiation in the solution of the equation of radiative transfer, nor does it include non-thermal collisional processes. Since PRM is the primary population mechanism for orthohelium in the quiet Sun, and since both the PRM and the CRM (as we demonstrate in this work) are both vital in flares, we used a modified version \texttt{RH} to model helium. Our implementation of coronal irradiation is described in \cite{2019ApJ...883...57K}. Non-thermal collisional ionisation rates for helium were saved for each snapshot of interest, written as a new input file to \texttt{RH}, and added to the thermal collisional ionisation rates. 

Snapshots of the 1F11 $\tau_{\mathrm{inj}} = 20$~s simulation at $0.1$~s cadence were processed through \texttt{RH}, solving the NLTE radiation transfer for a six-level-with-continuum hydrogen atom and nine-level-with-continuum helium atom (the same atomic model solved by \texttt{RADYN}) with several background species treated in LTE as sources of background opacity. For each snapshot, the hydrogen atomic level populations from \texttt{RADYN} were used in \texttt{RH} and fixed so that \texttt{RH} did not iterate to the statistical equilibrium solution for hydrogen, as in \cite{2019ApJ...883...57K}. This meant that non-equilibrium and non-thermal effects were included for hydrogen. This was done for (1) standard \texttt{RH}, (2) \texttt{RH} with coronal irradiation, and (3) \texttt{RH} with coronal irradiation and with non-thermal collisional ionisation of helium. Comparing the resulting He~\textsc{i} 10830~\AA\ emission in each case demonstrated that photoionisation via flare-enhanced He~\textsc{ii} 304~\AA\  was not as important in comparison to coronal emission $\lambda<504$~\AA\ \citep[similar to quiet Sun results of][]{1994ApJ...433..417W} and He~$C_{nt}$, from which we can conclude that omitting this in \texttt{RADYN} would not alter our overall conclusions.

%%%%%%%%%%%%%%%%%%%%%%%%%%%%%%%%%%%%%%%%%%%%%%%%%%%%%
%%%%%%%%%%%%%%%%%%%%%%%%     RESULTS     %%%%%%%%%%%%%%%%%%%%%
%%%%%%%%%%%%%%%%%%%%%%%%%%%%%%%%%%%%%%%%%%%%%%%%%%%%%

%%%%%
\section{He Ionisation Stratification \& Level Populations}
%%%%%
\begin{figure*}[ht]
	\centering 
	{\includegraphics[width = 0.85\textwidth, clip = true, trim = 0.cm 0.cm 0.cm 0.cm]{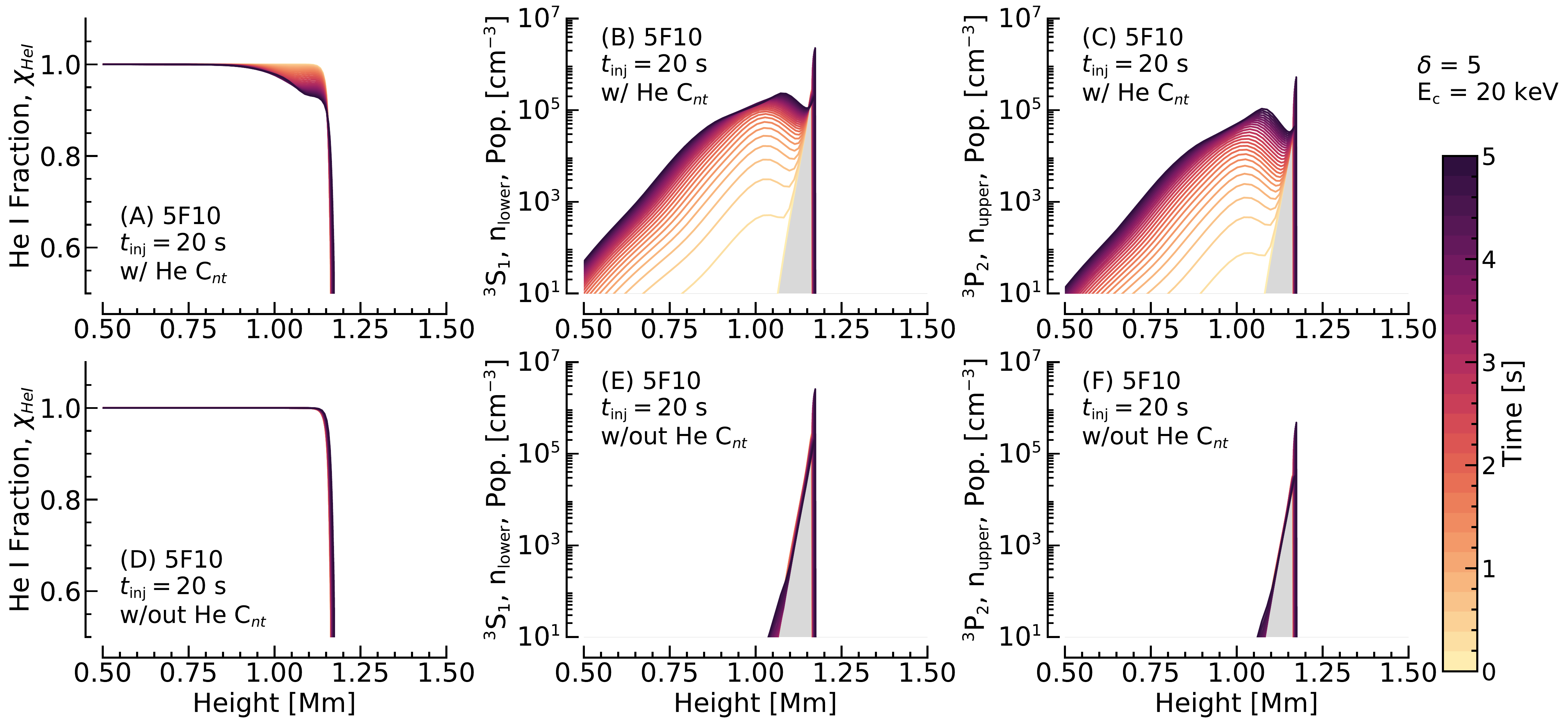}}
	\caption{\textsl{The He~\textsc{i} fraction (A,D), and the population densities of the lower (B,E) and upper (C,F) levels of the He~\textsc{i} 10830~\AA\ line as a function of time in the initial stages of one our flare simulations (5F10, $\delta = 5$, $E_{c} = 20$~keV, $\tau_{\mathrm{inj}} = 20$~s). The top row is the case with He~$C_{nt}$ and the bottom row is the case without He~$C_{nt}$.}}
	\label{fig:pop_evol_weaker}
\end{figure*}
\begin{figure*}[ht]
	\centering 
	{\includegraphics[width = 0.85\textwidth, clip = true, trim = 0.cm 0.cm 0.cm 0.cm]{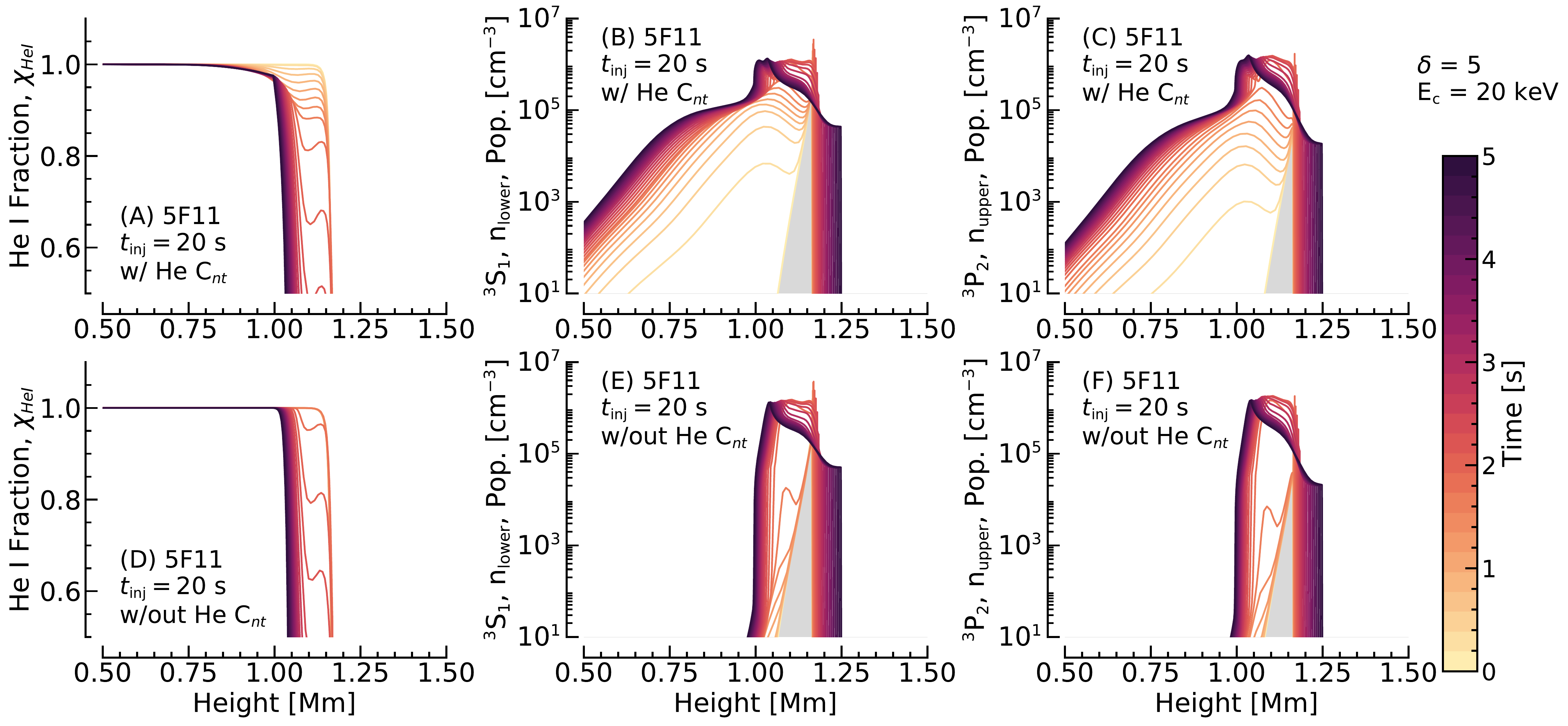}}
	\caption{\textsl{Same as Figure~\ref{fig:pop_evol_weaker} but for a stronger flare (5F11, $\delta = 5$, $E_{c} = 20$~keV, $\tau_{\mathrm{inj}} = 20$~s).}}
	\label{fig:pop_evol_stronger}
\end{figure*}
\begin{figure*}[ht]
	\centering 
	{\includegraphics[width = 0.85\textwidth, clip = true, trim = 0.cm 0.cm 0.cm 0.cm]{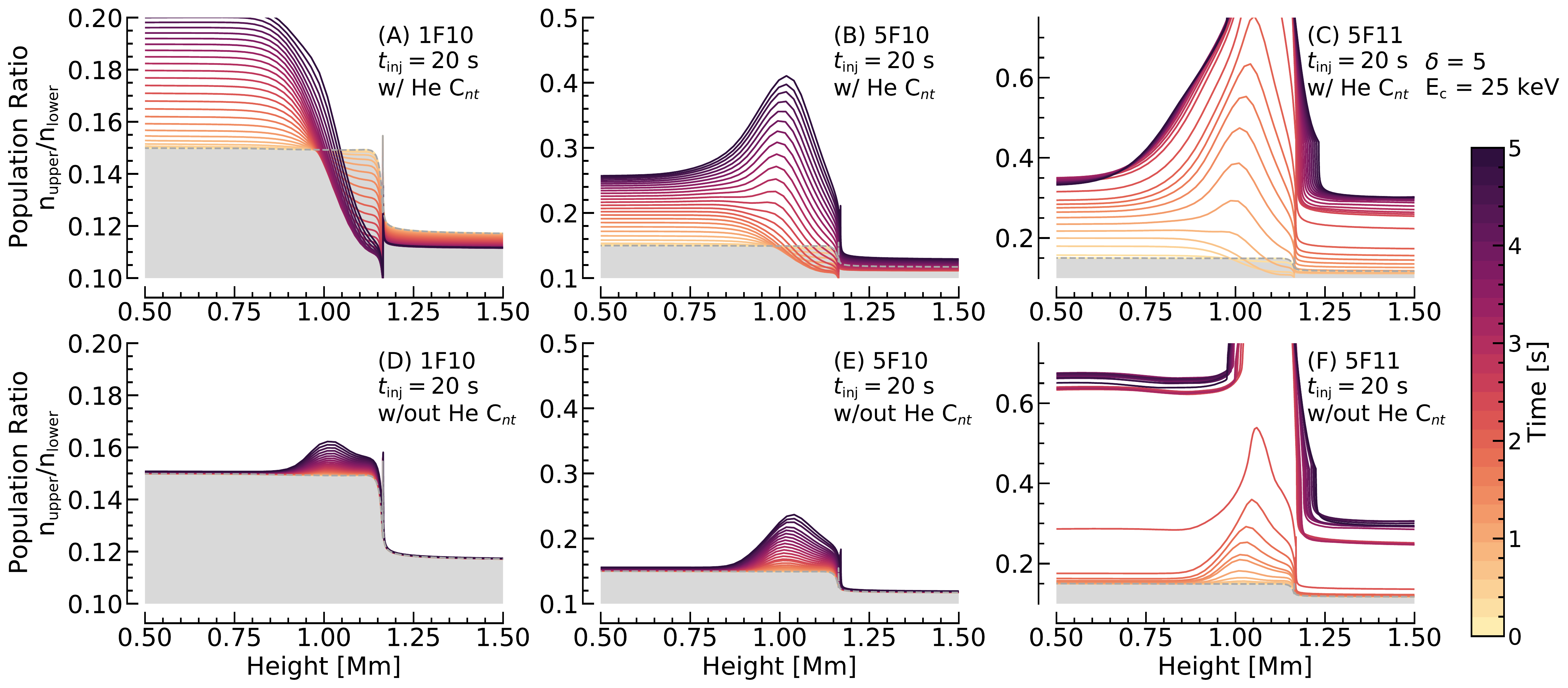}}
	\caption{\textsl{The ratio of the He~\textsc{i} 10830~\AA\ lines upper-to-lower level population densities in the initial stages of some flare simulations. The top row show simulations with He~$C_{nt}$, and the bottom row without He~$C_{nt}$. Colour represents time. In all experiments $\delta = 5$, $E_{c} = 25$~keV, and $\tau_{\mathrm{inj}} = 20$~s. The total injected energy fluxes are $1\times10$~erg~cm$^{-2}$ (A,D), $5\times10$~erg~cm$^{-2}$ (B,E), and $5\times11$~erg~cm$^{-2}$ (C,F). The greyscale is the pre-flare. }}
	\label{fig:popratio_evol}
\end{figure*}
Simulations that included He~$C_{nt}$ exhibited some important differences at early stages in the flare compared to those that did not. When He~$C_{nt}$ is present some fraction of helium quickly becomes ionised to He~\textsc{ii}, even in the relatively cool chromosphere between $z\sim$1~Mm and the TR. These locations were typically co-spatial with the peak of the electron beam heating rate. Within fractions of a second there can be as much as a $\sim10-20~\%$ He~\textsc{ii} fraction in the upper chromosphere, growing to an even larger fraction as the flare progresses and temperatures increases. In the absence of He~$C_{nt}$ there are substantially fewer He~\textsc{ii} ions at those heights, and those only form later in the simulation when the chromospheric temperature has risen and coronal irradiance increased. As time progresses the differences between the two scenarios reduce, indicating that He~$C_{nt}$ play a dominant role only in the early phase of the flare and that later thermal processes and PRM increase in importance. In the weaker simulations the smaller number of He~$C_{nt}$ results in correspondingly smaller fraction of He~\textsc{ii}. In the stronger simulations there is a larger fraction of He~\textsc{ii}, and the differences between including and omitting He~$C_{nt}$ reduce at a faster rate (presumably since the upper chromospheric temperature more rapidly increases). The same is true when a faster rate of flare energy injection is simulated. 

Since the temperature is still too cool to fully ionise helium, there are recombinations of free electrons to He~\textsc{ii}~$\rightarrow$He~\textsc{i}. These can take place to excited levels of helium, including the orthohelium states. Both $n_{\mathrm{lower}}$ and $n_{\mathrm{upper}}$ increase in population, with a strong peak in the mid-upper chromosphere, but with a wide tail to lower altitudes. Since penetration depth scales with electron energy, simulations with a larger $E_{c}$ exhibit a slightly deeper peak height, and a larger vertical extent of enhanced orthohelium populations. Softer distributions more efficiently heat the upper chromosphere, so that thermal effects occur more quickly in weaker flares. Populations scale also with flare strength due to the larger number of ionisations. 

These overpopulations of orthohelium would increase the opacity at $\lambda = 10830$~\AA\ due to increased absorption of photospheric photons by the $^{3}$S$_{1}$ ($n_{\mathrm{lower}}$) level. Without He~$C_{nt}$, there is no excess of recombinations through the excited levels at early times in the flare, and consequently no initial increase in $\lambda = 10830$~\AA\ opacity. Similar to the He ionisation fraction, as the flare develops the populations of orthohelium in the two scenarios of with and without He~$C_{nt}$ become more similar, in the mid-upper chromosphere (height $z \sim [1-1.25]$~Mm). This happens faster in stronger flares, and in flares with a softer spectrum (smaller $E_{c}$). This height range is where most of the non-thermal electron energy is deposited, and once the temperature has increased during the flare, thermal effects dominate the population of orthohelium at those altitudes. He~$C_{nt}$ clearly still play a dominant role at lower altitudes, and still act to populate orthohelium at  $z \sim [1-1.25]$~Mm even though thermal processes drive the simulations with and without He~$C_{nt}$ closer. Generally, the magnitude and widths of the peaks of orthohelium population densities and He~\textsc{i} fraction decrease with decreasing flare strength, and the time taken for the simulations with and without He~$C_{nt}$ to become similar decreases with increasing flare strength.

Inspecting Figure~\ref{fig:pop_evol_weaker}~\&~\ref{fig:pop_evol_stronger} clearly illustrates the differences that emerge in simulations with and without He~$C_{nt}$, for a weak and stronger flare, respectively. In the lower energy simulation, while the peak of the populations are similar, the case with He~$C_{nt}$ clearly exhibits a tail that would increase opacity of the He~\textsc{i} 10830 line at these times. In the higher energy simulations the the mid-upper chromosphere contains a peak population of orthohelium that is extended over several hundred km in simulations both with and without He~$C_{nt}$. This is due to the strong chromospheric temperature in those simulations leading to thermal collisional ionisation and recombination. Although they become similar rather quickly, there are a few seconds where the simulation with He~$C_{nt}$ has a larger orthohelium population than the simulation without He~$C_{nt}$. A similar figure is shown in Appendix~\ref{sec:extrafigs} for a smaller value of $E_{c}$ (Figure~\ref{fig:pop_evol_weaker_softer}).

It is, roughly speaking, the ratio of $n_{\mathrm{upper}}$ to $n_{\mathrm{lower}}$ that governs the intensity of the He~\textsc{i} 10830~\AA, and if the line is in absorption or emission. While both $n_{\mathrm{upper}}$ and $n_{\mathrm{lower}}$ increase during the flare, their ratio varies over time. \cite{2020ApJ...897L...6H} showed that in their flare this ratio increased during their flare simulation at the time that the line went into strong emission. We demonstrate here how the behaviour of this ratio changes over time in the scenario with and without He~$C_{nt}$. This is shown in Figure~\ref{fig:popratio_evol}, where we compare the ratios from three flare strengths with (top row) and without (bottom row) He~$C_{nt}$ (note the varying y-axis scale in each panel). This figure shows one of the harder non-thermal electron distributions ($E_{c} = 25$~keV) but results are typical of all simulations.

In all flares with He~$C_{nt}$ there is an initial decrease in the level population ratio, indicating that there is an increased amount of $n_{\mathrm{lower}}$ relative to $n_{\mathrm{upper}}$, compared to the pre-flare stratification, and therefore more absorptions. Stronger flares reverse this ratio quickly, eventually increasing significantly over the pre-flare. The line is in emission at these times. Flares with harder non-thermal electron distributions exhibit a stronger decrease, over a larger height range, for a longer period. See Figure~\ref{fig:popratio_evol_soft} in Appendix~\ref{sec:extrafigs} for the $E_{c} = 15$~keV case. 

At the locations where He~\textsc{i} 10380~\AA\ forms ($z\gtrsim1$~Mm), this population ratio in the 1F10 simulations does not actually increase over the pre-flare value for the simulations with $E_{c} > 20$~keV. It does at lower altitudes, but since the line is optically thick during the flare it is $z\gtrsim1$~Mm that really matters. As we will show in the next section, the He~\textsc{i} 10830~\AA\ line did not go into emission in those experiments. 

Omitting He~$C_{nt}$ results in the absence of this initial decrease in the ratio, with the ratio instead only increasing. Also, at times and in certain locations, the ratio in the scenario with He~$C_{nt}$ exceeds that of the scenario without  He~$C_{nt}$. 

Now that we have determined that inclusion or omission of He~$C_{nt}$ can result in notable differences to the stratification of the He~\textsc{i} ionisation fraction and orthohelium level population densities, and that those properties are also dependent on flare strength and non-thermal electron distribution spectral hardness, we demonstrate how they are manifested in the emergent He~\textsc{i} 10830~\AA\ line intensity.

%%%%%
\section{He~\textsc{i} 10830~\AA\ Response to Flares}
%%%%%
\subsection{Electron Beam Driven Flares}
\begin{figure}[htb]
	\centering 
	{\includegraphics[width = 0.5\textwidth, clip = true, trim = 0.cm 0.cm 0.cm 0.cm]{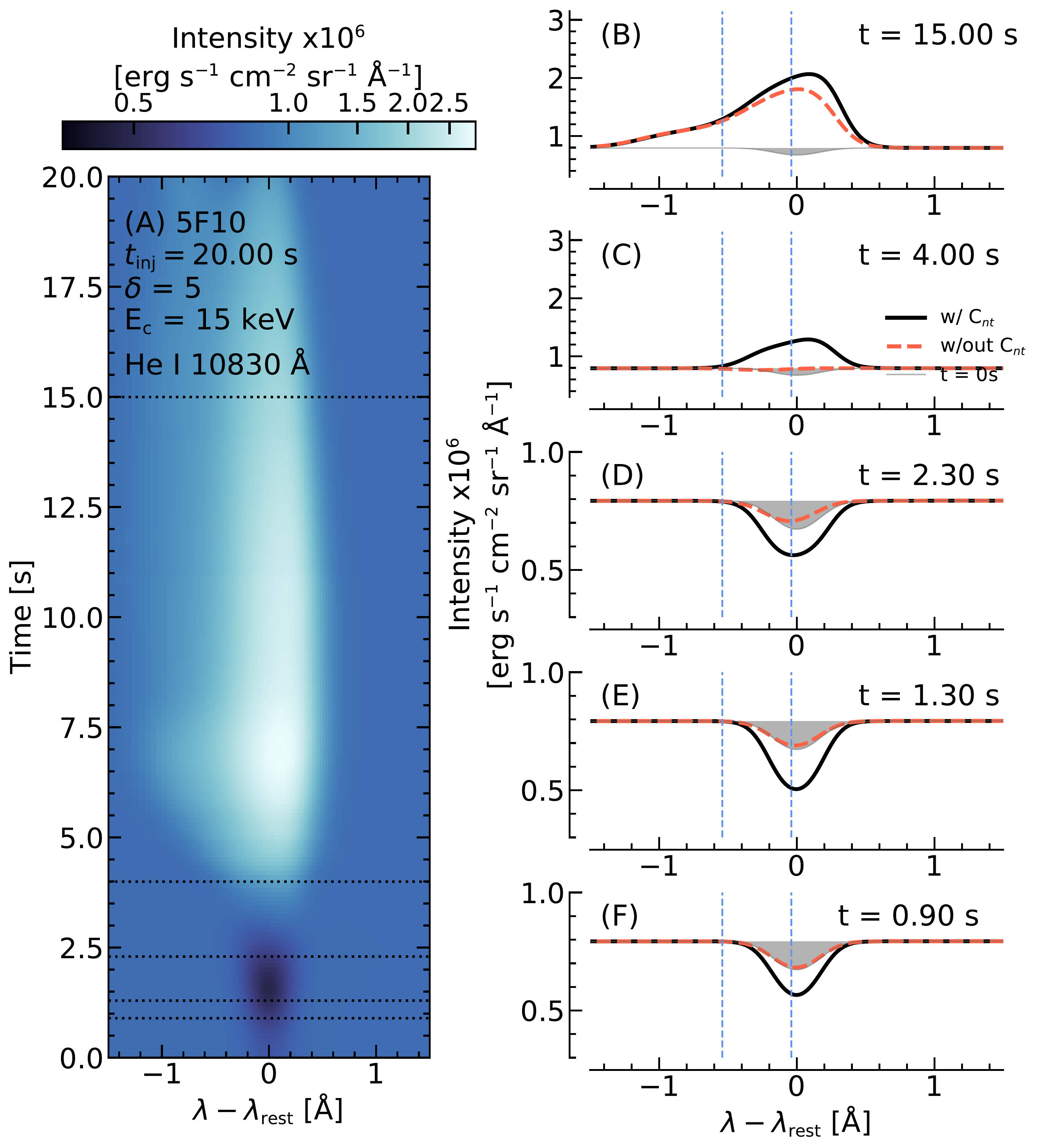}}
	\caption{\textsl{(A) The response of the He~\textsc{i} 10830~\AA\ line in a flare simulation (5F10, $\tau_{\mathrm{inj}} = 20$~s, $E_{c} = 15$~keV, $\delta = 5$) where He~$C_{nt}$ are included. The dashed lines indicate the times shown in panels (B-F), which compares the simulation in panel A to the equivalent without He~$C_{nt}$. In those panels the black lines include helium non-thermal collisional ionisation, and the red-dashed lines do not. The grey dashed line is the $t=0$~s profile. Dashed blue lines indicate the wavelength range imaged by \cite{2016ApJ...819...89X}.}}
	\label{fig:lineprofile_ex_softer}
\end{figure}

\begin{figure}[htb]
	\centering 
	{\includegraphics[width = 0.5\textwidth, clip = true, trim = 0.cm 0.cm 0.cm 0.cm]{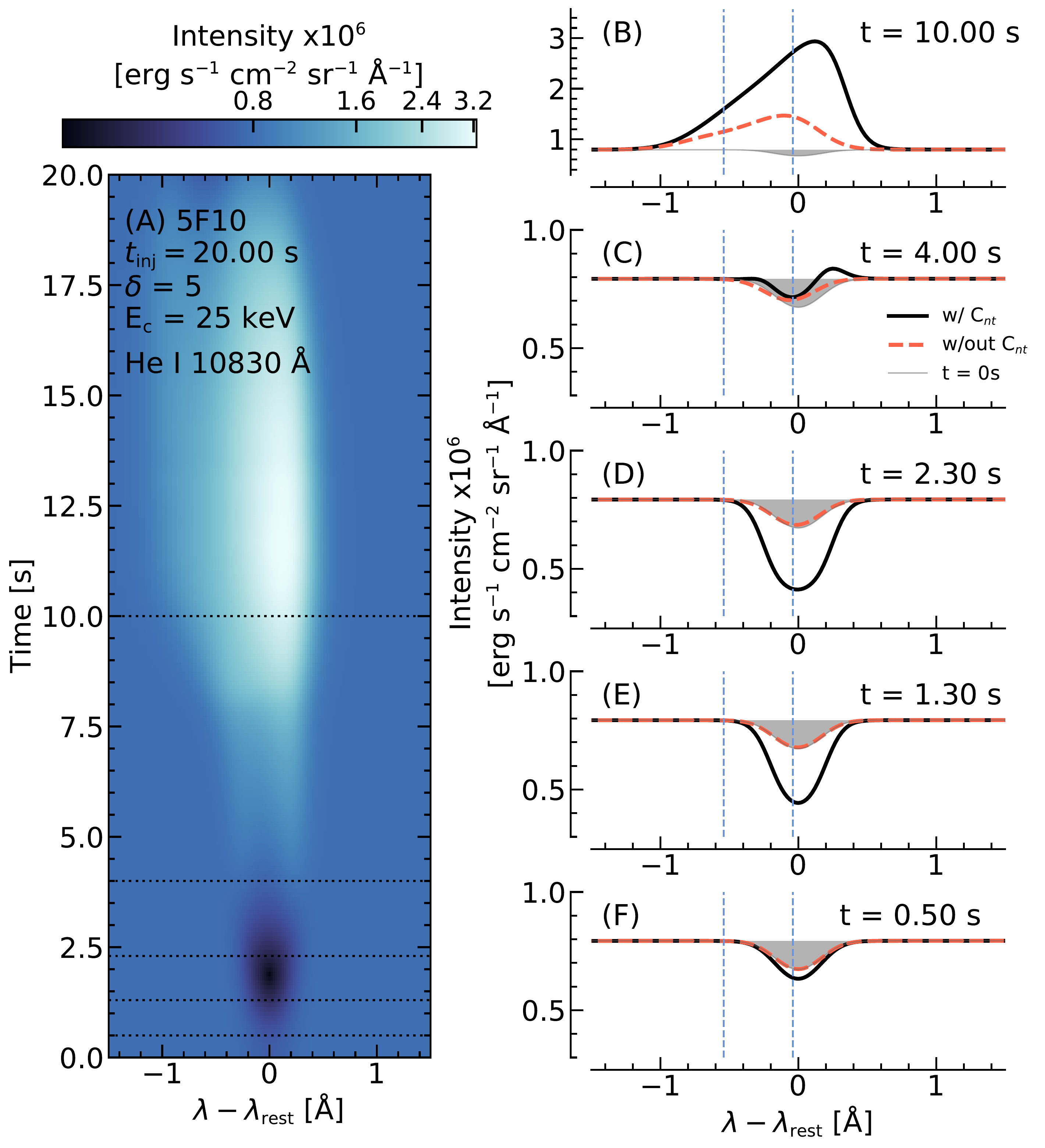}}
	\caption{\textsl{Same as Figure~\ref{fig:lineprofile_ex_softer} but for a harder non-thermal electron distribution, with $E_{c} = 25$~keV.}}
	\label{fig:lineprofile_ex_harder}
\end{figure}

\begin{figure*}[htb]
\begin{center}
        \vbox{
	\hbox{
	\subfloat{\includegraphics[width = .5\textwidth, clip = true, trim = 0.cm 0.cm 0.cm 0.cm]{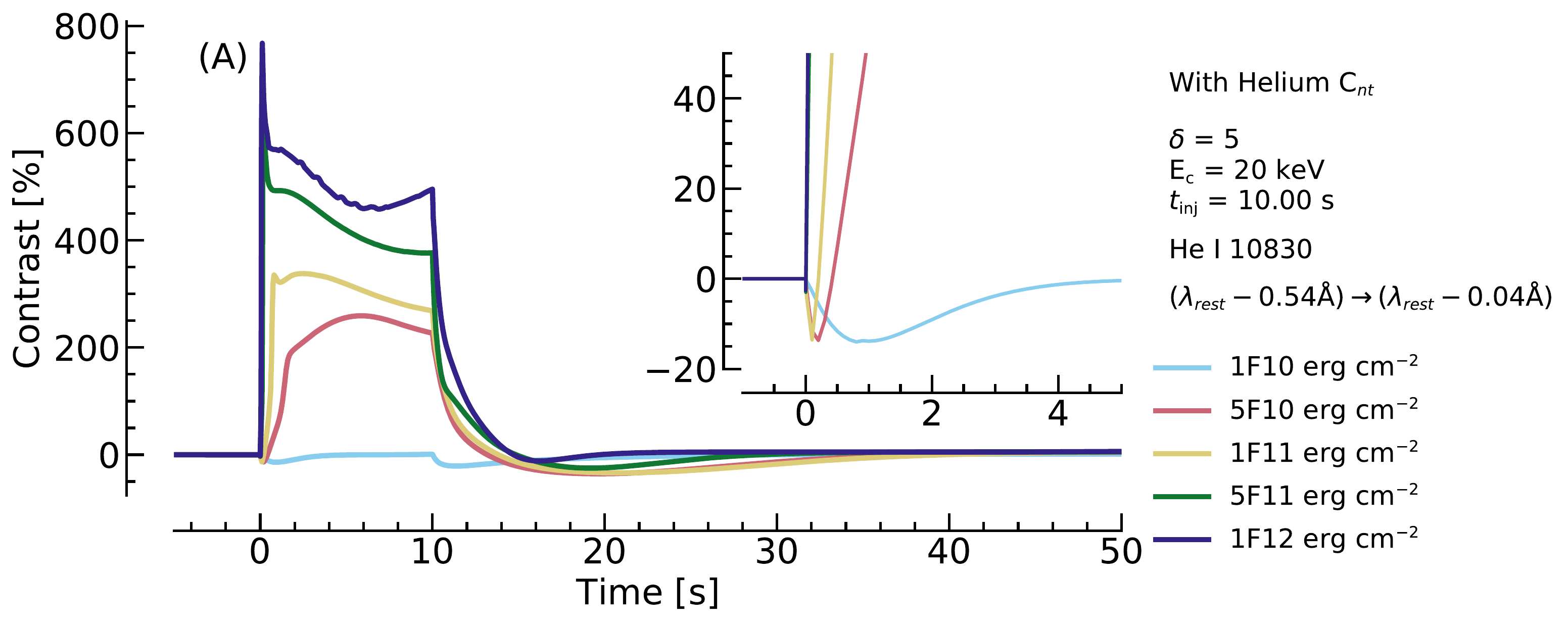}}	
	\subfloat{\includegraphics[width = .5\textwidth, clip = true, trim = 0.cm 0.cm 0.cm 0.cm]{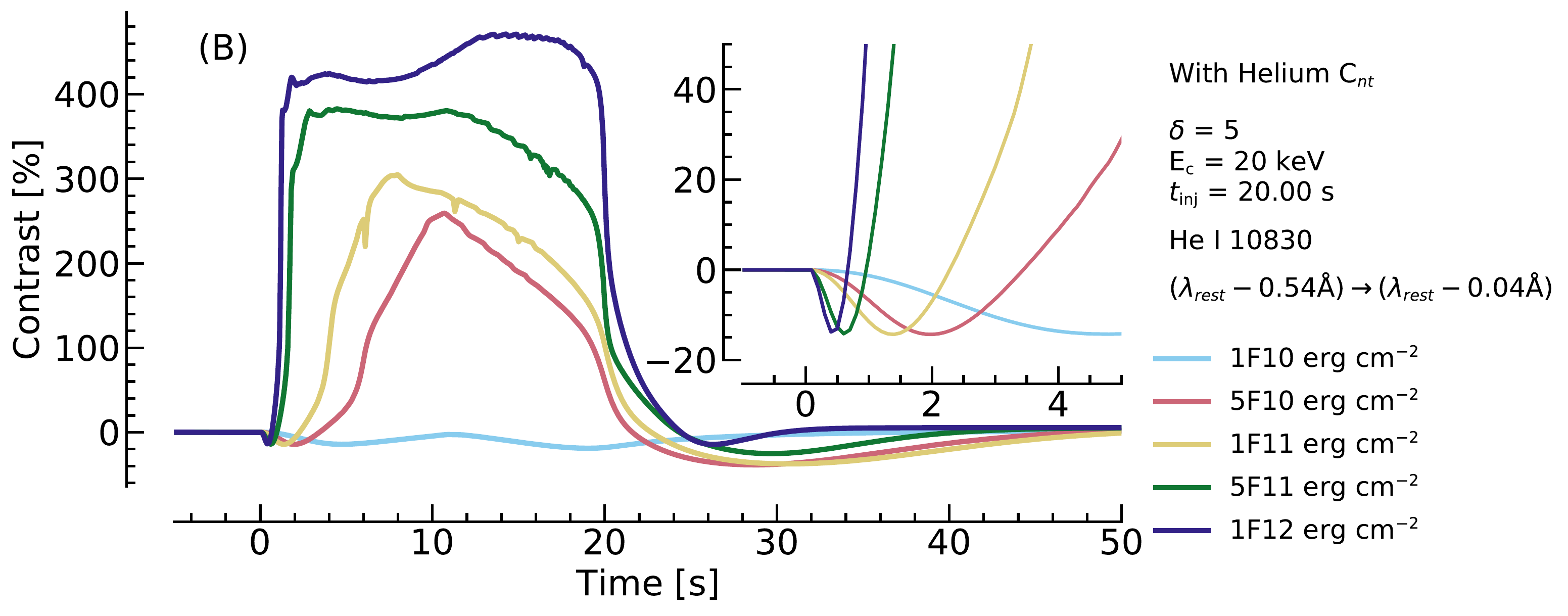}}	
	}
	}
	\vbox{
	\hbox{
	\subfloat{\includegraphics[width = .5\textwidth, clip = true, trim = 0.cm 0.cm 0.cm 0.cm]{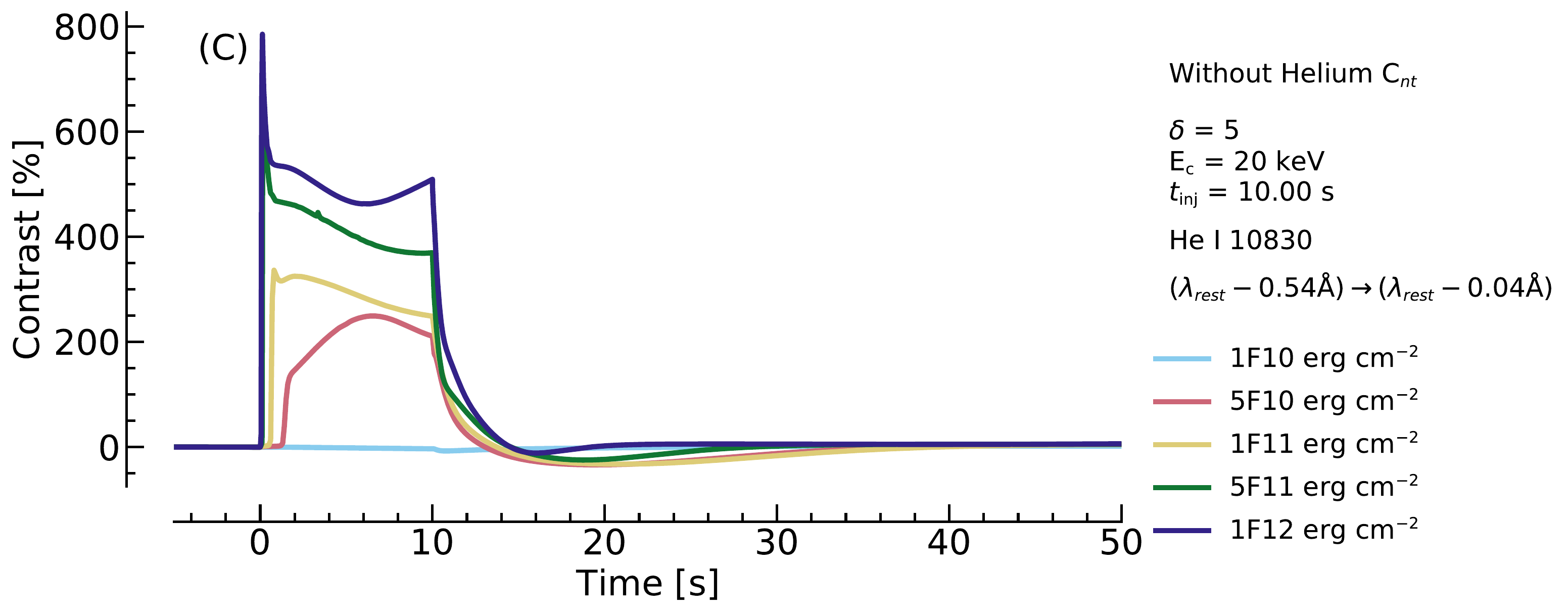}}	
	\subfloat{\includegraphics[width = .5\textwidth, clip = true, trim = 0.cm 0.cm 0.cm 0.cm]{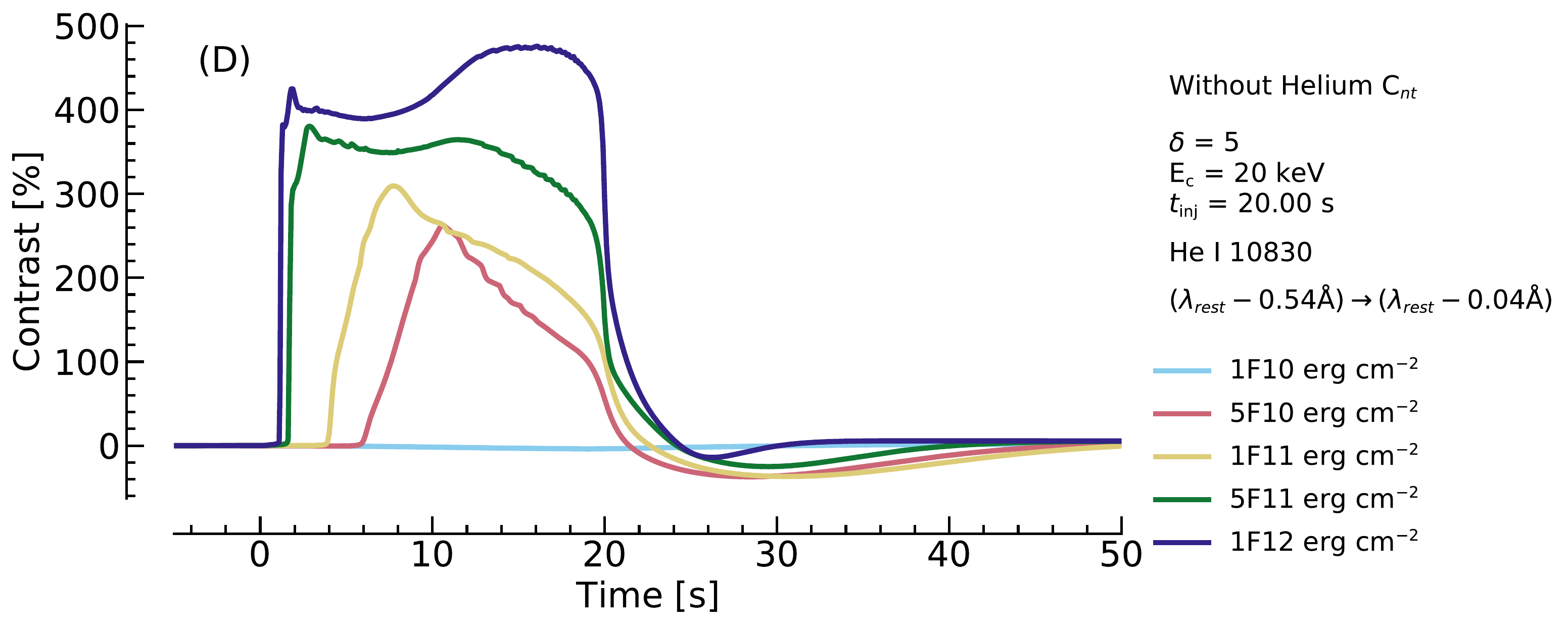}}	
	}
	}
	\caption{\textsl{Lightcurves of the He~\textsc{i} 10830~\AA\ line, integrated over $\lambda_{\mathrm{rest}}-0.54$~\AA\ to $\lambda_{\mathrm{rest}}-0.04$~\AA. All simulations have $\delta = 5$, $E_{c} = 20$~keV, and colour represents total energy flux. Panels (A) and (B) are simulations including He~$C_{nt}$, with $\tau_{\mathrm{inj}} = 10$~s \& $\tau_{\mathrm{inj}} = 20$~s, respectively. Dimmings at flare-onset are present.  Panels (C) and (D) show simulation omitting He~$C_{nt}$ with $\tau_{\mathrm{inj}} = 10$~s \& $\tau_{\mathrm{inj}} = 20$~s, respectively. Dimmings at flare-onset are no longer present. }}
	\label{fig:lcurves_flux}
\end{center}
\end{figure*}

\begin{figure*}[htb]
	\centering 
	{\includegraphics[width = 0.75\textwidth, clip = true, trim = 0.cm 0.cm 0.cm 0.cm]{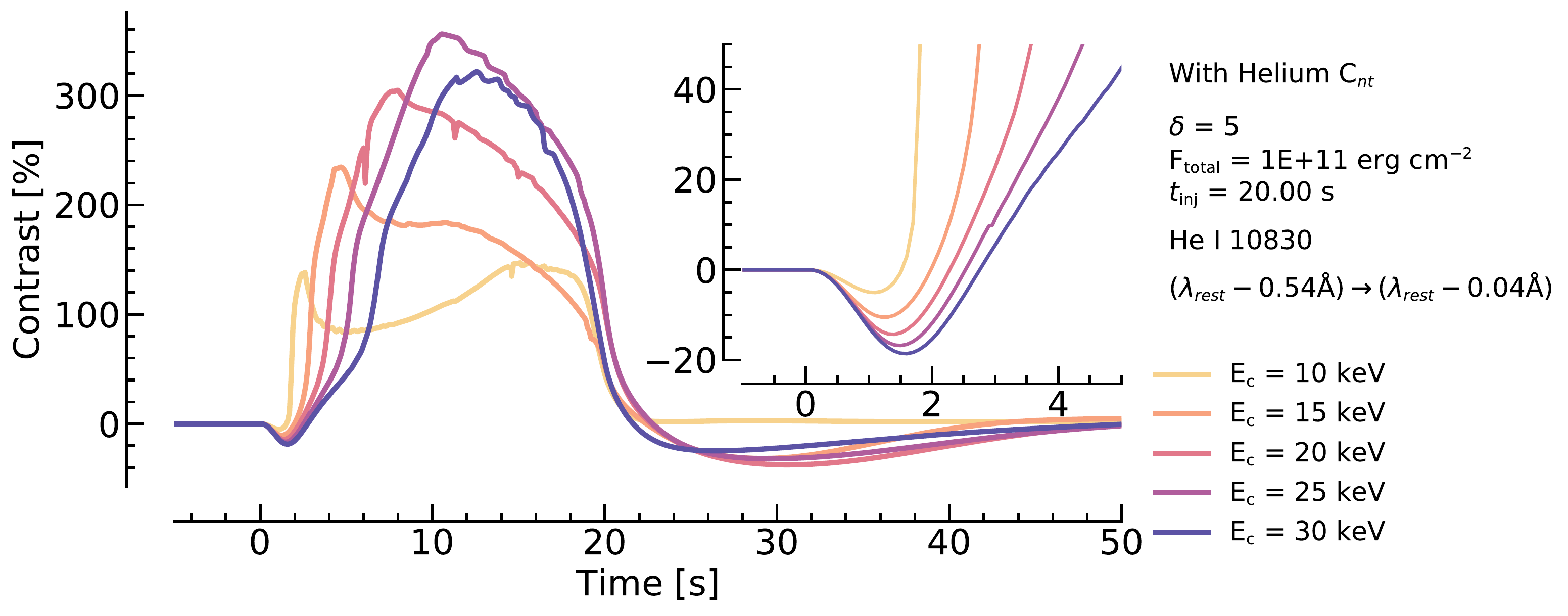}}
	\caption{\textsl{Lightcurves of He~\textsc{i} 10830~\AA\ for simulations with $\delta = 5$, total energy flux 1F11, $\tau_{\mathrm{inj}}= 20$~s, and a range of low-energy cutoff values $E_{c} = [10,15,20,25,30]$~keV. He~$C_{nt}$ are included.}}
	\label{fig:lcurves_ec}
\end{figure*}

\begin{figure*}[htb]
	\centering 
	{\includegraphics[width = 0.65\textwidth, clip = true, trim = 0.cm 0.cm 0.cm 0.cm]{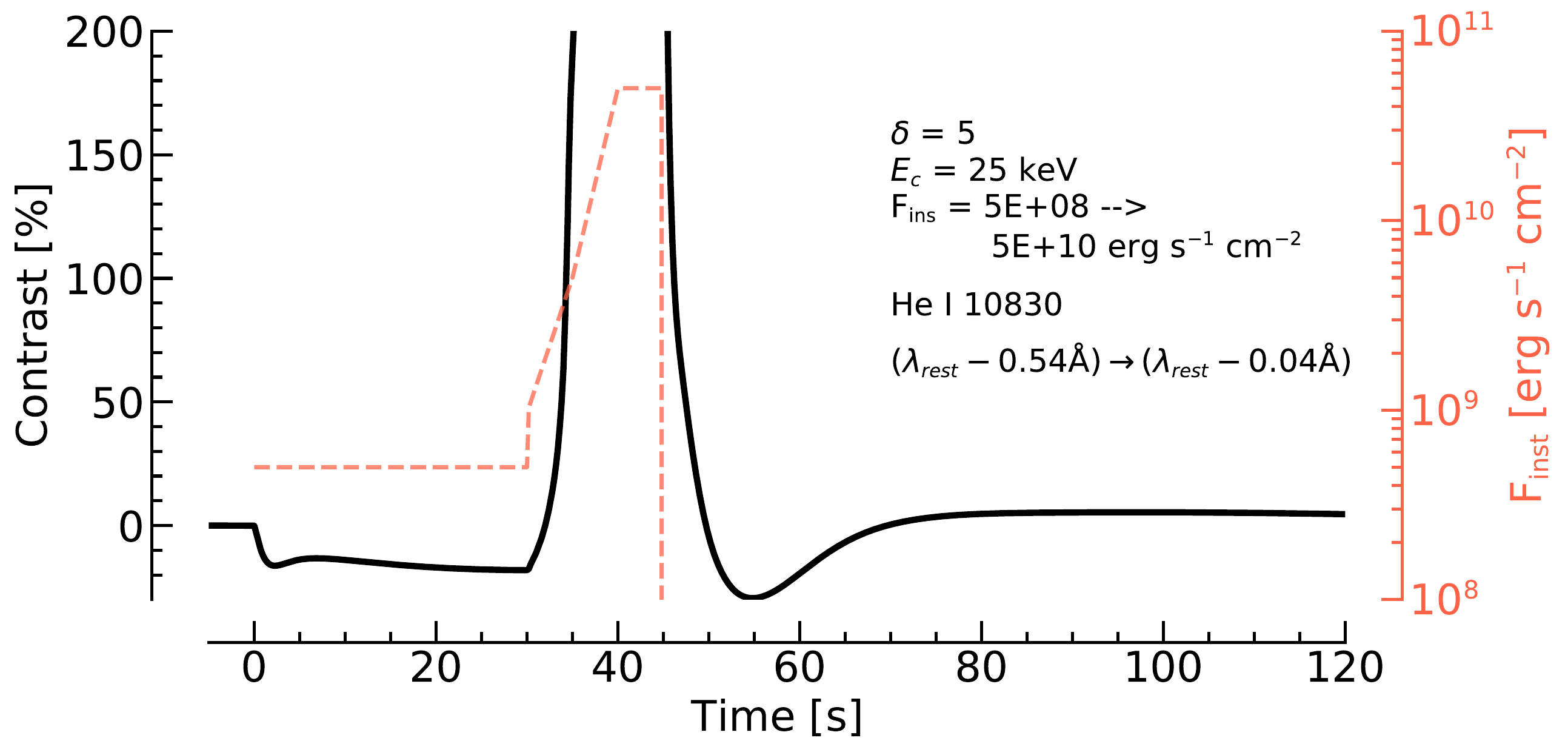}}
	\caption{\textsl{Lightcurve of He~\textsc{i} 10830~\AA\ for the simulation with $\delta = 5$, $E_{c} = 25$~keV, which initially had weak energy flux of $5\times10^{8}$~erg~cm$^{-2}$~s$^{-1}$ before ramping up to $5\times10^{10}$~erg~cm$^{-2}$~s$^{-1}$. He~$C_{nt}$ are included. The injected instantaneous energy flux is shown as the red dashed line for reference. To focus on the initial dimming the contrast scale has been clipped.}}
	\label{fig:lcurves_smalltolarge}
\end{figure*}

Every electron beam driven flare simulation that included He~$C_{nt}$ exhibited dimming in the initial phase of the flare, that is enhanced absorption over the pre-flare producing a negative contrast. With the exception of the weakest flares (1F10, $E_{c} > 20$~keV) the line subsequently went into emission, producing a positive contrast.  Simulations that did not include He~$C_{nt}$ \textsl{did not exhibit dimming during the heating phase}, they only contained emission over the pre-flare.  

Figures~\ref{fig:lineprofile_ex_softer}~\&~\ref{fig:lineprofile_ex_harder} show representative cases, the 5F10 $t_{\mathrm{inj}} = 20$~s simulation with $E_{c} = 15$~keV and $E_{c} = 25$~keV, respectively. Panel (A) shows the line intensity (colour) as a function of wavelength for the heating phase of the flare (colormap scaled to the 2/5 power). The dark coloured patch at early times is the line dimming, that is a deeper absorption profile. After $t\sim3-4$~s the line switches to being more intense than the pre-flare (though still in absorption, near the continuum level) before going strongly into emission a short time later. Both of these periods would be observed as a positive contrast. The line broadens, is Doppler shifted, and exhibits asymmetries. Though the line appears redshifted (the peak intensity is in the red wing), the line actually forms in upflowing plasma. Being optically thick, this results in the absorption profile shifting to the blue also, which preferentially absorbs blue wing photons, giving rise to a red peak. This phenomenon has been described in detail by, e.g., \cite{2015ApJ...813..125K,2016ApJ...832..147K} \& \cite{2016ApJ...827..101K} as concerns the H$\alpha$ and Mg II h \& k lines. Central reversal features are present in the line once it starts to go into emission. The duration and depth of the enhanced absorption feature both scale with increasing $E_{c}$.  

Comparisons to the simulations without He~$C_{nt}$ are shown in panels (B-F) of each figure, which show line profiles from five snapshots in the flare. Black lines are profiles including He~$C_{nt}$, red dashed are those without He~$C_{nt}$. The $t=0$~s is shown for context also (grey shaded area). It is clear that He~$C_{nt}$ has a strong effect on the line. Not only is the dimming of the line absent when omitting He~$C_{nt}$, but the overall intensity of the line is generally weaker, though by varying degrees through the duration of the flare. 

Lightcurves of the integrated intensity show that the duration of the dimming varies strongly with flare strength and energy injection timescale. Weaker flares, and flares with more gradual energy injection, exhibit more sustained dimming. Some of the stronger flares, particularly those with $t_{\mathrm{inj}} = 10$~s have dimmings that last only fractions of a second (unobservable with current instrumentation). Flares that include He~$C_{nt}$ generally have stronger total emission than those without. The variation of dimmings as a function of total flare energy and injection timescale is shown in Figure~\ref{fig:lcurves_flux}. Shown in that figure are lightcurves (presented as contrasts over the pre-flare, $C = (I_{flare} - I_{t=0})/I_{t=0} \times 100$) from flares with $E_{c} = 20$~keV, for the range of flare energies simulated. Panel (A) shows $\tau_{\mathrm{inj}} = 10$~s, and panel (B) shows $\tau_{\mathrm{inj}} = 20$~s. The latter case contains more sustained enhanced absorption. Panels (C) and (D) show the same but with He~$C_{nt}$ omitted, where no enhanced absorption is present. In that figure the intensity was integrated over $\lambda_{\mathrm{rest}}-0.54$~\AA\ to $\lambda_{\mathrm{rest}}-0.04$~\AA, the same passband as the BBSO observations reported by \cite{2016ApJ...819...89X}. Mass flows do Doppler shift the line, and lead to asymmetries, so the choice of passband will affect the lightcurve, but did not artificially produce or remove the enhanced absorption feature. Integrating over a passband centered on line core gave qualitatively similar results, but of course with somewhat different numerical values. 

As well as varying with flare strength, both the strength and duration of the absorption feature varies as a function of spectral hardness of the non-thermal electron distribution. A larger $E_{c}$ produced deeper, more sustained absorptions. Figure~\ref{fig:lcurves_ec} illustrates for a fixed energy flux 1F11 $\tau_{\mathrm{inj}} = 20$~s the effect of varying the low-energy cutoff. See also Section~\ref{sec:absorptionprops}.

The weakest set of flares in our study, the 1F10 simulations, stay in absorption for the entirety of the heating phase when He~$C_{nt}$ is included, apart from the softest distributions $E_{c} = [10, 15]$~keV, presumably because in those simulations the upper chromosphere became hot enough to drive thermal collisions later in the flare. In the harder simulations the temperature rise would have been more modest and not enough to combat the greater increase in opacity present since a larger swathe of the chromsophere produced additional orthohelium. Interestingly, if He~$C_{nt}$ were omitted then these weakest simulations do go into emission, likely because there is no additional opacity to overcome so that even modest temperature enhancements can increase He~\textsc{i} 10830~\AA\ emission. 

While we can qualitatively match observations in the sense that a period of enhanced absorption is produced at flare onset, and while we obtain negative contrasts similar to those recently observed by \cite{2016ApJ...819...89X}, our simulations predict a substantially shorter period of dimming than observed and a larger positive contrast when the line is in emission. Our modelling predicts only several seconds of dimming, compared to several tens to more than 100s of dimming in the observations of \cite{2016ApJ...819...89X}. A follow up to this work will focus on formation properties of the line, and attempt to identify reasons behind this discrepancy. For now we note that the only way that we have thus far been able to extend the period of enhanced absorption was to initially inject a very weak beam, before ramping up the energy flux to what we generally consider to be standard flare values. If we inject a beam with $\delta = 5$, $E_{c} = 25$~keV, and instantaneous energy flux $F_{\mathrm{ins}} = 5\times10^{8}$~erg~cm$^{-2}$~s$^{-1}$ for a period of 30~s before increasing the energy flux to $F_{\mathrm{ins}} = 1\times10^{9}$~erg~cm$^{-2}$~s$^{-1}$ for 5~s, then to $F_{\mathrm{ins}} = 5\times10^{9}$~erg~cm$^{-2}$~s$^{-1}$  for 5~s, and finally to $F_{\mathrm{ins}} = 5\times10^{10}$~erg~cm$^{-2}$~s$^{-1}$ for 10~s, then we obtain the lightcurve shown in Figure~\ref{fig:lcurves_smalltolarge}. The enhanced absorption is present for the duration of the weak beam heating, and only when the energy flux increases, and the chromospheric temperature rises, does the line go into emission. The time profile of the energy flux is shown on the right-hand axis of Figure~\ref{fig:lcurves_smalltolarge} for reference.

%%%%%%%
% Thermal Conduction Exps
%%%%%%
\subsection{Flares driven by Thermal Conduction}

\begin{figure}[htb]
	\centering 
	{\includegraphics[width = 0.5\textwidth, clip = true, trim = 0.cm 0.cm 0.cm 0.cm]{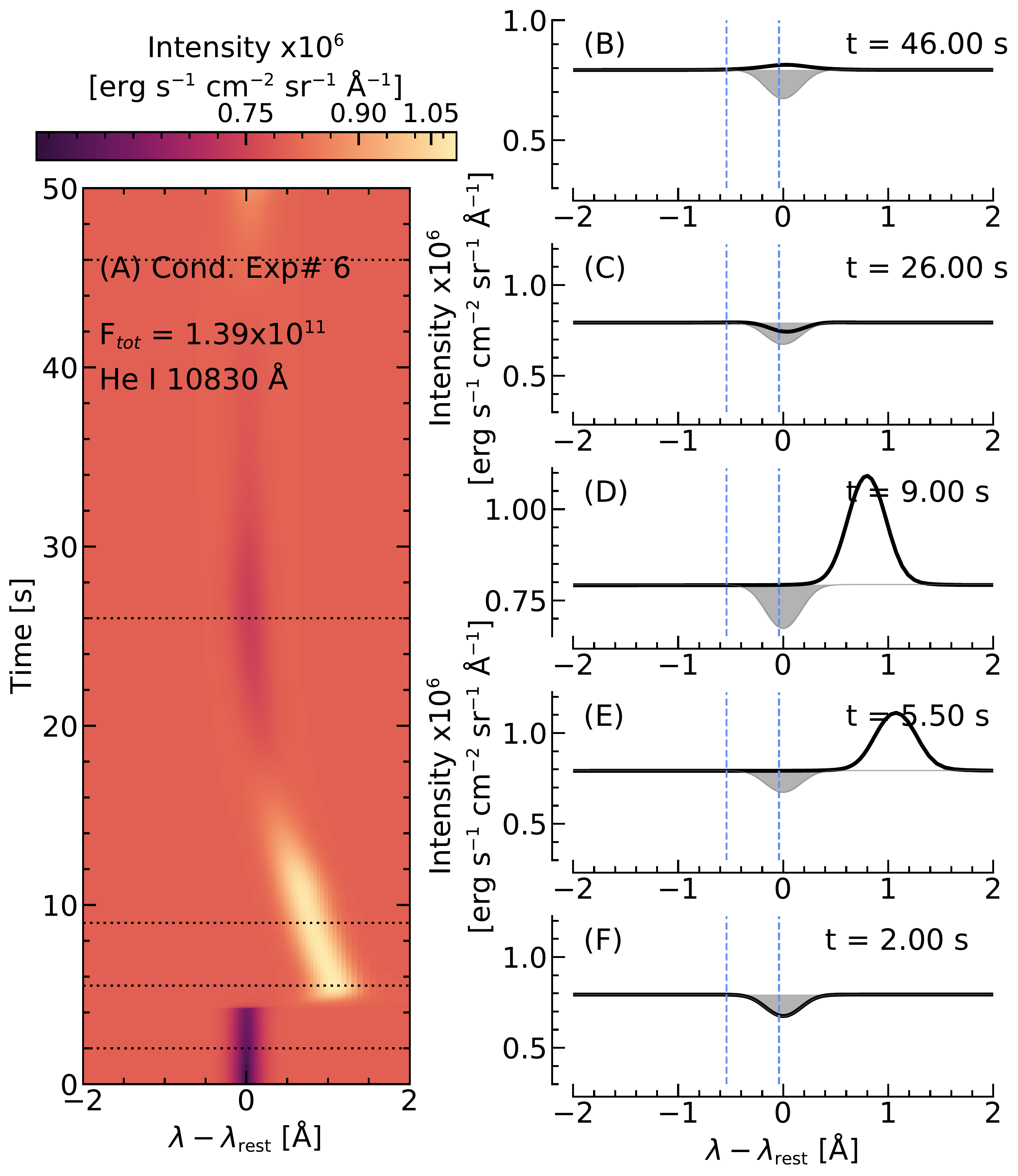}}
	\caption{\textsl{Same as Figure~\ref{fig:lineprofile_ex_softer} but for direct heating experiment \#6.}}
	\label{fig:lineprofile_cond}
\end{figure}
\begin{figure*}[htb]
	\centering 
	{\includegraphics[width = 0.75\textwidth, clip = true, trim = 0.cm 0.cm 0.cm 0.cm]{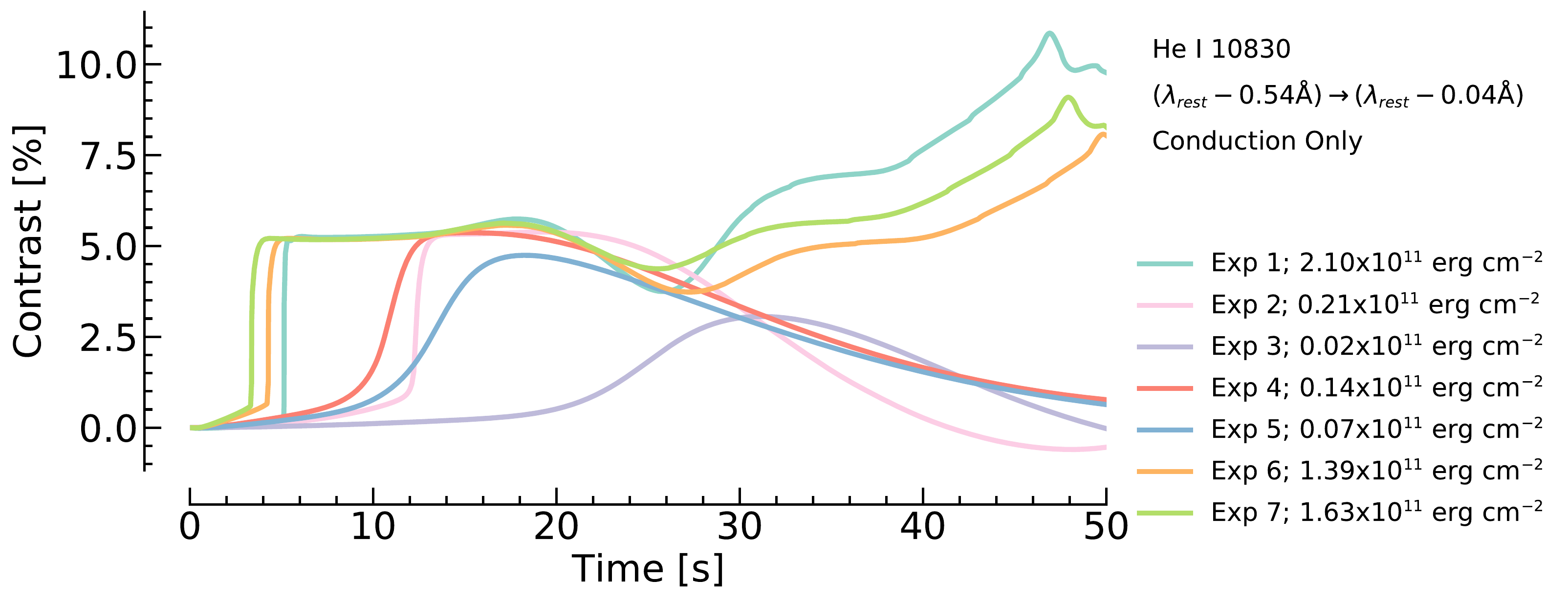}}
	\caption{\textsl{Lightcurves of He~\textsc{i} 10830~\AA\ for simulations simulations driven by \textsl{in situ} coronal heating and thermal conduction only. Wavelength is integrated over $\lambda_{\mathrm{rest}}-0.54$~\AA\ to $\lambda_{\mathrm{rest}}-0.04$~\AA.}}
	\label{fig:lcuves_conduction}
\end{figure*}
It has been established that in electron beam driven flares that CRM dominates over PRM in overpopulating orthohelium in the initial stages, leading to enhanced absorption. The role of the PRM in the later stages of the flare will be explored in more detail in a follow up work, but it seems that it takes some time for coronal irradiance to increase sufficiently drive additional photoionisation of helium. By the time the corona has been heated and mass loaded, the temperature in the chromosphere is high enough that the line is in emission.

A reasonable question is then, can direct heating of the corona before the chromosphere is heated increase irradiance and photoionisations of helium and lead to enhanced absorption? Direct \textsl{in situ} heating of plasma at loop tops has been suggested as necessary to explain certain flare observations \citep[e.g.][and references therein]{2009A&A...498..891B,2016JGRA..12111667H,2020A&A...644A.172W}, and recent simulation results suggest that energy release from reconnection \citep{2019NatAs...3..160C} and retraction of flare loops \citep{2015ApJ...813..131L,2016ApJ...833..211L,2018ApJ...868..148L} can deposit energy at the loop tops. This energy would then be transported to the lower atmosphere via the conductive flux. 

We modelled purely thermal flares, depositing energy directly into the corona. In those experiments the coronal temperature increase preceded the chromospheric temperature increase. However, even in this circumstance, there was an insufficient increase in coronal irradiation to produce an overpopulation of orthohelium, and therefore enhanced absorption. Figure~\ref{fig:lineprofile_cond} shows the He~\textsc{i} 10830~\AA\ line profile evolution for one of our conduction experiments, and Figure~\ref{fig:lcuves_conduction} shows the temporal evolution of the contrast for all seven experiments. No initial period of enhanced absorption is present. 

This is likely because while the coronal temperature increased early in the flare, the coronal density did not increase until ablation of chromospheric plasma began to mass load the loop. The coronal radiation field would scale strongly with the emission measure of the plasma. By the time this happens the chromosphere has began to be heated. 

These simulations suggest that flares with little or no non-thermal Hard X-ray emission would be unlikely to exhibit enhanced absorption of orthohelium lines. Observations of enhanced absorption early in the flare before mass flows are generated are therefore indicative of the presence of the accelerated particles. 

%%%%%%%
% Statistical Overview
%%%%%%
\section{Properties of Absorption Feature}\label{sec:absorptionprops}
\begin{figure}[htb]
	\centering 
	{\includegraphics[width = 0.5\textwidth, clip = true, trim = 0.cm 0.cm 0.cm 0.cm]{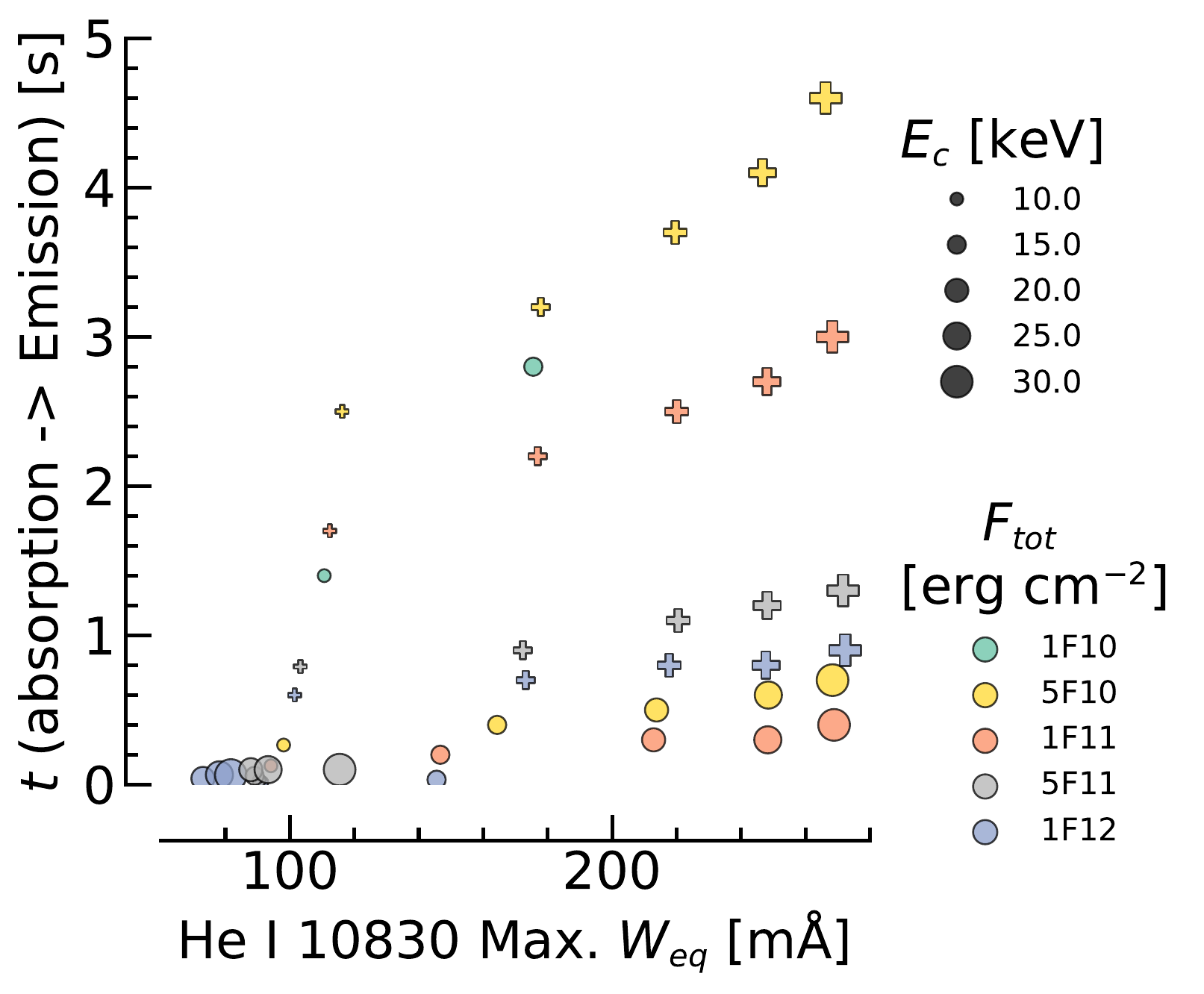}}
	\caption{\textsl{The relationship between absorption properties and non-thermal electron distribution. The maximum equivalent width of the absorption profile in each simulation is shown against the time taken for the line to go into emission. Colours refer to flare strength (total injected energy flux) and symbol size is low-energy cutoff $E_{c}$. Circles are those simulations with $\tau_{\mathrm{inj}} = 10$~s, and crosses are those with $\tau_{\mathrm{inj}} = 20$~s.}}
	\label{fig:scatterplots}
\end{figure}
\begin{figure}[htb]
\begin{center}
        \vbox{
	\hbox{
	\subfloat{\includegraphics[width = .5\textwidth, clip = true, trim = 0.cm 0.cm 0.cm 0.cm]{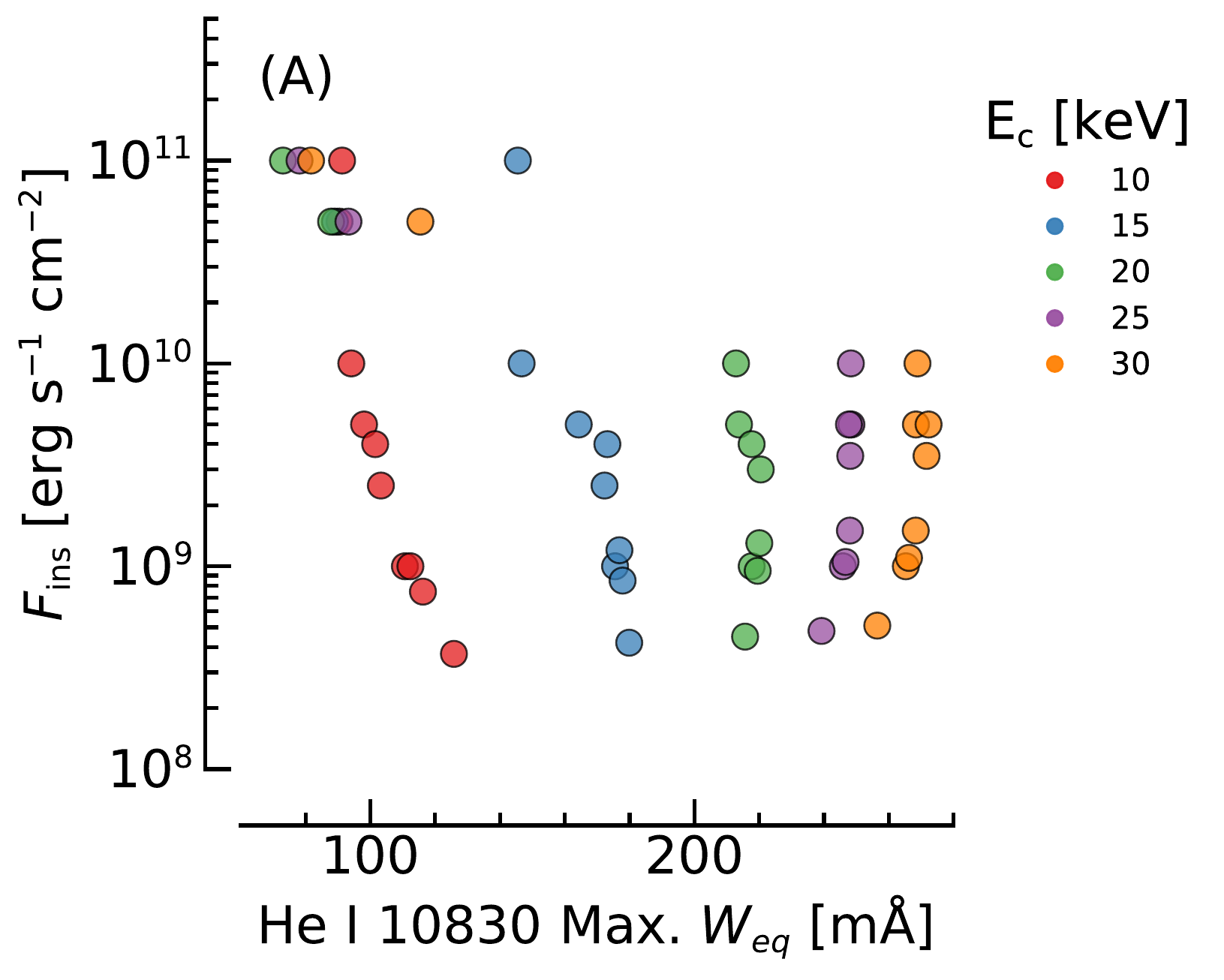}}	
	}
	\hbox{
	\subfloat{\includegraphics[width = .5\textwidth, clip = true, trim = 0.cm 0.cm 0.cm 0.cm]{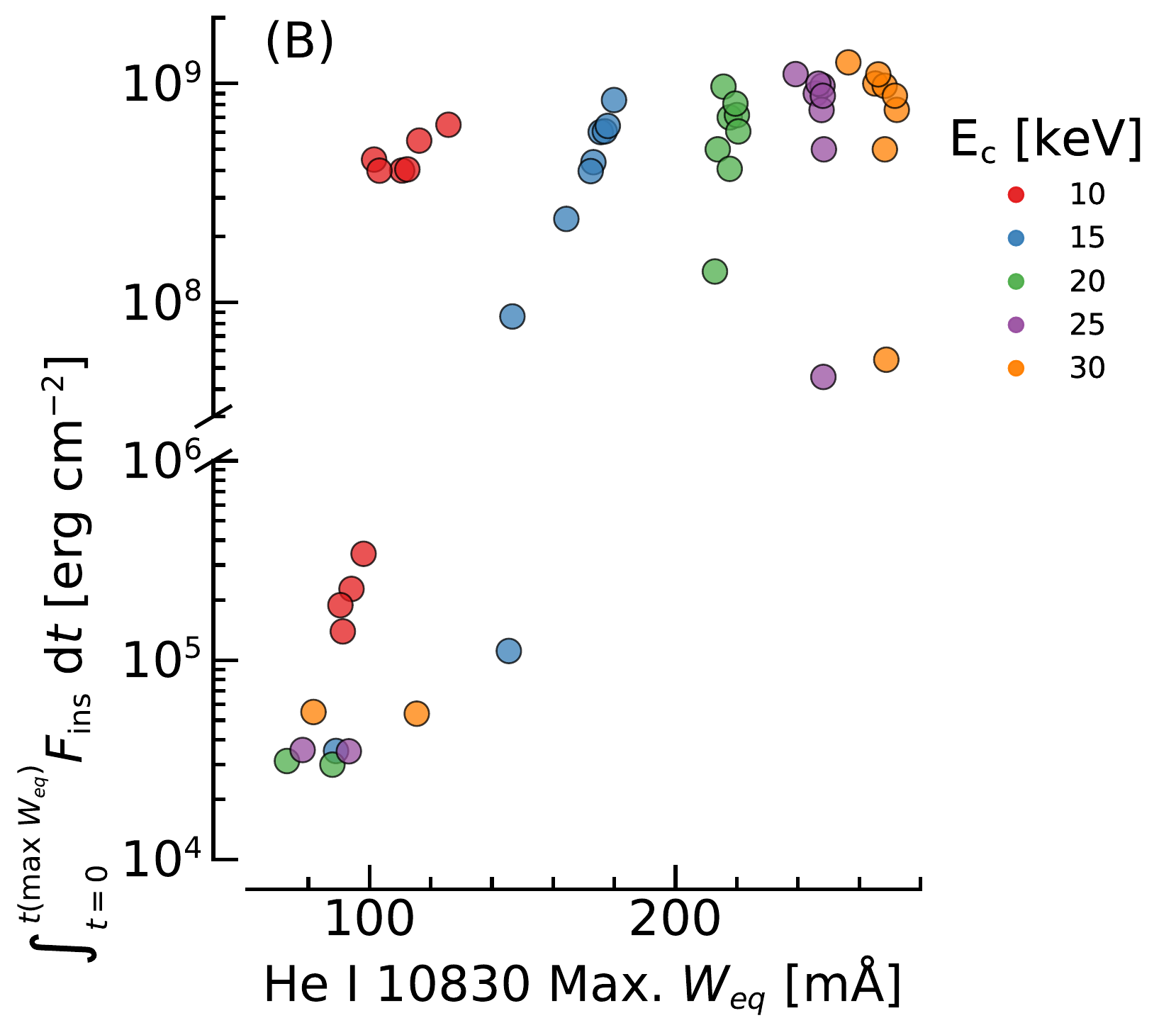}}	
	}
	}
	\caption{\textsl{The maximum equivalent width of the He~\textsc{i} 10830~\AA\ line, $W_{\mathrm{eq,max}}$, as a function of (A) instantaneous injected energy flux at that time, and (B) the injected energy flux integrated from $t=0$ to the time of $W_{\mathrm{eq,max}}$. Colour is low-energy cutoff $E_{c}$. }}
	\label{fig:scatterplots_2}
\end{center}
\end{figure}
\begin{figure}[htb]
\begin{center}
        \vbox{
	\hbox{
	\subfloat{\includegraphics[width = .5\textwidth, clip = true, trim = 0.cm 0.cm 0.cm 0.cm]{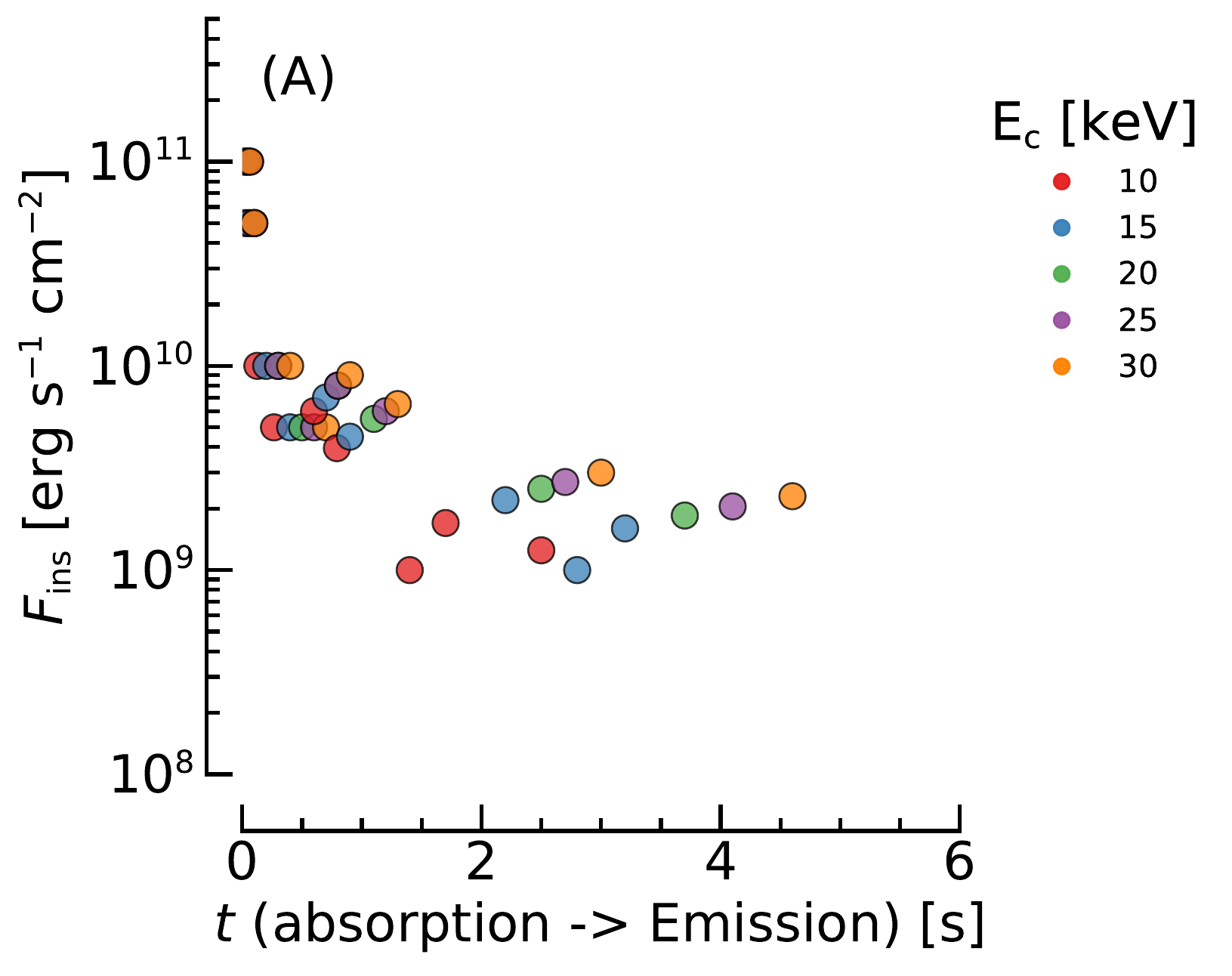}}	
	}
	\hbox{
	\subfloat{\includegraphics[width = .5\textwidth, clip = true, trim = 0.cm 0.cm 0.cm 0.cm]{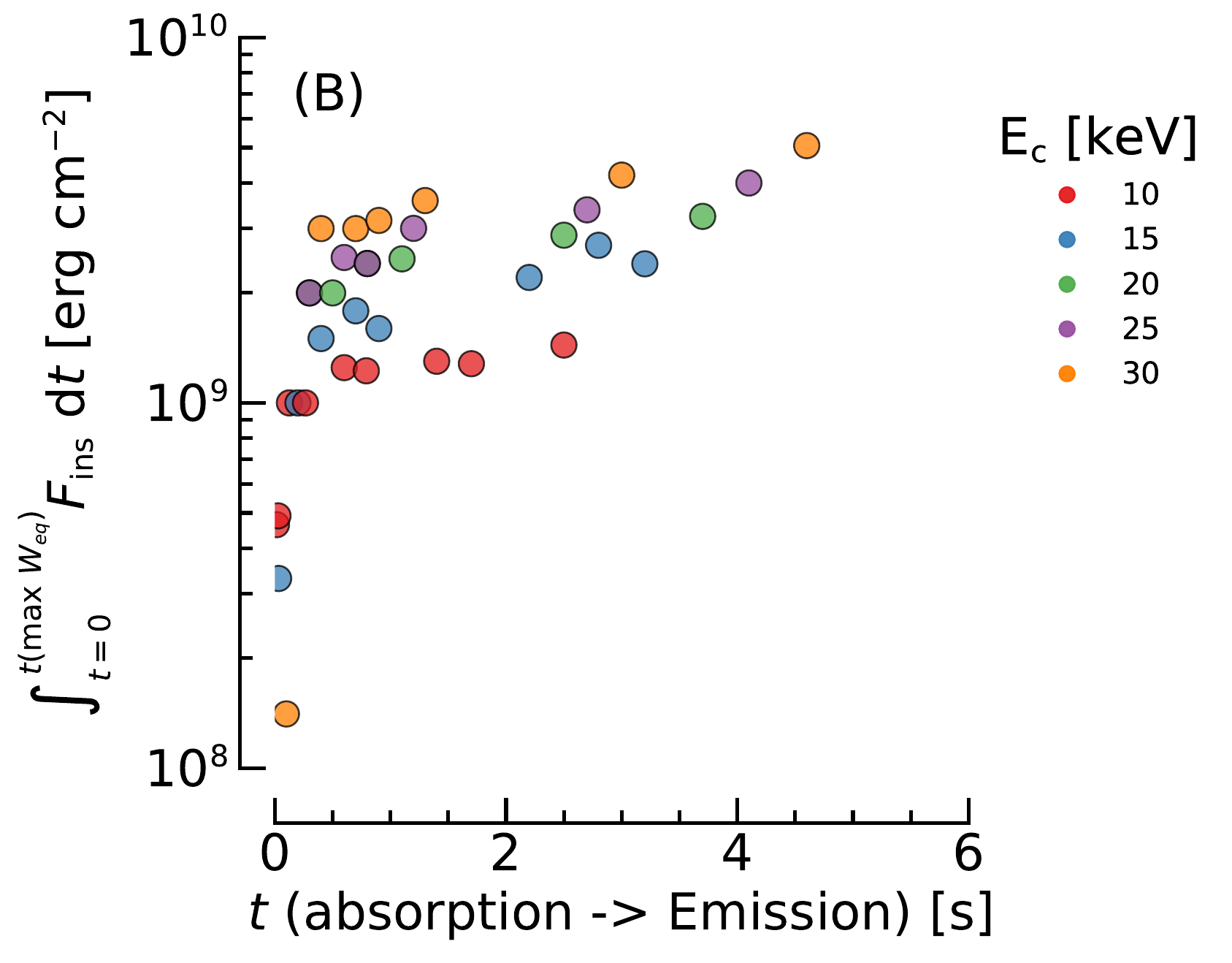}}	
	}
	}
	\caption{\textsl{The time taken for He~\textsc{i} 10830~\AA\ to reach emission, $t_{\mathrm{emiss}}$, as a function of (A) instantaneous injected energy flux at that time, and (B) the injected energy flux integrated from $t=0$ to $t_{\mathrm{emiss}}$. Colour is low-energy cutoff $E_{c}$. }}
	\label{fig:scatterplots_3}
\end{center}
\end{figure}

In the preceding sections we have demonstrated that enhanced absorption of He~\textsc{i} 10830~\AA\ takes place when non-thermal collisional ionisations of helium are present, and that the duration of the absorption varies with flare strength and spectral distribution of the impacting non-thermal electron distribution. Here we discuss how properties of the absorption line can be related to the the properties of the non-thermal electron distribution. 

The equivalent width, $W_{eq}$, gives a measure of a the strength of the absorption line

\begin{equation}\label{eq:width}
   W_{eq} = \int\frac{I_{c} - I_{\lambda}}{I_{c}} \mathrm{d}\lambda,
\end{equation} 

\noindent where $I_{c}$ is the intensity of the continuum, $I_{\lambda}$ is the intensity at wavelength $\lambda$. The maximum value of $W_{eq}$ in each simulated flare, $W_{eq,max}$, was used to define the maximum strength of enhanced absorption for that flare. As a measure of the duration of absorption we find the time taken for the line to go into emission, $t_{\mathrm{emiss}}$, as the time at which $W_{eq}$ becomes negative (since in our definition of Equation~\ref{eq:width} an emission line returns a negative value). Figure~\ref{fig:scatterplots} shows a scatter plot of $W_{eq,max}$ as a function of $t_{\mathrm{emiss}}$. The symbol size is scaled to represent increasing $E_{c}$, and symbol colour represents the total injected energy flux. Circles are flares with $\tau_{\mathrm{inj}}  = 10$~s, and crosses are $\tau_{\mathrm{inj}}  =20$~s. In that figure, some of the weakest flares (green symbols) are not present as they never go into emission. There is a cluster of simulations that show small $W_{eq,max}$ and very short $t_{\mathrm{emiss}}$. Those are the strongest flares. Of those with longer duration periods of absorption, for a fixed flare strength a larger $E_{c}$ results in a larger $t_{\mathrm{emiss}}$, and a larger $W_{eq,max}$. Decreasing flare strength lengthens the period of absorption, giving a larger $t_{\mathrm{emiss}}$. Flares with a more gradual energy injection timescales have longer duration period of absorption. While the magnitude of absorption depends quite strongly on $E_{c}$, it does not really show a strong relationship to flare strength.

Exploring the seeming lack of relationship between $W_{eq}$ and flare strength further we show in Figure~\ref{fig:scatterplots_2} $W_{eq,max}$ as a function of the instantaneous injected energy flux (erg~s$^{-1}$~cm$^{-2}$) at the time of maximum $W_{eq,max}$ (panel A) and as a function of the total energy flux injected between $t=0$~s and time of maximum $W_{eq,max}$. Colour is the low-energy cutoff. The former (panel A) shows us the strength of the bombarding electron flux at the time when $W_{eq}$ is largest, and the latter shows us the cumulative amount of energy deposited during the flare up to the point that $W_{eq}$ is largest. There is little relation between the energy flux and the strength of the absorption feature. The dependence on $E_{c}$ is again very clear, however. The clusters in the top left of panel (A) and bottom left of panel (B) are the strongest flares with $\tau_{\mathrm{inj}} = 10$~s. Future work will investigate this further, determining if the number flux of electrons in certain energy ranges can produce a more clear relationship. 

In Figure~\ref{fig:scatterplots_3} we show $t_{\mathrm{emiss}}$ as a function of the instantaneous injected energy flux (erg~s$^{-1}$~cm$^{-2}$) at $t_{\mathrm{emiss}}$ (panel A) and as a function of the total energy flux injected between $t=0$~s and $t_{\mathrm{emiss}}$. Colour is again the low-energy cutoff. Here it seems that the smaller the instantaneous flux the longer the period of absorption. Note that in that figure the $\tau_{\mathrm{inj}} = 10$~s simulations result in the horizontal groupings (since in those simulations have a constant $F_{ins}$). Panel (B) shows that a longer period of absorption implies a larger total injected flux, for a fixed $E_{c}$. 

Taking all of this information together we have the following picture: (1) Flares with harder $E_{c}$ penetrate more deeply, producing orthohelium over a wider vertical extent, thus building up greater opacity; (2) The absorption of He~\textsc{i} 10830~\AA\ photons is increased, giving a deeper, wider absorption profile. These harder non-thermal electron distributions heat the upper atmosphere less efficiently than softer distributions, so that more time is allowed to pass (and energy flux be deposited) before the upper chromosphere temperature and electron density reach the threshold for thermal processes drive the line into emission; (3) Increasing the flare strength means these thresholds are reached sooner, as does increasing the rate of energy deposition. This picture of chromospheric temperature driving the line into emission is consistent with simulations that only include thermal collisional ionisation and recombination as a pathway to populate orthohelium. \cite{2012ApJ...745...14S} modelled stellar flares using the \texttt{RH} code, but without including He~$C_{nt}$ or coronal irradiation, finding that the line did go into emission without inclusion of CRM or PRM, but that it was weaker than observed (the ratio of He~\textsc{i} 10830~\AA\ to other infrared lines was too small). So, without the CRM and PRM, the line does go into emission when the temperature increases, but those two mechanisms increase the line intensity.

%%%%%%%%%%%%%%%%%%%%%%%%%%%%%%%%%%%%%%%%%%%%%%%%%%%%%
%%%%%%%%%%%%%%%%%%%%%     CONCLUSIONS     %%%%%%%%%%%%%%%%%%%%%
%%%%%%%%%%%%%%%%%%%%%%%%%%%%%%%%%%%%%%%%%%%%%%%%%%%%%

%%%%%
\section{Summary \& Conclusions}\label{sec:conc}
%%%%%
Using the radiation hydrodynamics code \texttt{RADYN} we performed a large parameter study of electron beam driven solar flares, covering a variety of flare strengths and low-energy cutoffs. From that grid of simulations we analysed the behaviour of the He~\textsc{i} 10830~\AA\ line, with a particular focus on the early phase which has been observed to undergo a period of enhanced absorption - a so-called negative flare. We found that:

\begin{enumerate}

\item To produce enhanced absorption features of He~\textsc{i} 10830~\AA\ it is necessary for non-thermal collisional ionisation of helium to be present. Experiments that omit He~$C_{nt}$ resulted exclusively in emission, as did simulations of flares driven only by thermal conduction;

\item In most of our flare simulations the He~\textsc{i} 10830~\AA\ went into emission, but the weakest flares in our sample did not, despite including He~$C_{nt}$. Therefore, observations such as \cite{2018JASTP.173...50K} of weak flares that exhibit absorption for the duration, are still compatible with the non-thermal collisional ionisation-recombination mechanism. 

\item The duration of the emission is a function of both the flare strength and the hardness of the beam (here explored through the low-energy cutoff $E_{c}$). Strong flares exhibit a vanishingly short period of enhanced absorption. Enhanced absorption persists for longer in weaker flares. Increasing $E_{c}$ results in longer duration enhanced absorption. Together, these point to the fact that stronger and softer beams heat the upper atmosphere more quickly producing conditions favourable to drive He~\textsc{i} 10830~\AA\ into emission;

\item The strength of the absorption feature scales with low-energy cutoff. This can be understood as a larger swathe of the chromosphere undergoing a higher rate of non-thermal collisional ionisations, increasing the opacity at 10830~\AA\ and absorbing more photospheric radiation;

\item We were unable to match the observed duration of enhanced absorption, with our simulations predicting only a few seconds compared to the observed several tens to hundreds of seconds. Injecting an electron beam with very low energy flux for $30$~s before ramping up the energy flux to a more typical level did produce an extended period of absorption that persisted as long as the weak beam was bombarding the chromosphere. This may indicate that a weak flux of non-thermal particles is present in the mid-upper chromosphere before the bulk of flare energy is deposited into each source. Follow up work will also investigate different temporal properties of energy injection, such as a series of very short pulses, or multi-threaded modelling 
\citep[e.g.][]{2018ApJ...856..149R}.

\end{enumerate}

Our results demonstrate that, at least in the early phase of the flare, non-thermal collisional ionisations dominate the formation of orthohelium. In a forthcoming paper we will elucidate the formation properties orthohelium in more detail, investigating if non-thermal collisions dominate only in the early phase or throughout the flare duration. That work will also discuss more comprehensively the conditions necessary to bring the line into emission. This latter point is important to both understand the plasma properties that we observe, but also to being to resolve the discrepancy as regards the shorter-than-observed duration of enhanced absorption. 

That these He~\textsc{i} 10830~\AA\ dimmings occur after the initial deposition of flare energy into the chromosphere, having been observed to be present at the narrow leading edge of flare ribbons, is particularly interesting given other recent work regarding the early phases of flare sources. \cite{2018ApJ...861...62P} analysed Interface Region Imaging Spectrograph \citep[IRIS;][]{2014SoPh..289.2733D} Mg~\textsc{ii} h \& k spectroscopic observations of the leading edge of flare ribbons, finding that unique profiles were present. These profiles were extremely broad and double peaked with blueshifted central reversals, differing from the single peaked profiles typically associated with flare ribbons. Also using IRIS, \cite{Jeffreyeaav2794} studied a small flare (B class) with a very high cadence of 1.7~s. They found that the increase and peak of the non-thermal line width of the Si~\textsc{iv} 1402.77~\AA\ line preceded the rise and peak of line intensity. This was interpreted as a sign that MHD turbulence was present in flare footpoints before the plasma was strongly heated and that this turbulence likely contributed towards the heating. Our simulation results also hint that non-thermal processes may be occurring for some time before the chromospheric temperature increases significantly (conclusion \#4). Observations of the leading edge of flare ribbons  with subsecond-to-second cadence, and high spectral and spatial resolution, should be made to shed light on non-thermal processes and energy transport in solar flares. 

Observations of orthohelium have the potential to act as a diagnostic of energy input via non-thermal particle precipitation during solar flares, particularly weaker events. Future modelling efforts must aim to determine to what extent we can exploit such observations, including modelling the He~\textsc{i} D3 lines in solar flares. \cite{2013ApJ...774...60L} observed a flare in which the He~\textsc{i} D3 was undergoing enhanced absorption, while the He~\textsc{i} 10830~\AA\ line was already in emission.  Differences in the formation and response in flares from these lines that share energy levels are likely a fruitful route to diagnostics of flare energy transport. We have already begun the process of including the He~\textsc{i} D3 lines in our simulations. Spectroscopic observations are important to obtain a more comprehensive understanding of the response of the lines during flares, and to better interrogate the models. These will be available via the Daniel K. Inouye Solar Telescope (DKIST), BBSO/GST, and the Swedish Solar Telescope (SST) during the current solar cycle.  Further, while we have discussed only electron beams as a source of non-thermal particles that drive non-thermal collisional ionisation, there has been  speculation regarding the role of Alfv\'enic waves in flares, including their ability to locally accelerate particles in the chromosphere \citep[e.g.][]{2008ApJ...675.1645F}. The rate of non-thermal collisions that would result from such a distribution of electrons should be explored and compared to the electron beam case. \\

%%%%%%%%%%%%%%%%%%%%%%%%%%%%%%%%%%%%%%%%%%%%%%%%%%%%%
%%%%%%%%%%%%%%%%%%%%     ACKN and BIBLIO     %%%%%%%%%%%%%%%%%%%%%
%%%%%%%%%%%%%%%%%%%%%%%%%%%%%%%%%%%%%%%%%%%%%%%%%%%%%

\textsc{Acknowledgments:} \small{GSK, YX, VP, and HW acknowledge financial support from the NASA ROSES Heliophysics Supporting Research program (Grant\# NASA 80NSSC19K0859). J.C.A. acknowledges funding from NASA’s Heliophysics Innovation Fund and NASA’s Heliophysics Supporting Research program. VS acknowledges financial support from a NSF FDSS grant. We acknowledge NSF grant ATM-1954737. We thank the referee for comments that improved the clarity of our manuscript.}

\appendix

\section{Comparing different experimental set-ups}\label{sec:extrafigs}
Here we illustrate the differences that occur in the orthohelium populations when different non-thermal electron distributions are injected to the flare loops. These should be compared to the figures as indicated in the main text. 

\begin{figure*}[ht]
	\centering 
	{\includegraphics[width = 0.85\textwidth, clip = true, trim = 0.cm 0.cm 0.cm 0.cm]{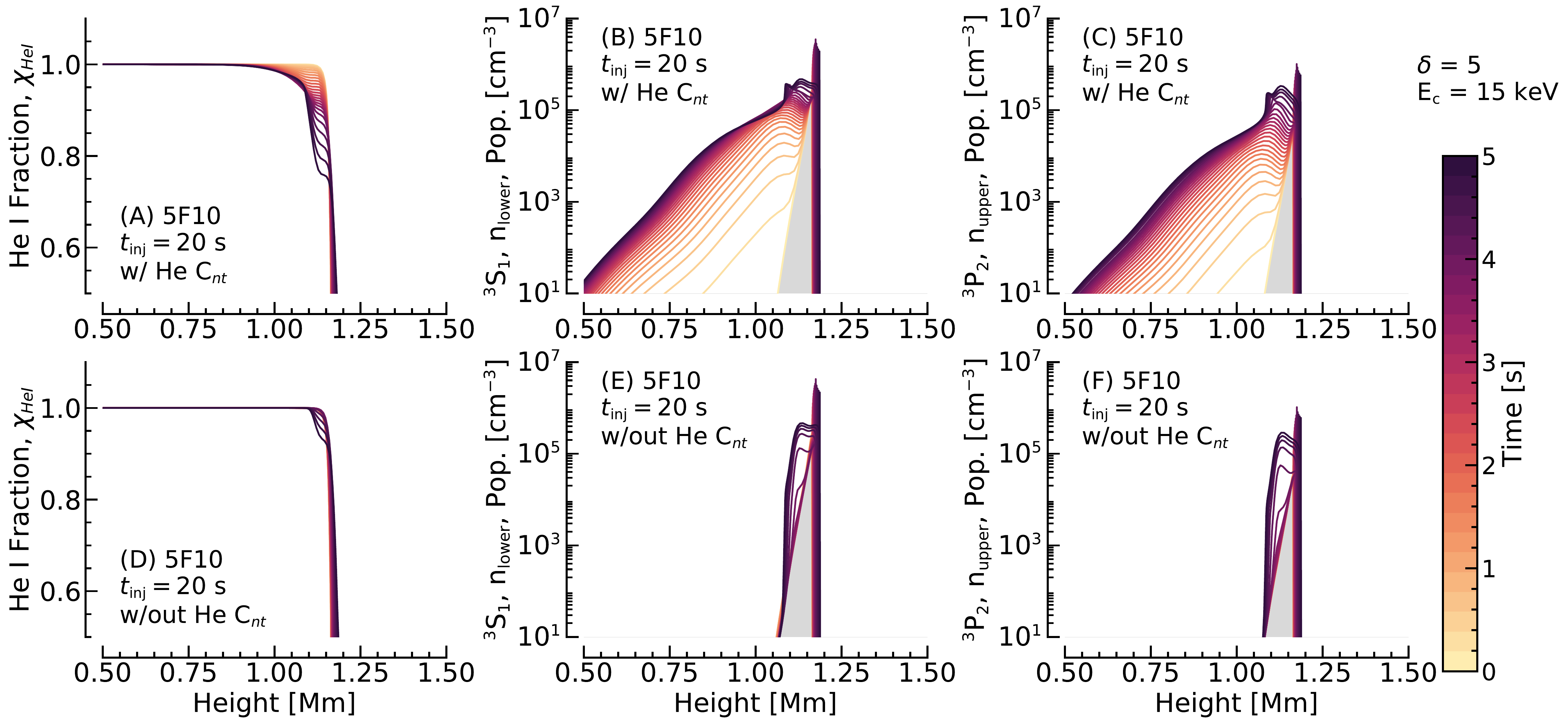}}
	\caption{\textsl{Same as Figure~\ref{fig:pop_evol_weaker} but for a softer non-thermal electron distribution, with $E_{c} = 15$~keV.}}
	\label{fig:pop_evol_weaker_softer}
\end{figure*}

\begin{figure*}[ht]
	\centering 
	{\includegraphics[width = 0.85\textwidth, clip = true, trim = 0.cm 0.cm 0.cm 0.cm]{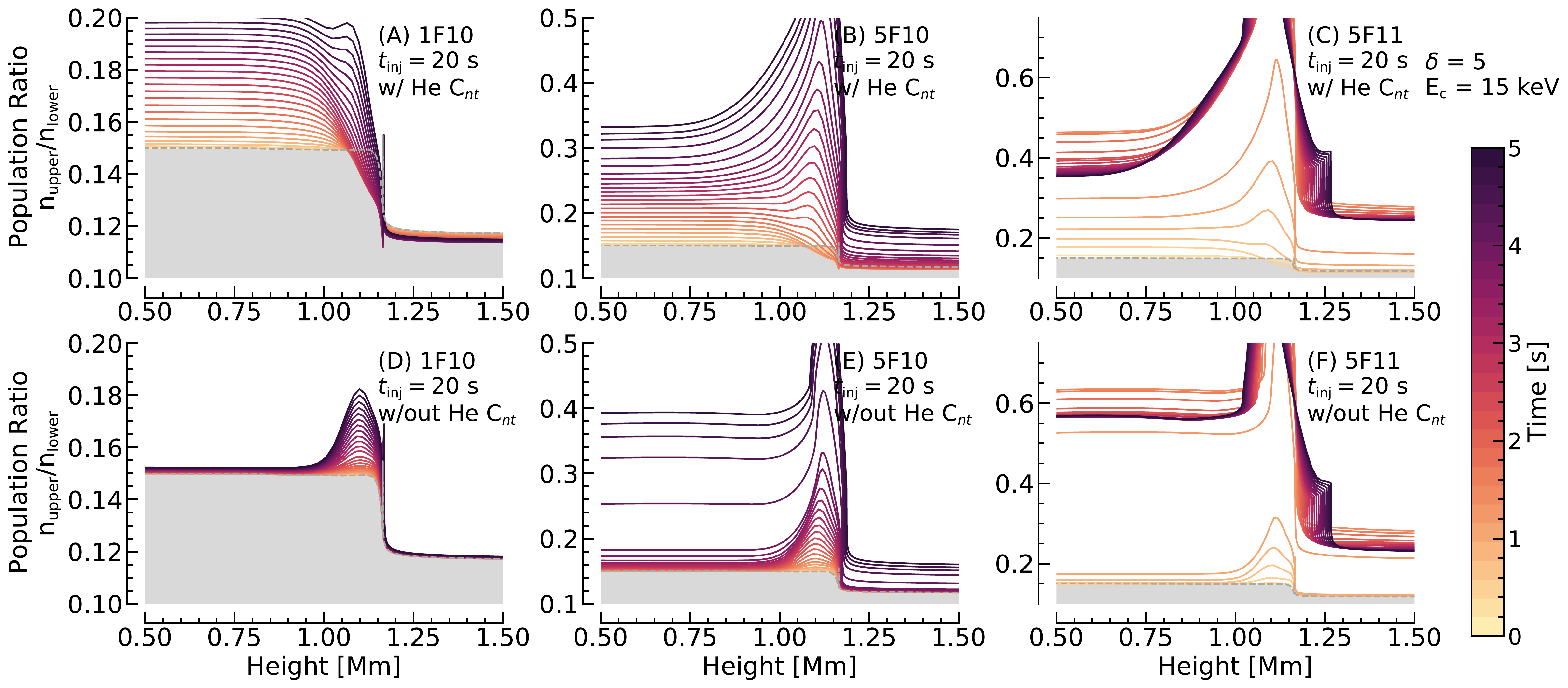}}
	\caption{\textsl{Same as Figure~\ref{fig:popratio_evol} but for a softer non-thermal electron distribution, with $E_{c} = 15$~keV.}}
	\label{fig:popratio_evol_soft}
\end{figure*}

\bibliographystyle{aasjournal}
\bibliography{Kerr_etal_He10830_NonThermCol_Paper1}

\begin{thebibliography}{}
\expandafter\ifx\csname natexlab\endcsname\relax\def\natexlab#1{#1}\fi
\providecommand{\url}[1]{\href{#1}{#1}}
\providecommand{\dodoi}[1]{doi:~\href{http://doi.org/#1}{\nolinkurl{#1}}}
\providecommand{\doeprint}[1]{\href{http://ascl.net/#1}{\nolinkurl{http://ascl.net/#1}}}
\providecommand{\doarXiv}[1]{\href{https://arxiv.org/abs/#1}{\nolinkurl{https://arxiv.org/abs/#1}}}

\bibitem[{{Abbett} \& {Hawley}(1999)}]{1999ApJ...521..906A}
{Abbett}, W.~P., \& {Hawley}, S.~L. 1999, \apj, 521, 906,
  \dodoi{10.1086/307576}

\bibitem[{{Aboudarham} \& {Henoux}(1986)}]{1986A&A...168..301A}
{Aboudarham}, J., \& {Henoux}, J.~C. 1986, \aap, 168, 301

\bibitem[{{Allred} {et~al.}(2020){Allred}, {Alaoui}, {Kowalski}, \&
  {Kerr}}]{2020ApJ...902...16A}
{Allred}, J.~C., {Alaoui}, M., {Kowalski}, A.~F., \& {Kerr}, G.~S. 2020, \apj,
  902, 16, \dodoi{10.3847/1538-4357/abb239}

\bibitem[{{Allred} {et~al.}(2005){Allred}, {Hawley}, {Abbett}, \&
  {Carlsson}}]{2005ApJ...630..573A}
{Allred}, J.~C., {Hawley}, S.~L., {Abbett}, W.~P., \& {Carlsson}, M. 2005,
  \apj, 630, 573, \dodoi{10.1086/431751}

\bibitem[{{Allred} {et~al.}(2015){Allred}, {Kowalski}, \&
  {Carlsson}}]{2015ApJ...809..104A}
{Allred}, J.~C., {Kowalski}, A.~F., \& {Carlsson}, M. 2015, \apj, 809, 104,
  \dodoi{10.1088/0004-637X/809/1/104}

\bibitem[{{Andretta} \& {Jones}(1997)}]{1997ApJ...489..375A}
{Andretta}, V., \& {Jones}, H.~P. 1997, \apj, 489, 375, \dodoi{10.1086/304760}

\bibitem[{{Arnaud} \& {Rothenflug}(1985)}]{1985A&AS...60..425A}
{Arnaud}, M., \& {Rothenflug}, R. 1985, \aaps, 60, 425

\bibitem[{{Avrett} {et~al.}(1994){Avrett}, {Fontenla}, \&
  {Loeser}}]{1994IAUS..154...35A}
{Avrett}, E.~H., {Fontenla}, J.~M., \& {Loeser}, R. 1994, in Infrared Solar
  Physics, ed. D.~M. {Rabin}, J.~T. {Jefferies}, \& C.~{Lindsey}, Vol. 154, 35

\bibitem[{{Battaglia} {et~al.}(2009){Battaglia}, {Fletcher}, \&
  {Benz}}]{2009A&A...498..891B}
{Battaglia}, M., {Fletcher}, L., \& {Benz}, A.~O. 2009, \aap, 498, 891,
  \dodoi{10.1051/0004-6361/200811196}

\bibitem[{{Brown} {et~al.}(2018){Brown}, {Fletcher}, {Kerr}, {Labrosse},
  {Kowalski}, \& {De La Cruz Rodr{\'\i}guez}}]{2018ApJ...862...59B}
{Brown}, S.~A., {Fletcher}, L., {Kerr}, G.~S., {et~al.} 2018, \apj, 862, 59,
  \dodoi{10.3847/1538-4357/aacc29}

\bibitem[{{Cao} {et~al.}(2010){Cao}, {Gorceix}, {Coulter}, {Ahn}, {Rimmele}, \&
  {Goode}}]{2010AN....331..636C}
{Cao}, W., {Gorceix}, N., {Coulter}, R., {et~al.} 2010, Astronomische
  Nachrichten, 331, 636, \dodoi{10.1002/asna.201011390}

\bibitem[{{Carlsson} \& {Stein}(1992)}]{1992ApJ...397L..59C}
{Carlsson}, M., \& {Stein}, R.~F. 1992, \apjl, 397, L59, \dodoi{10.1086/186544}

\bibitem[{{Carlsson} \& {Stein}(1995)}]{1995ApJ...440L..29C}
---. 1995, \apjl, 440, L29, \dodoi{10.1086/187753}

\bibitem[{{Carlsson} \& {Stein}(1997)}]{1997ApJ...481..500C}
---. 1997, \apjl, 481, 500

\bibitem[{{Centeno} {et~al.}(2008){Centeno}, {Trujillo Bueno}, {Uitenbroek}, \&
  {Collados}}]{2008ApJ...677..742C}
{Centeno}, R., {Trujillo Bueno}, J., {Uitenbroek}, H., \& {Collados}, M. 2008,
  \apj, 677, 742, \dodoi{10.1086/528680}

\bibitem[{{Cheung} {et~al.}(2019){Cheung}, {Rempel}, {Chintzoglou}, {Chen},
  {Testa}, {Mart{\'\i}nez-Sykora}, {Sainz Dalda}, {DeRosa}, {Malanushenko},
  {Hansteen}, {De Pontieu}, {Carlsson}, {Gudiksen}, \&
  {McIntosh}}]{2019NatAs...3..160C}
{Cheung}, M.~C.~M., {Rempel}, M., {Chintzoglou}, G., {et~al.} 2019, Nature
  Astronomy, 3, 160, \dodoi{10.1038/s41550-018-0629-3}

\bibitem[{{De Pontieu} {et~al.}(2014){De Pontieu}, {Title}, {Lemen}, {Kushner},
  {Akin}, {Allard}, {Berger}, {Boerner}, {Cheung}, {Chou}, {Drake}, {Duncan},
  {Freeland}, {Heyman}, {Hoffman}, {Hurlburt}, {Lindgren}, {Mathur}, {Rehse},
  {Sabolish}, {Seguin}, {Schrijver}, {Tarbell}, {W{\"u}lser}, {Wolfson},
  {Yanari}, {Mudge}, {Nguyen-Phuc}, {Timmons}, {van Bezooijen}, {Weingrod},
  {Brookner}, {Butcher}, {Dougherty}, {Eder}, {Knagenhjelm}, {Larsen},
  {Mansir}, {Phan}, {Boyle}, {Cheimets}, {DeLuca}, {Golub}, {Gates}, {Hertz},
  {McKillop}, {Park}, {Perry}, {Podgorski}, {Reeves}, {Saar}, {Testa}, {Tian},
  {Weber}, {Dunn}, {Eccles}, {Jaeggli}, {Kankelborg}, {Mashburn}, {Pust},
  {Springer}, {Carvalho}, {Kleint}, {Marmie}, {Mazmanian}, {Pereira}, {Sawyer},
  {Strong}, {Worden}, {Carlsson}, {Hansteen}, { aarts}, {Wiesmann}, {Aloise},
  {Chu}, {Bush}, {Scherrer}, {Brekke}, {Martinez-Sykora}, {Lites}, {McIntosh},
  {Uitenbroek}, {Okamoto}, {Gummin}, {Auker}, {Jerram}, {Pool}, \&
  {Waltham}}]{2014SoPh..289.2733D}
{De Pontieu}, B., {Title}, A.~M., {Lemen}, J.~R., {et~al.} 2014, \solphys, 289,
  2733, \dodoi{10.1007/s11207-014-0485-y}

\bibitem[{{Dere} {et~al.}(1997){Dere}, {Landi}, {Mason}, {Monsignori Fossi}, \&
  {Young}}]{1997A&AS..125..149D}
{Dere}, K.~P., {Landi}, E., {Mason}, H.~E., {Monsignori Fossi}, B.~C., \&
  {Young}, P.~R. 1997, \aaps, 125, \dodoi{10.1051/aas:1997368}

\bibitem[{{Ding} {et~al.}(2005){Ding}, {Li}, \& {Fang}}]{2005A&A...432..699D}
{Ding}, M.~D., {Li}, H., \& {Fang}, C. 2005, \aap, 432, 699,
  \dodoi{10.1051/0004-6361:20041366}

\bibitem[{{Ding} {et~al.}(2003){Ding}, {Liu}, {Yeh}, \&
  {Li}}]{2003A&A...403.1151D}
{Ding}, M.~D., {Liu}, Y., {Yeh}, C.~T., \& {Li}, J.~P. 2003, \aap, 403, 1151,
  \dodoi{10.1051/0004-6361:20030428}

\bibitem[{{Dorfi} \& {Drury}(1987)}]{1987JCoPh..69..175D}
{Dorfi}, E.~A., \& {Drury}, L.~O. 1987, Journal of Computational Physics, 69,
  175, \dodoi{10.1016/0021-9991(87)90161-6}

\bibitem[{Du \& Li(2008)}]{Du_2008}
Du, Q.-S., \& Li, H. 2008, Chinese Journal of Astronomy and Astrophysics, 8,
  723, \dodoi{10.1088/1009-9271/8/6/12}

\bibitem[{{Fang} {et~al.}(1993){Fang}, {Henoux}, \&
  {Gan}}]{1993A&A...274..917F}
{Fang}, C., {Henoux}, J.~C., \& {Gan}, W.~Q. 1993, \aap, 274, 917

\bibitem[{{Fletcher} \& {Hudson}(2008)}]{2008ApJ...675.1645F}
{Fletcher}, L., \& {Hudson}, H.~S. 2008, \apj, 675, 1645,
  \dodoi{10.1086/527044}

\bibitem[{{Fuhrmeister} {et~al.}(2020){Fuhrmeister}, {Czesla}, {Hildebrandt},
  {Nagel}, {Schmitt}, {Jeffers}, {Caballero}, {Hintz}, {Johnson},
  {Sch{\"o}fer}, {Zechmeister}, {Reiners}, {Ribas}, {Amado}, {Quirrenbach},
  {Nortmann}, {Bauer}, {B{\'e}jar}, {Cort{\'e}s-Contreras}, {Dreizler},
  {Galad{\'\i}-Enr{\'\i}quez}, {Hatzes}, {Kaminski}, {K{\"u}rster}, {Lafarga},
  \& {Montes}}]{2020A&A...640A..52F}
{Fuhrmeister}, B., {Czesla}, S., {Hildebrandt}, L., {et~al.} 2020, \aap, 640,
  A52, \dodoi{10.1051/0004-6361/202038279}

\bibitem[{{Goldberg}(1939)}]{1939ApJ....89..673G}
{Goldberg}, L. 1939, \apj, 89, 673, \dodoi{10.1086/144092}

\bibitem[{{Golding} {et~al.}(2014){Golding}, {Carlsson}, \&
  {Leenaarts}}]{2014ApJ...784...30G}
{Golding}, T.~P., {Carlsson}, M., \& {Leenaarts}, J. 2014, \apj, 784, 30,
  \dodoi{10.1088/0004-637X/784/1/30}

\bibitem[{{Golding} {et~al.}(2016){Golding}, {Leenaarts}, \&
  {Carlsson}}]{2016ApJ...817..125G}
{Golding}, T.~P., {Leenaarts}, J., \& {Carlsson}, M. 2016, \apj, 817, 125,
  \dodoi{10.3847/0004-637X/817/2/125}

\bibitem[{{Goode} \& {Cao}(2012)}]{2012ASPC..463..357G}
{Goode}, P.~R., \& {Cao}, W. 2012, in Astronomical Society of the Pacific
  Conference Series, Vol. 463, Second ATST-EAST Meeting: Magnetic Fields from
  the Photosphere to the Corona., ed. T.~R. {Rimmele}, A.~{Tritschler},
  F.~{W{\"o}ger}, M.~{Collados Vera}, H.~{Socas-Navarro}, R.~{Schlichenmaier},
  M.~{Carlsson}, T.~{Berger}, A.~{Cadavid}, P.~R. {Gilbert}, P.~R. {Goode}, \&
  M.~{Kn{\"o}lker}, 357

\bibitem[{{Graham} {et~al.}(2020){Graham}, {Cauzzi}, {Zangrilli}, {Kowalski},
  {Sim{\~o}es}, \& {Allred}}]{2020ApJ...895....6G}
{Graham}, D.~R., {Cauzzi}, G., {Zangrilli}, L., {et~al.} 2020, \apj, 895, 6,
  \dodoi{10.3847/1538-4357/ab88ad}

\bibitem[{{Harvey} \& {Recely}(1984)}]{1984SoPh...91..127H}
{Harvey}, K.~L., \& {Recely}, F. 1984, \solphys, 91, 127,
  \dodoi{10.1007/BF00213619}

\bibitem[{{Henoux} {et~al.}(1990){Henoux}, {Aboudarham}, {Brown}, {van den
  Oord}, \& {van Driel-Gesztelyi}}]{1990A&A...233..577H}
{Henoux}, J.~C., {Aboudarham}, J., {Brown}, J.~C., {van den Oord}, G.~H.~J., \&
  {van Driel-Gesztelyi}, L. 1990, \aap, 233, 577

\bibitem[{{Holman}(2016)}]{2016JGRA..12111667H}
{Holman}, G.~D. 2016, Journal of Geophysical Research (Space Physics), 121,
  11,667, \dodoi{10.1002/2016JA022651}

\bibitem[{{Holman} {et~al.}(2011){Holman}, {Aschwanden}, {Aurass}, {Battaglia},
  {Grigis}, {Kontar}, {Liu}, {Saint-Hilaire}, \&
  {Zharkova}}]{2011SSRv..159..107H}
{Holman}, G.~D., {Aschwanden}, M.~J., {Aurass}, H., {et~al.} 2011, \ssr, 159,
  107, \dodoi{10.1007/s11214-010-9680-9}

\bibitem[{{Huang} {et~al.}(2020){Huang}, {Sadykov}, {Xu}, {Jing}, \&
  {Wang}}]{2020ApJ...897L...6H}
{Huang}, N., {Sadykov}, V.~M., {Xu}, Y., {Jing}, J., \& {Wang}, H. 2020, \apjl,
  897, L6, \dodoi{10.3847/2041-8213/ab9b7a}

\bibitem[{Jeffrey {et~al.}(2018)Jeffrey, Fletcher, Labrosse, \&
  Sim{\~o}es}]{Jeffreyeaav2794}
Jeffrey, N. L.~S., Fletcher, L., Labrosse, N., \& Sim{\~o}es, P. J.~A. 2018,
  Science Advances, 4, \dodoi{10.1126/sciadv.aav2794}

\bibitem[{{Ka{\v{s}}parov{\'a}} {et~al.}(2009){Ka{\v{s}}parov{\'a}}, {Varady},
  {Heinzel}, {Karlick{\'y}}, \& {Moravec}}]{2009A&A...499..923K}
{Ka{\v{s}}parov{\'a}}, J., {Varady}, M., {Heinzel}, P., {Karlick{\'y}}, M., \&
  {Moravec}, Z. 2009, \aap, 499, 923, \dodoi{10.1051/0004-6361/200811559}

\bibitem[{{Kerr} {et~al.}(2019{\natexlab{a}}){Kerr}, {Allred}, \&
  {Carlsson}}]{2019ApJ...883...57K}
{Kerr}, G.~S., {Allred}, J.~C., \& {Carlsson}, M. 2019{\natexlab{a}}, \apj,
  883, 57, \dodoi{10.3847/1538-4357/ab3c24}

\bibitem[{{Kerr} {et~al.}(2020){Kerr}, {Allred}, \&
  {Polito}}]{2020ApJ...900...18K}
{Kerr}, G.~S., {Allred}, J.~C., \& {Polito}, V. 2020, \apj, 900, 18,
  \dodoi{10.3847/1538-4357/abaa46}

\bibitem[{{Kerr} {et~al.}(2019{\natexlab{b}}){Kerr}, {Carlsson}, {Allred},
  {Young}, \& {Daw}}]{2019ApJ...871...23K}
{Kerr}, G.~S., {Carlsson}, M., {Allred}, J.~C., {Young}, P.~R., \& {Daw}, A.~N.
  2019{\natexlab{b}}, \apj, 871, 23, \dodoi{10.3847/1538-4357/aaf46e}

\bibitem[{{Kerr} {et~al.}(2016){Kerr}, {Fletcher}, {Russell}, \&
  {Allred}}]{2016ApJ...827..101K}
{Kerr}, G.~S., {Fletcher}, L., {Russell}, A.~J.~B., \& {Allred}, J.~C. 2016,
  \apj, 827, 101, \dodoi{10.3847/0004-637X/827/2/101}

\bibitem[{{Kobanov} {et~al.}(2018){Kobanov}, {Chelpanov}, \&
  {Pulyaev}}]{2018JASTP.173...50K}
{Kobanov}, N., {Chelpanov}, A., \& {Pulyaev}, V. 2018, Journal of Atmospheric
  and Solar-Terrestrial Physics, 173, 50, \dodoi{10.1016/j.jastp.2018.04.007}

\bibitem[{{Kowalski} {et~al.}(2017){Kowalski}, {Allred}, {Daw}, {Cauzzi}, \&
  {Carlsson}}]{2017ApJ...836...12K}
{Kowalski}, A.~F., {Allred}, J.~C., {Daw}, A., {Cauzzi}, G., \& {Carlsson}, M.
  2017, \apj, 836, 12, \dodoi{10.3847/1538-4357/836/1/12}

\bibitem[{{Kowalski} {et~al.}(2015){Kowalski}, {Hawley}, {Carlsson}, {Allred},
  {Uitenbroek}, {Osten}, \& {Holman}}]{2015SoPh..290.3487K}
{Kowalski}, A.~F., {Hawley}, S.~L., {Carlsson}, M., {et~al.} 2015, \solphys,
  290, 3487, \dodoi{10.1007/s11207-015-0708-x}

\bibitem[{{Kuridze} {et~al.}(2015){Kuridze}, {Mathioudakis}, {Sim{\~o}es},
  {Rouppe van der Voort}, {Carlsson}, {Jafarzadeh}, {Allred}, {Kowalski},
  {Kennedy}, {Fletcher}, {Graham}, \& {Keenan}}]{2015ApJ...813..125K}
{Kuridze}, D., {Mathioudakis}, M., {Sim{\~o}es}, P.~J.~A., {et~al.} 2015, \apj,
  813, 125, \dodoi{10.1088/0004-637X/813/2/125}

\bibitem[{{Kuridze} {et~al.}(2016){Kuridze}, {Mathioudakis}, {Christian},
  {Kowalski}, {Jess}, {Grant}, {Kawate}, {Sim{\~o}es}, {Allred}, \&
  {Keenan}}]{2016ApJ...832..147K}
{Kuridze}, D., {Mathioudakis}, M., {Christian}, D.~J., {et~al.} 2016, \apj,
  832, 147, \dodoi{10.3847/0004-637X/832/2/147}

\bibitem[{{Landi} {et~al.}(2013){Landi}, {Young}, {Dere}, {Del Zanna}, \&
  {Mason}}]{2013ApJ...763...86L}
{Landi}, E., {Young}, P.~R., {Dere}, K.~P., {Del Zanna}, G., \& {Mason}, H.~E.
  2013, \apj, 763, 86, \dodoi{10.1088/0004-637X/763/2/86}

\bibitem[{{Leenaarts} {et~al.}(2016){Leenaarts}, {Golding}, {Carlsson},
  {Libbrecht}, \& {Joshi}}]{2016A&A...594A.104L}
{Leenaarts}, J., {Golding}, T., {Carlsson}, M., {Libbrecht}, T., \& {Joshi}, J.
  2016, \aap, 594, A104, \dodoi{10.1051/0004-6361/201628490}

\bibitem[{{Li} {et~al.}(2006){Li}, {You}, \& {Du}}]{2006SoPh..235..107L}
{Li}, H., {You}, J., \& {Du}, Q. 2006, \solphys, 235, 107,
  \dodoi{10.1007/s11207-006-2094-x}

\bibitem[{{Li} {et~al.}(2007){Li}, {You}, {Yu}, \& {Du}}]{2007SoPh..241..301L}
{Li}, H., {You}, J., {Yu}, X., \& {Du}, Q. 2007, \solphys, 241, 301,
  \dodoi{10.1007/s11207-007-0282-y}

\bibitem[{{Libbrecht} {et~al.}(2020){Libbrecht}, {Bj{\o}rgen}, {Leenaarts}, {de
  la Cruz Rodr{\'\i}guez}, {Hansteen}, \& {Joshi}}]{2020arXiv201015946L}
{Libbrecht}, T., {Bj{\o}rgen}, J.~P., {Leenaarts}, J., {et~al.} 2020, arXiv
  e-prints, arXiv:2010.15946.
\newblock \doarXiv{2010.15946}

\bibitem[{{Libbrecht} {et~al.}(2019){Libbrecht}, {de la Cruz Rodr{\'\i}guez},
  {Danilovic}, {Leenaarts}, \& {Pazira}}]{2019A&A...621A..35L}
{Libbrecht}, T., {de la Cruz Rodr{\'\i}guez}, J., {Danilovic}, S., {Leenaarts},
  J., \& {Pazira}, H. 2019, \aap, 621, A35, \dodoi{10.1051/0004-6361/201833610}

\bibitem[{{Liu} {et~al.}(2013){Liu}, {Xu}, {Deng}, {Lee}, {Zhang}, {Prasad
  Choudhary}, \& {Wang}}]{2013ApJ...774...60L}
{Liu}, C., {Xu}, Y., {Deng}, N., {et~al.} 2013, \apj, 774, 60,
  \dodoi{10.1088/0004-637X/774/1/60}

\bibitem[{{Longcope} {et~al.}(2016){Longcope}, {Qiu}, \&
  {Brewer}}]{2016ApJ...833..211L}
{Longcope}, D., {Qiu}, J., \& {Brewer}, J. 2016, \apj, 833, 211,
  \dodoi{10.3847/1538-4357/833/2/211}

\bibitem[{{Longcope} {et~al.}(2018){Longcope}, {Unverferth}, {Klein},
  {McCarthy}, \& {Priest}}]{2018ApJ...868..148L}
{Longcope}, D., {Unverferth}, J., {Klein}, C., {McCarthy}, M., \& {Priest}, E.
  2018, \apj, 868, 148, \dodoi{10.3847/1538-4357/aaeac4}

\bibitem[{{Longcope} \& {Klimchuk}(2015)}]{2015ApJ...813..131L}
{Longcope}, D.~W., \& {Klimchuk}, J.~A. 2015, \apj, 813, 131,
  \dodoi{10.1088/0004-637X/813/2/131}

\bibitem[{{Machado} {et~al.}(1980){Machado}, {Avrett}, {Vernazza}, \&
  {Noyes}}]{1980ApJ...242..336M}
{Machado}, M.~E., {Avrett}, E.~H., {Vernazza}, J.~E., \& {Noyes}, R.~W. 1980,
  \apj, 242, 336, \dodoi{10.1086/158467}

\bibitem[{{Panos} {et~al.}(2018){Panos}, {Kleint}, {Huwyler}, {Krucker},
  {Melchior}, {Ullmann}, \& {Voloshynovskiy}}]{2018ApJ...861...62P}
{Panos}, B., {Kleint}, L., {Huwyler}, C., {et~al.} 2018, \apj, 861, 62,
  \dodoi{10.3847/1538-4357/aac779}

\bibitem[{{Penn}(2000)}]{2000SoPh..197..313P}
{Penn}, M.~J. 2000, \solphys, 197, 313, \dodoi{10.1023/A:1026510025378}

\bibitem[{{Penn} \& {Kuhn}(1995)}]{1995ApJ...441L..51P}
{Penn}, M.~J., \& {Kuhn}, J.~R. 1995, \apjl, 441, L51, \dodoi{10.1086/187787}

\bibitem[{{Polito} {et~al.}(2018){Polito}, {Testa}, {Allred}, {De Pontieu},
  {Carlsson}, {Pereira}, {Go{\v s}i{\'c}}, \& {Reale}}]{2018ApJ...856..178P}
{Polito}, V., {Testa}, P., {Allred}, J., {et~al.} 2018, \apj, 856, 178,
  \dodoi{10.3847/1538-4357/aab49e}

\bibitem[{{Polito} {et~al.}(2019){Polito}, {Testa}, \& {De
  Pontieu}}]{2019ApJ...879L..17P}
{Polito}, V., {Testa}, P., \& {De Pontieu}, B. 2019, \apjl, 879, L17,
  \dodoi{10.3847/2041-8213/ab290b}

\bibitem[{{Reep} {et~al.}(2018){Reep}, {Polito}, {Warren}, \&
  {Crump}}]{2018ApJ...856..149R}
{Reep}, J.~W., {Polito}, V., {Warren}, H.~P., \& {Crump}, N.~A. 2018, \apj,
  856, 149, \dodoi{10.3847/1538-4357/aab273}

\bibitem[{{Rubio da Costa} \& {Kleint}(2017)}]{2017ApJ...842...82R}
{Rubio da Costa}, F., \& {Kleint}, L. 2017, ApJ, 842, 82,
  \dodoi{10.3847/1538-4357/aa6eaf}

\bibitem[{{Rubio da Costa} {et~al.}(2016){Rubio da Costa}, {Kleint},
  {Petrosian}, {Liu}, \& {Allred}}]{2016ApJ...827...38R}
{Rubio da Costa}, F., {Kleint}, L., {Petrosian}, V., {Liu}, W., \& {Allred},
  J.~C. 2016, \apj, 827, 38, \dodoi{10.3847/0004-637X/827/1/38}

\bibitem[{{Schmidt} {et~al.}(2012){Schmidt}, {Kowalski}, {Hawley}, {Hilton},
  {Wisniewski}, \& {Tofflemire}}]{2012ApJ...745...14S}
{Schmidt}, S.~J., {Kowalski}, A.~F., {Hawley}, S.~L., {et~al.} 2012, \apj, 745,
  14, \dodoi{10.1088/0004-637X/745/1/14}

\bibitem[{{Sim{\~o}es} {et~al.}(2017){Sim{\~o}es}, {Kerr}, {Fletcher},
  {Hudson}, {Gim{\'e}nez de Castro}, \& {Penn}}]{2017A&A...605A.125S}
{Sim{\~o}es}, P. J.~A., {Kerr}, G.~S., {Fletcher}, L., {et~al.} 2017, \aap,
  605, A125, \dodoi{10.1051/0004-6361/201730856}

\bibitem[{{Uitenbroek}(2001)}]{2001ApJ...557..389U}
{Uitenbroek}, H. 2001, \apj, 557, 389, \dodoi{10.1086/321659}

\bibitem[{{van Driel-Gesztelyi} {et~al.}(1994){van Driel-Gesztelyi}, {Hudson},
  {Anwar}, \& {Hiei}}]{1994SoPh..152..145V}
{van Driel-Gesztelyi}, L., {Hudson}, H.~S., {Anwar}, B., \& {Hiei}, E. 1994,
  \solphys, 152, 145, \dodoi{10.1007/BF01473197}

\bibitem[{{Vernazza} {et~al.}(1981){Vernazza}, {Avrett}, \&
  {Loeser}}]{1981ApJS...45..635V}
{Vernazza}, J.~E., {Avrett}, E.~H., \& {Loeser}, R. 1981, \apjs, 45, 635,
  \dodoi{10.1086/190731}

\bibitem[{{Wahlstrom} \& {Carlsson}(1994)}]{1994ApJ...433..417W}
{Wahlstrom}, C., \& {Carlsson}, M. 1994, \apj, 433, 417, \dodoi{10.1086/174654}

\bibitem[{{Warmuth} \& {Mann}(2020)}]{2020A&A...644A.172W}
{Warmuth}, A., \& {Mann}, G. 2020, \aap, 644, A172,
  \dodoi{10.1051/0004-6361/202039529}

\bibitem[{{Xu} {et~al.}(2016){Xu}, {Cao}, {Ding}, {Kleint}, {Su}, {Liu}, {Ji},
  {Chae}, {Jing}, {Cho}, {Cho}, {Gary}, \& {Wang}}]{2016ApJ...819...89X}
{Xu}, Y., {Cao}, W., {Ding}, M., {et~al.} 2016, \apj, 819, 89,
  \dodoi{10.3847/0004-637X/819/2/89}

\bibitem[{{You} \& {Oertel}(1992)}]{1992ApJ...389L..33Y}
{You}, J.~Q., \& {Oertel}, G.~K. 1992, \apjl, 389, L33, \dodoi{10.1086/186342}

\bibitem[{{Zeng} {et~al.}(2014){Zeng}, {Qiu}, {Cao}, \&
  {Judge}}]{2014ApJ...793...87Z}
{Zeng}, Z., {Qiu}, J., {Cao}, W., \& {Judge}, P.~G. 2014, \apj, 793, 87,
  \dodoi{10.1088/0004-637X/793/2/87}

\bibitem[{{Zhu} {et~al.}(2019){Zhu}, {Kowalski}, {Tian}, {Uitenbroek},
  {Carlsson}, \& {Allred}}]{2019ApJ...879...19Z}
{Zhu}, Y., {Kowalski}, A.~F., {Tian}, H., {et~al.} 2019, \apj, 879, 19,
  \dodoi{10.3847/1538-4357/ab2238}

\bibitem[{{Zirin}(1975)}]{1975ApJ...199L..63Z}
{Zirin}, H. 1975, \apjl, 199, L63, \dodoi{10.1086/181849}

\bibitem[{{Zirin}(1980)}]{1980ApJ...235..618Z}
---. 1980, \apj, 235, 618, \dodoi{10.1086/157667}

\end{thebibliography}

\end{document}